\documentclass[sigplan,10pt,manuscript,nonacm]{acmart}
\acmISBN{} 
\acmDOI{} 
\startPage{1}

\usepackage{fancyhdr}

\definecolor{shadecolor}{rgb}{.9, .9, .9}
\newenvironment{verbatimblock}%
   {\snugshade\verbatim}%
   {\endverbatim\endsnugshade}

\setcopyright{none}

\usepackage[parfill]{parskip}

\PassOptionsToPackage{table}{xcolor}

\usepackage{colortbl}
\usepackage{amsmath}

\usepackage{color}
\usepackage{amssymb}
\usepackage{algorithmic}
\usepackage{endnotes}
\usepackage{epsfig}
\usepackage{epstopdf}
\usepackage{caption}
\usepackage{subcaption}
\usepackage{url}
\usepackage{svg}
\usepackage{color}
\colorlet{shadecolor}{gray!30}
\usepackage{graphicx}
\usepackage{multirow}
\usepackage[breakable]{tcolorbox}
\usepackage{framed}

\usepackage{pgfplots,tikz}
\usepackage{tikz-network}
\usetikzlibrary{arrows,automata}
\usetikzlibrary{external}
\pgfplotsset{compat=1.16}
\usepackage{wasysym}
\usepackage{listings}
\usepackage{bm}
\usepackage{relsize}
\usepackage{paralist}
\usepackage{url}
\usepackage{comment}
\usepackage{mdframed}
\usepackage{booktabs}
\usepackage{wrapfig}
\usepackage[toc,page]{appendix}
\usepackage{wasysym}
\usepackage{listings}
\usepackage{setspace}
\usepackage{float}
\usepackage{longtable}
\usepackage[nodayofweek]{datetime}
\usepackage{setspace}
\usepackage{tabularx}
\usepackage[inline,shortlabels]{enumitem}

\definecolor{lbcolor}{rgb}{0.9,0.9,0.9} 
\usetikzlibrary{backgrounds}
\usetikzlibrary{patterns}

\newcommand*\wrapletters[1]{\wr@pletters#1\@nil}
\def\wr@pletters#1#2\@nil{#1\allowbreak\if&#2&\else\wr@pletters#2\@nil\fi}

\newcommand{\code}[1]{\sloppy{\texttt{#1}}}

\usepackage{xargs}

\newcommandx{\changethis}[2][1=]{\todo[linecolor=red,backgroundcolor=red!25,bordercolor=red,#1]{#2}}
\newcommandx{\changefixed}[2][1=]{\todo[linecolor=green,backgroundcolor=green!25,bordercolor=green,#1]{#2}}
\newcommandx{\unsure}[2][1=]{\todo[linecolor=purple,backgroundcolor=purple!25,bordercolor=purple,#1]{#2}}
\newcommandx{\thiswillnotshow}[2][1=]{\todo[disable,#1]{#2}}

\begin{document}

\title{Multi-Lingual Development \& Programming Languages Interoperability: An Empirical Study}
\titlenote{Supported by Len Blavatnik and the Blavatnik Family foundation}

\author{Tsvi Cherny-Shahar}
\affiliation{
  \department{Blavatnik School of Computer Science} 
  \institution{Tel Aviv University}                 
  \country{Israel}
}
\email{tsvic@mail.tau.ac.il}                        
\author{Amiram Yehudai}
\affiliation{
  \department{Blavatnik School of Computer Science} 
  \institution{Tel Aviv University}                 
  \country{Israel}
}
\email{amiramy@tau.ac.il}

\begin{abstract}
As part of a research on a novel in-process multiprogramming-
language interoperability system, this study investigates the interoperability and usage of multiple programming languages within a large dataset of GitHub projects and Stack Overflow Q\&A. It addresses existing multi-lingual development practices and interactions between programming languages, focusing on in-process multi-programming language interoperability. The research examines a dataset of 414,486 GitHub repositories, 22,156,001 Stack Overflow questions from 2008-2021 and 173 interoperability tools. The paper's contributions include a comprehensive dataset, large-scale analysis, and insights into the prevalence, dominant languages, interoperability tools, and related issues in multi-language programming.  The paper presents the research results, shows that C is a central pillar in programming language interoperability, and outlines \emph{simple interoperability} guidelines. These findings and guidelines contribute to our multi-programming language interoperability system research, also laying the groundwork for other systems and tools by suggesting key features for future interoperability tools.
\end{abstract}

\keywords{multi-lingual, cross-language, interoperability, foreign-function-interface, empiric}

\maketitle

\section{Introduction}
multi-lingual development and interoperability of programming languages have existed for many years and are a known practices \cite{issue_challenges_solutions} \cite{xlang_survey} \cite{behind_the_scenes} \cite{multilingual_systems_constructed}  \cite{understanding_lang_selection} \cite{polyglot}.  As part of a research on a novel in-process multi-programming-language interoperability system, we want to survey the existing usage and interactions between programming languages. While these topics have been studied 
before in multiple articles (e.g. \cite{issue_challenges_solutions} \cite{multilingual_systems_constructed} \cite{vulnerability_proneness_multilingual} \cite{impact_of_interlanguage_dependencies}), our research focuses on programming languages (instead of any software language) and in-process interoperability. Also, this research is done on a large dataset of GitHub \cite{github} projects and StackOverflow \cite{stackoverflow} Q\&A questions. Both GitHub and StackOverflow are known sources and have been used for empirical studies in the past (e.g. \cite{large_scale_study_github}, \cite{impact_of_interlanguage_dependencies}, \cite{issue_challenges_solutions}). The discussion in this paper addresses the related work. 

The \textbf{first contribution} is a dataset of 414,486 open source projects (also called repositories interchangeably) metadata from GitHub, and 22,156,001 Q\&A questions from StackOverflow website. The projects and questions dating from 2008 to the end of 2021, as the data acquisition was done at the beginning of 2022. The datasets are available using links shown in section \ref{sec:data_collection}.

The \textbf{second contribution} is the analysis we have performed on a large dataset and the answer to the research questions detailed below. As part of the research, we want to check if the problem is really wide spread on a large-scale dataset to validate the relevant finding of previous studies conducted by others on smaller scale datasets. Although we focus on findings for an interoperability system research, we extract statistics and discuss the results of previous work to compare to the large dataset findings. As the dataset is huge, in some metrics we cannot perform fine-tune analysis as done in previous work. Therefore, we are taking different approaches to mitigate the scale of the dataset.

This study answers the following research questions on a large-scale dataset:

\textbf{$RQ_1$:} \emph{How common is multi-lingual and multi-PL development?}

\textbf{$RQ_2$:} \emph{What are the dominant languages and programming languages?}

\textbf{$RQ_3$:} \emph{Which programming languages are mostly used together? and which binding mechanisms?}

\textbf{$RQ_4$:} \emph{What are the common interoperability tools?}

\textbf{$RQ_5$:} \emph{How many issues and discussions relate to multi-PL?}

By answering the research questions, we can validate the results done by previous work and understand better how developers use multi-PL development, which PLs and tools they use, and how many issues and discussions are there on this topic. These results allow us to better define the required features for future interoperability tools, in order to provide simpler and more intuitive multi-PL programming.

As noted above, in order to answer our research questions, we have conducted an empirical study on GitHub\cite{github} source repository and Stack Overflow \cite{stackoverflow} Q\&A website. GitHub and Stack Overflow were chosen as they are prevalent in their fields \cite{github_rank}\cite{stackoverflow_rank}. We have analyzed 414,486 GitHub projects (also called repositories interchangeably), 22,156,001 Stack Overflow questions, dating from 2008 to 2021, and 173 interoperability tools to understand the current usage of multi-PL. Also, we analyze the GitHub projects' issue boards and discussion groups to detect issues relating to multi-PL. Our focus is on programming languages and techniques of interoperability, as opposed to other languages used in software development, not strictly for writing the program instructions and logic. Hence, we discuss and classify a list of programming languages while
taking into account existing classifications \cite{list_of_pl_wiki}\cite{github_linguist}  and we catalog different interoperability techniques of programming languages based on our findings.\\
As mentioned earlier, although similar studies have been conducted in the past, our novelty lies in 
\begin{itemize*}
    \item a significantly larger dataset over a longer period
    \item focusing the study on multi-PL programming, which is different from "multi-lingual"
    \item classification of PL compared to other existing classifications \cite{github_linguist}\cite{list_of_pl_wiki} used in previous work (e.g. \cite{polyglot}, \cite{100k_opensource} \cite{large_scale_study_github})
    \item  classifying binding types of interoperability tools on a large scale dataset using a different approach from previous work (e.g. \cite{multilingual_systems_constructed} \cite{polyfax}) 
    \item showing that C is a central pillar in programming language interoperability
    \item defining a guideline for interoperability tools named \emph{simple interoperability}
\end{itemize*}

Section \ref{sec:related_work} discusses previous and related work. Section \ref{sec:data_collection} details the datasets and how we have collected them. Section \ref{sec:prog_lang_classification} details how we classify which languages are programming languages. Section \ref{sec:terminology} explains the terminology used throughout the paper. Section \ref{sec:multilang_result} presents the finding and answers to the research questions. Section \ref{sec:threats_to_validity} shows threats to validity. Section \ref{sec:discussion} discusses the implications of the research questions' answers and presents some insights. Section \ref{sec:conclusions} concludes the study, its results and the discussion of the paper.

\section{Related Work}\label{sec:related_work}

As we perform the data acquisition and analysis at the beginning of 2022, several relevant papers have been published after we conducted our study.

In 2024, \cite{multilingual_systems_constructed}, Li et al. analyzed a curated dataset of 7,113 GitHub projects. As part of the paper, the authors analyze language interface mechanisms using PolyFax \cite{polyfax}, a tool that detects, among other things, language interface mechanisms in open-source projects. In order to detect interface mechanisms, PolyFax uses a predefined patterns of "top languages" to detect the different mechanisms. \cite{multilingual_systems_constructed} shows the type of language interface used based on the combinations of programming languages that are selected within the analyzed projects. PolyFax defines four types of interfaces:
\begin{enumerate*}
\item Implicit Invocation (IMI) - out-of-process interface (e.g. shell, Remote Procedure Call (RPC) via network, etc.)
\item Foreign Function Interface (FFI) - a programming language calls a function in another programming language (e.g. CTypes \cite{python_ctypes}, JNI \cite{jni})
\item Embodiment (EDB) - Languages co-exist and interdependent (e.g. css \cite{css_rfc} with HTML \cite{html_rfc})
\item Hidden Interaction (HIT) - No interaction found
\end{enumerate*}.


Note that excluding combinations of a PL with "shell" language (e.g., python-shell) from the findings, all remaining combinations listed utilize FFI either by itself or along in conjunction with out-of-process interfaces (except ruby-swift, which no interaction mechanism was detected). Reintroducing the PL-"shell" combinations reveals an out-of-process interface type, which is logical since the application typically runs the shell as a separate process or the shell runs applications as separate processes. Although the paper addresses the question of "who is the main language in the combination", it does not indicate which programming language initiates the binding. It is important as it implies the direction of the interoperability, which we want to explore to understand if there is an intermediate language or all languages are inter-connected directly.  As we perform a similar analysis, besides analyzing a larger dataset, unlike PolyFax, our research  differentiates between RPC and shell-binding (i.e. runs on a single computer). PolyFax tags both bindings as \emph{implicit invocation} (IMI) to signify "out-of-process invocation", while they are very different. RPC might run on multiple computers on which shell binding remains on the same computer. In addition, we look for specific usage of 173 different interoperability tools we have manually analyzed, which indicates the direction of the interoperability and on a much larger scale (details in section \ref{sec:multilang_result}. Specifically $RQ_3$ and $RQ_4$)

\cite{multilingual_systems_constructed} presents a list of the top language combinations in 7,113 GitHub projects (TLCO). We have performed a similar analysis, but in addition to being much larger, we also differentiate by the scale of the data and how we look at the combinations. While \cite{multilingual_systems_constructed} looks at the projects combinations as a whole, our focus is on the existing combinations of each pair of languages called friendships. For example, if a project $P_1$ contains languages $L_1$, $L_2$ and $L_3$ and $P_2$ contains $L_2$ and $L_3$, in \cite{multilingual_systems_constructed} these are different combinations. In our calculation, we would count two connections (two edges) from $L_2$ and $L_3$ rather than combinations of all the languages.  This insight shows the most connected languages (i.e. pairs), rather the combination (i.e. all languages in the project). We do so because we do not think that the combinations of the whole project are the main factor for interoperability, but the pairs of languages. We also realize that a pair of languages residing in the same project does not mean they interoperate, therefore we perform the tool usage code search. The tool usage analysis also adds direction to the interoperability. For example, if $P_3$ contains $L_1$ and $L_2$ and the analysis detects that $L_1$ calls $L_2$, then we add a directed edge $L_1 \rightarrow  L_2$ ($L_1 \rightarrow L_2 \neq L_1 \rightarrow L_2$).

In \cite{demystifying_issues} and  \cite{issue_challenges_solutions}  586 StackOverflow questions that are highly related to multi-lingual development are analyzed. The results relates to our $RQ_5$, but while the papers provide an in-depth analysis of the posts and insight into their suggested solutions, we are interested in the percentage of posts relating to multiple programming languages and interoperability tools to understand how common it is compared to the other issues in Stack Overflow. Also, we have collected GitHub issues and discussions from the projects in the GitHub repository to get a glimpse of the percentage of multi-lingual related issues compared to others. 

In 2013, Bissyandé et al. \cite{100k_opensource} surveyed 100k projects from GitHub and asked several research questions regarding programming languages and their implications. The paper's $RQ_2$ is relevant to our study, and it asks, "How many projects are written in more than one programming language, and what is the degree of interoperability of each language towards the others?". \cite{100k_opensource} concludes that C plays a central role in interoperability, which our study confirms on a larger dataset and a longer period. A noticeable limitation in \cite{100k_opensource} is the programming languages analyzed within the 100k projects, since "only" 30 popular programming languages were selected, while the number of programming languages is much larger. Another limitation is the usage of GitHub as the only data source.

A survey taken in 2017 by Mayer et al. \cite{xlang_survey} questioning professional software developers about cross-language development stated, "Over 90\% of respondents reported problems related to cross-language linking" and concluded "We suggest that future practical as well as research efforts..." will develop "...better techniques for cross-language linking for improved changeability and understandability". The survey also shows that "more than one language" in a project is quite common. It is essential to point out that \cite{xlang_survey} does not clearly distinguish between languages and programming languages.
Our study validates this conclusion on programming languages using large datasets from GitHub and StackOverflow.

Tomassetti and Torchiano \cite{polyglot} performed, in 2014, an analysis on 15k GitHub projects focusing on multi-lingual (i.e. polyglot-ism) projects. The research classifies the languages into their types, among them is programming. The classification mainly follows \cite{github_linguist}, which we have some concerns about (detailed in section \ref{sec:prog_lang_classification}). \cite{polyglot} answers the questions "What is the level of polyglot-ism of open source projects?" and "Which are the typical pairs of interacting languages?". Tomassetti and Torchiano found that more than 90\% of the projects use more than one language, and more than 50\% of the projects employ more than one programming language. Although our study validates \cite{polyglot} on a larger dataset, our PL classification set is more strict, as detailed in section \ref{sec:prog_lang_classification}. In addition, our study further investigates the usage of multi-PL.

In \cite{lang_interaction}, Tomassetti et al. classify semantic interactions between languages in the multi-lingual project, identifying six categories. Whereas our study also classifies the interaction between languages, \cite{lang_interaction} discusses semantic interaction, as our study interactions discuss the mechanisms which bind PLs together.

\section{Datasets and Collection}\label{sec:data_collection}
This study involves two different datasets and a sub-dataset. In some research questions, we reduce the dataset due to the effort required to acquire and analyze the complete dataset.

Links to the material are available in section \ref{sec:data_availability}.

\subsection{Stack Overflow Dataset}
To acquire Stack Overflow questions from 2014 to 2021, we have downloaded the Stack Overflow data dump from archive.org \cite{stackoverflow_data_dump}. To acquire data from 2008 to 2014, we use Stack Exchange Data Explorer (SEDE) \cite{sede}. From a list of 54,466,905 acquired posts, we have extracted 22,156,001 questions. We store each question's title and other metadata (detailed in the Supplements). We extracted languages, PLs, and interoperability-related tags for each question. Some tags are aliases for others (e.g., \code{C++11} and \code{C++}). Therefore, we are mapping alias tags to their original tag (e.g., \code{C++11} $\rightarrow$ \code{C++}) to avoid duplications. The list of interoperability-related tags is crafted manually, and it contains tags relating to multi-lingual and interoperability tools (e.g., \code{ctypes}, \code{jni}). The lists are detailed in the accompanying supplements document.

\subsection{GitHub Dataset}
Using GitHub API \cite{github_repo_api}, we have acquired the metadata of all public GitHub projects with more than 50 stars. Due to the large GitHub dataset, we try to remove "noise projects", such as personal test projects. We decided to filter the projects by having at least 50 stars. Notice that any filtering methods that GitHub API does not support become almost infeasible due to the number of projects and GitHub's rate limiting. As stars might indicate some public recognition, we believe that it filters out "noise" and reduces the size of the dataset. In any way, even if our filtering filters out more than "noise", the remaining dataset is still substantial relative to previous studies. Filtering using stars is a known practice in other papers to filter by popularity (e.g. \cite{multilingual_systems_constructed}, \cite{vulnerability_proneness_multilingual}) 
We have acquired 415,736 projects' metadata from 2008 to December 2021.\\
The projects' metadata returned from the GitHub API contains the languages used in the project. Using this data, we are filtering projects with more than 20 languages. Although this filters a small number of projects (0.3\% of projects) and does not affect the results, it does remove "fun projects" noise, such as "Hello world in every computer language" \cite{hello_world_in_every_lang}. After applying these filters, the dataset has 414,486 GitHub projects with more than 50 stars and up to 20 languages. Finally, we remove 27,879 projects without any language. These are usually projects that contain only markdowns or other files that GitHub does not recognize.\\
After applying all filters, the GitHub dataset contains 386,607 projects with at least one language, up to 20 languages, and more than 50 stars.

\subsection{In-Code Analysis GitHub Dataset}
Cloning and indexing 414,486 projects is a time and space consuming effort. The website sourcegraph.com \cite{sourcegraph} is a service that indexes public repositories of GitHub, and this index our dataset. Using \cite{sourcegraph}, we can search the source code of the projects in the dataset using terms and regular expressions. We have filtered the results from source graph to repositories until end of 2021 and over 50 stars to match the GitHub dataset.

\subsection{Issues \& Discussions Sub-Dataset}
To find issues related to the interoperability of programming languages, we have acquired issues and discussion groups of the projects we have acquired. Not all projects have discussion and issues. In total, we have acquired issues and discussion groups from 251,025 projects using the GitHub API.

\section{Programming Language Classification}\label{sec:prog_lang_classification}
GitHub linguist project \cite{github_linguist} has classified many languages, but we have many concerns with its classification, as programming languages are overclassified. For example, \cite{github_linguist} classifies \texttt{dockerfile} \cite{dockerfile}, a script to define a container, as a programming language, while we classify it as a non-programming language, as it is not used to write the logic layer of an application. Another example is \texttt{Makefile} \cite{makefile}, which is a script usually used as part of a build process of an application and not as the logic layer of the application.

Another list of programming languages is available on Wikipedia \cite{list_of_pl_wiki}, which according to the authors contains "... index to notable programming languages, in current or historical use. The dialects of BASIC, esoteric programming languages, and markup languages are not included...". From the text and using this list, it is noticeable that many programming languages are missing, such as \texttt{Objective C++}, \texttt{Visual Basic}, and others.

As discussed in section \ref{sec:related_work}, previous work that requires classification of programming languages either used GitHub linguist project \cite{github_linguist} or picked a smaller "well-known" programming language set. Due to that and the concerns we raised above, we have created a more strict list of programming languages, where the general guideline is that \textit{a program uses at least one programming language. A programming language is used to write the business logic of the program.}

GitHub API lists languages used within each project based on the file's extensions. We have constructed a list of 451 languages by unifying all the languages in the acquired projects. For Stack Overflow, we manually construct the set of \emph{all languages} from the list of all Stack Overflow tags available in SEDE \cite{sede}. Every language tag is added to the languages set, which at the end of the process contains 124 languages.

We follow the following procedure separately for the 451 languages from GitHub and the 124 languages from Stack Overflow. We define a subset of languages recognized as \emph{programming languages}, including only languages used to specify the logic of the application. This definition excludes markup languages (e.g., HTML, CSS), data languages (e.g., JSON, YAML), and build-oriented languages (e.g., Make, CMake). We exclude shell languages (e.g., bash, tcsh), often used as external tools and not part of the application's logic. Two researchers individually perform the PL classification process, and only languages that both researchers agree to classify as PL are added to the set of programming languages. We took this conservative approach to ensure that we do not overestimate, e.g., the number of multi-PL projects. Of the 451 languages on GitHub, only 243 are classified as PL. In Stack Overflow, of the 124 languages, 86 were accepted as PL. The existence of two programming language lists is technical due to the different "text" used for the same language.

Comparing our GitHub PL lists to \cite{github_linguist} shows that both lists agree on classifying 299 languages, which is  66.7\% of the GitHub programming languages we have detected. 147 (32.8\%) languages were not classified the same, and two languages were not found in the GitHub file. If we used \cite{github_linguist} definitions of PLs, 50\% of the analyzed GitHub projects would be multi-PL projects (as \cite{polyglot} found using a smaller dataset), while using our definition, it is 35\%. We prefer the more "conservative" definition not to overestimate the multi-PL projects.

Comparing our GitHub programming language classification to Wikipedia's \cite{list_of_pl_wiki} shows that both lists agree on 131 languages to be classified as PL, which is 53.6\% of the programming languages we have detected. On the other hand, the authors of \cite{list_of_pl_wiki} state that the list excludes PLs. Therefore, we expected our list to include some PLs that do not appear in \cite{list_of_pl_wiki}.

Notably, the mismatched classifications are mostly esoteric, as in the worst case, only 2.45\% of the multi-PL projects we detected in GitHub would become non-multi-PL if we chose a different PL classification. Therefore, in terms of the impact on the findings of this study, the list used is insignificant.

\section{Terminology}\label{sec:terminology}
The following terms are used throughout the rest of the paper:
\begin{itemize}
    \item \emph{Single language project} is a GitHub project with exactly one language
    \item \emph{Multi-lingual project} is a GitHub project with at least two languages\vspace{-2mm}
    \item \emph{Multi-PL project} is a GitHub project with at least two programming languages\vspace{-2mm}
    \item \emph{Host language} is a programming language initiating a call to a different programming language\vspace{-2mm}
    \item \emph{Guest/Foreign language} is a programming language implementing code called by a host language\vspace{-2mm}
    \item \emph{Foreign entities} are entities, such as functions, methods, classes etc, implemented in the foreign language\vspace{-2mm}
\end{itemize}

We define \textit{project Friendship} as languages included in the same GitHub project, but to filter out potentially rare friendships, we consider $L_1\rightarrow L_2$ to be a friendship only if at least 10\% of $L_1$'s projects include $L_2$. This definition allows us to understand which languages are used to build a common goal regardless of the technology used to perform the binding.

Next, we define four different types and one subtype of binding mechanisms between languages.
Notice that these definitions differ from the language interactions presented by Tomassetti et al. \cite{lang_interaction} as we define binding mechanisms while \cite{lang_interaction} defines semantic language interaction. The definitions are as follows:
\begin{itemize}
\item \textit{Client/Server binding} defined as languages that communicate using message passing, where the most popular one is over the network, and usually not specific for in-process interoperability\vspace{-2mm}
\item \textit{Interoperate binding} is defined as a language that calls another language. When the binding is in the same OS process, the binding type is found in foreign functions and virtual machines. In many cases, this binding requires loading the target language runtime. When the binding is in different OS processes, the processes communicate using shared memory or message passing.
interoperability binding has an important sub-category:
\vspace{-2mm}
\begin{itemize}
    \item \textit{Indirect interoperability binding} is when language $L_1$ access $L_2$ using an intermediate language $L_i$. For example, if Java accesses Python running as the popular interpreter CPython \cite{cpython}, Java uses JNI \cite{jni} to interoperate with C. C loads the Python interpreter using CPython API. In this case, $L_1$ is Java, $L_2$ is Python, and $L_i$ is C. 
\end{itemize}
\item \textit{Shell binding} is defined as a language calling code in another language by running another OS process via a shell command line.\vspace{-2mm}
\item \textit{Manual binding} is a binding formed when a user (human) executes code in different languages manually, passing data from one program to another. Programs using this binding type are usually called "tools"
\end{itemize}

\section{Study Methodology \& Findings}
\label{sec:multilang_result}
\subsection*{\bm{$RQ_1$}: How common is multi-lingual and multi-PL development?}\label{sec:how_common_is_multi-lingual_development}

\begin{table*}[]
\centering
\begin{tabular}{|c|ll|ll|}
\hline
\multirow{3}{*}{GitHub Projects} & \multicolumn{2}{l|}{Size of Dataset}                            & \multicolumn{2}{l|}{386,607}                                    \\ \cline{2-5} 
                                 & \multicolumn{2}{l|}{Multi-Lingual}                              & \multicolumn{2}{l|}{256,560 (66.4\%)}                           \\ \cline{2-5} 
                                 & \multicolumn{2}{l|}{Multi-PL}                                   & \multicolumn{2}{l|}{146,778 (38.0\%)}                          
                                 \\ \hline \hline
\multirow{4}{*}{Stack Overflow}  & \multicolumn{2}{l|}{}                                           & \multicolumn{1}{l|}{Questions}          & Views                 \\ \cline{2-5} 
                                 & \multicolumn{2}{l|}{Size of Dataset}                            & \multicolumn{1}{l|}{22,156,001}         & 56,534,822,211        \\ \cline{2-5} 
                                 & \multicolumn{2}{l|}{Multi-Lingual}                              & \multicolumn{1}{l|}{2,915,695 (13.2\%)} & 7,396,732,397 (13\%)  \\ \cline{2-5} 
                                 & \multicolumn{2}{l|}{Multi-PL}                                   & \multicolumn{1}{l|}{667,791 (3\%)}      & 1,346,219,383 (2.3\%) \\ \hline
\end{tabular}
\caption{$RQ_1$ Findings Statistics}
\label{tbl:rq1_stats}
\end{table*}

\begin{figure}[h]
    \centering
    \includegraphics[scale=1,width=.9\textwidth]{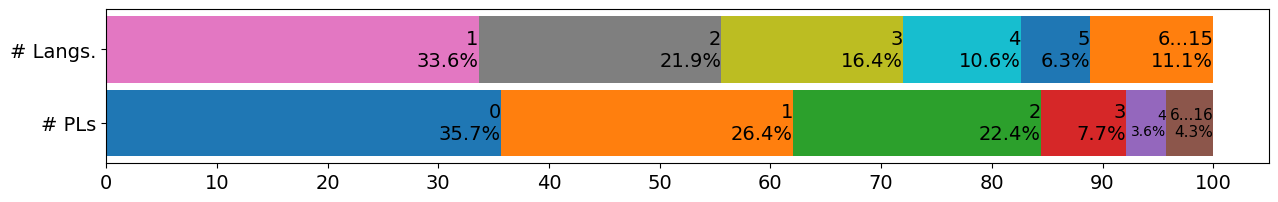}
    
    \caption{GitHub Language Statistics}
    \label{fig:github-number-of-languages}
\end{figure}

To answer this question, in GitHub, we count the languages and programming languages in each GitHub project. Each project with at least two languages is counted as multi-lingual. Every project with at least two programming languages is counted as multi-PL.

In Stack Overflow, we count the tags representing language, programming languages, and those related to PL interoperability (detailed in supplements). Each question containing at least two language tags or interoperability-related is counted as multi-lingual. Each question containing at least two programming languages or a PL interoperability tag is counted as multi-PL.

Table \ref{tbl:rq1_stats} shows statistics on $RQ_1$ findings. In GitHub, 66.4\% of projects are multi-lingual. The average is three languages, and half of the projects have at least two languages. Multi-PL projects consist of 38\% of projects. Figures \ref{fig:github-number-of-languages} break down the percentage of projects by their languages and PLs usage. Interestingly, the percentage of projects with one PL (26.4\%) is almost the same as projects with two PLs (22.4\%). This finding indicates that multi-lingual projects are the most common type of project.

In Stack Overflow, table \ref{tbl:rq1_stats} shows that 13.2\% of the questions contain multiple languages or interoperability-related tags, whereas only 3\% of questions contain multiple PLs or interoperability-related tags. These findings show that Stack Overflow users have questions and challenges revolving around multi-lingual issues. However, we also learn that there aren't many questions regarding multi-PL issues. Also, it does not align with the amount of multi-PL projects in GitHub. We should point out that GitHub and Stack Overflow are not necessarily the same users, nor do the questions reflect issues in GitHub projects.


\subsection*{\bm{$RQ_2$}: What are the dominant languages?} \label{sec:what_are_the_dominant_languages} 

$RQ_1$ shows that multi-lingual and multi-PL development is common. In this $RQ$, we further investigate which languages are used most in GitHub and asked most in Stack Overflow. 

The Figures in this $RQ$ show the percentage of language use in single, multi-lingual, and multi-PL projects. For each project type, we count the number of times each language is used and the total sum of bytes each language uses (as returned by the GitHub API). It is important to note both the count and the size of each language to better understand the language used. Assume language $L_1$ appears in many projects (high count), but its usage in every project is very small (small size). Language $L_2$ appears in fewer projects (low count) but is used heavily in the projects it appears in (big size). If we mind only the count, we overlook the importance of $L_2$. If we mind only the size, we overlook the importance of $L_1$. Therefore, we present both size and count.

\begin{figure}[h]
\centering
    \includegraphics[scale=1,width=.9\textwidth]{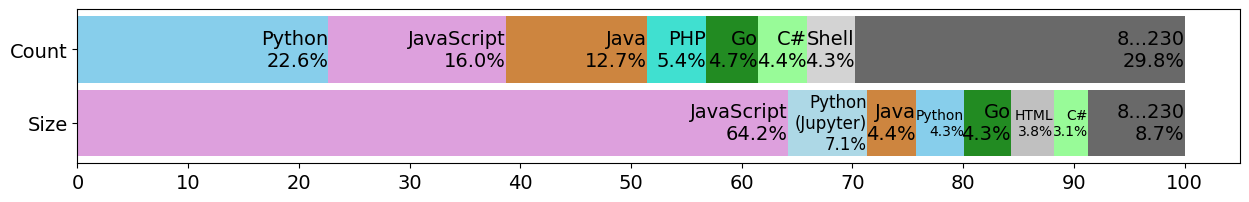}
    \caption{GitHub single language statistics}
    \label{fig:github-languages-percent-single}
\end{figure}

Most single-language projects use programming languages (figure \ref{fig:github-languages-percent-single}), which is expected as a programming language is required to write the application or library logic. Only 8.8\% of the projects have no PLs. The most common language is Python (22.6\%). Checking single-language projects by size reveals that although Python projects are the most common, most of the source code is written in JavaScript (64.2\%), while Python is far behind with 11.4\% (by combining Python and Python in Jupyter). This finding indicates that although developers more commonly pick Python in single-language projects, The largest projects (with a single language) are implemented in JavaScript.

\begin{figure}[h]
\centering
    \includegraphics[scale=1,width=.9\textwidth]{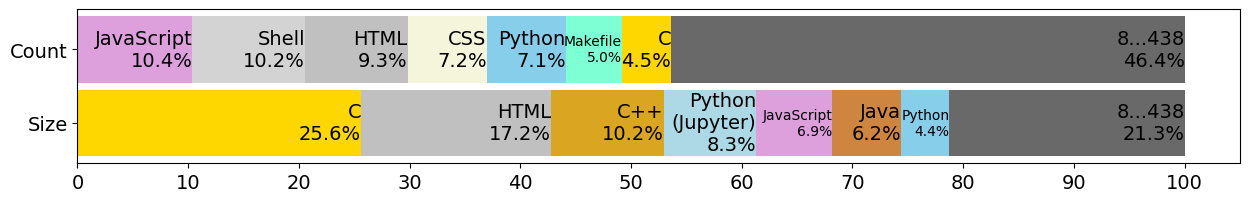}
    \caption{GitHub multi-lingual statistics}
    \label{fig:github-languages-percent-multi-lingual}
\end{figure}

Unlike single-language projects, in multi-lingual projects, the usage of non-programming languages becomes more apparent (figure \ref{fig:github-languages-percent-multi-lingual}), and most prominently in web development. HTML (9.3\%) and CSS (7.2\%) are widely used, as web development projects are usually multi-lingual. When examining language usage in multi-lingual projects by source-code size, we find that C and C++ are heavily used (25.6\% and 10.2\%, respectively). Interestingly, C and C++ projects are less common in a single language, which might align with their relatively low popularity in Stack Overflow developers surveys \cite{stack_overflow_developer_survey_2021}\cite{stack_overflow_developer_survey_2020}\cite{stack_overflow_developer_survey_2019}\cite{stack_overflow_developer_survey_2018}\cite{stack_overflow_developer_survey_2017}. C and C++ together account for more than a third in multi-lingual projects. Our GitHub API counts languages by file extensions, so any embedded/inline usage of C/C++ is not considered. For example, Go files can embed C code.

\begin{figure}[h]
\centering
    \includegraphics[scale=1,width=.9\textwidth]{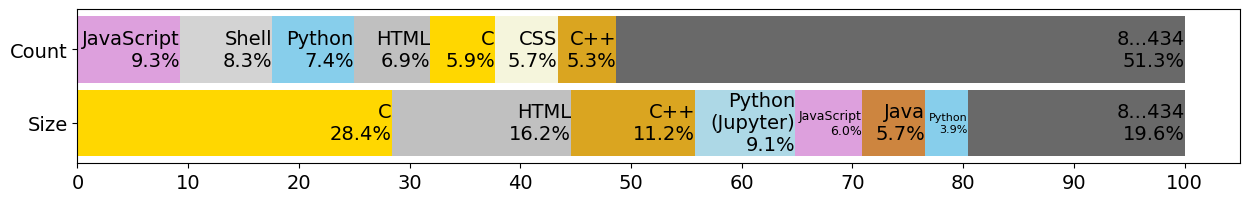}
    \caption{GitHub multi-PL statistics}
    \label{fig:github-languages-percent-multi-pl}
\end{figure}

Investigating further into the multi-PL projects (figure \ref{fig:github-languages-percent-multi-pl}), a subset of multi-lingual projects, we can see a drop in HTML and CSS, from 9.3\% and 7.2\% to 6.9\% and 5.7\% respectively. At the same time, we see an increase in C from 4.5\% to 5.9\%. Analyzing by size, C, and C++ are still heavily used in the projects they appear, and their percentage grows from 25.6\% and 10.2\% to 28.4\% and 11.2\%, respectively, which means almost 40\% of the code written in multi-PL projects in C or C++.

\begin{figure}[h]
\centering
    \includegraphics[scale=1,width=.9\textwidth]{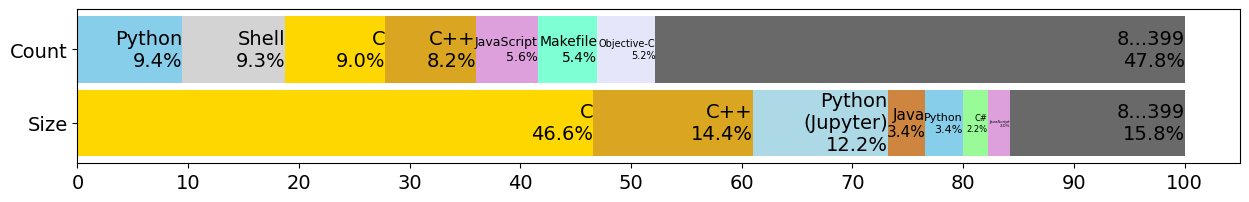}
    \caption{GitHub multi-PL without Web projects programming-languages statistics}
    \label{fig:github-languages-percent-multi-pl-no-web}
\end{figure}

Due to the high count of JavaScript and the size of HTML, JavaScript is popular only because of web-oriented projects. To validate this claim, we create another subset of a project called \emph{Multi-PL without web}, which excludes all the projects that contain either HTML or CSS from the multi-PL set. The set \emph{Multi-PL without web} contains 89,392 projects (i.e., 60.9\% of the multi-PL projects). The statistics are presented in figure \ref{fig:github-languages-percent-multi-pl-no-web}.

The data in figure \ref{fig:github-languages-percent-multi-pl-no-web} shows that in non-web multi-PL projects, C and C++ can be found in many more projects with a count of 9\% and 8.2\% respectively out of all languages. Calculating the C and C++ appearance percentage from all multi-PL without web projects, C and C++ appear in 34.9\% and 31.6\% of the projects, respectively. It shows that C and C++ are very popular in those projects.

Inspecting the languages by size, C is 46\% of the code, and C++ is 14.4\%. After C and C++, we can find Python in Jupyter Notebook with 12.2\%. Right after that, the percentage dives to Java with 3.4\%. This finding clearly shows the importance of C, in particular, and C++ in non-web multi-lingual projects.

Another interesting finding is that Python is the most picked language in multi-PL projects (9.4\%) but with a relatively low percentage of the amount of code (only 3.4\%). This finding may indicate that although Python is used in many multi-lingual projects, the Python part of these projects is relatively small. Last, we also see that shell scripting is quite popular with 10.2\% alongside Makefile with 5.4\% in non-web multi-PL projects. Although we do not investigate and validate further, in multi-lingual projects, the building step becomes more complex and requires much more shell and Makefile scripting.

\begin{figure}[h]
\centering
    \includegraphics[scale=1,width=.9\textwidth]{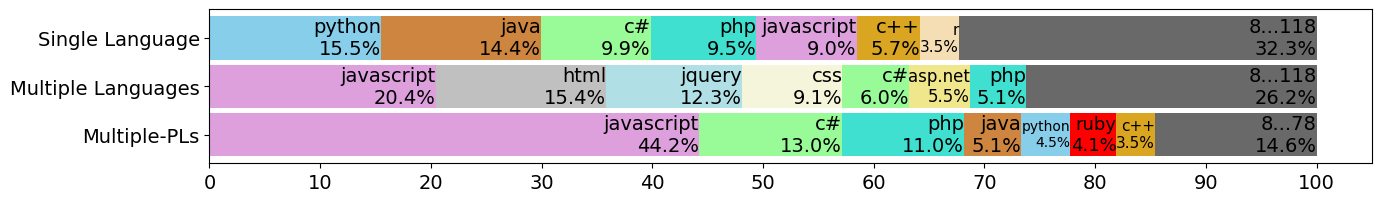}
    \caption{Stack Overflow languages statistics}
    \label{fig:stackoverflow-languages-statistics}
\end{figure}

Stack Overflow language statistics (figure \ref{fig:stackoverflow-languages-statistics}) shows web development-related languages are the most popular questions in multi-lingual questions. According to the 2011 Stack Overflow developers survey \cite{stack_overflow_developer_survey_2021}, 76.8\% of respondents are full-stack or front-end developers. If the survey reflects Stack Overflow's users, it is not surprising that web development is very popular in the Stack Overflow website.

Another notable finding is the appearance and percentage of Microsoft languages in the top 7 languages, like C\# and ASP.NET, that seems less popular in GitHub open source projects. This finding indicates that although GitHub public projects are an excellent and valuable data source, it does not truly reflect the industry, as we believe these questions reflect projects in closed-source environments.

\subsection*{\bm{$RQ_3$}: Which programming languages are mostly used together? and which binding mechanisms?} \label{sec:which_programming_languages_are_mostly_used_together} 

$RQ_1$ and $RQ_2$ show that multi-lingual and multi-PL development is a common practice where C is the most popular programming language among multi-PL projects. This $RQ$ investigates which programming languages are mainly used together. Notice that we are focusing on programming languages since we aim to investigate the logical layer of a program.

\begin{figure}[h]
\centering
    \begin{tikzpicture}[scale=0.8]
    \Vertices[size=0, style={inner sep=0pt}]{figures/github_project_binding_vertices.csv}
    \Edges[color=lightgray]{figures/github_project_binding_edges.csv}
    \end{tikzpicture}
    \caption{GitHub Friendship bindings}
    \label{fig:project-friendship}
\end{figure}
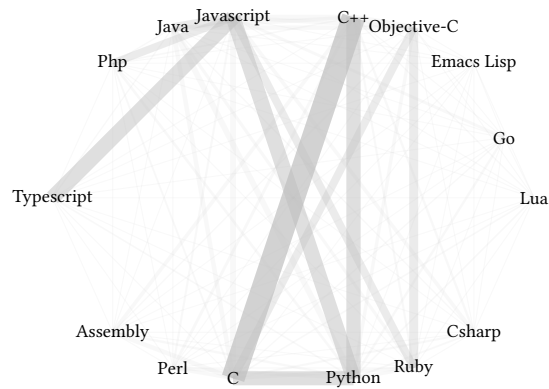

Project friendships presented in figure \ref{fig:project-friendship} show friendship bindings in GitHub. Line thickness indicates the popularity of the binding, the higher the popularity the thicker the line.

The most common friendship is between C and C++, with 13\% of all friendships. This finding is not much of a surprise, as the interoperability between these languages is quite popular and straightforward using \texttt{extern} \code{"C"}. Another expected friendship-binding seen in the data is between JavaScript and TypeScript (8.2\%). The two languages interoperate well together.

The next notable friendship bindings are with Python. In 8\% of the projects, Python shares with C, and 8.11\% with C++. Also, Python friendship binds with JavaScript in 8.3\% of the projects. We claim that Python's friendship with C and C++ is mainly in non-web projects,(i.e., projects that do not contain HTML or CSS files)
while Python's friendship with JavaScript is mainly in web projects. To validate this claim (not prove as we do not check the source code), we check the percentage of web projects. 32.4\% of projects with C and Python friendships are web-projects, 29.3\% of projects with C++ and Python friendships are web-project and 85.4\% of projects with JavaScript and Python friendships are web projects. Therefore, in most cases, Python appears with JavaScript. 

As this study focuses on interoperability, we wish to detect interoperability bindings. In order to determine if a project is using interoperability binding or not, we have to analyze the project's source code. In order to detect and analyze the bindings, we searched snippets of code on GitHub source code indexed by \emph{sourcegraph} \cite{sourcegraph} to detect code that matches different interoperability binding libraries between the languages. For example, to detect C calling to Python using CPython API, we search in C source files for \texttt{"Py\_InitializeEx("} and make sure it is not in a C (single-line) comment.

Text search is not an optimal methodology compared to analyzing the source code abstract syntax trees, but due to the size of the data source and the high cost of acquiring and analyzing such vast amounts of code, searching for snippets in an index database allows us to detect the snippets at a more reasonable cost. As we search for interoperate library, we craft specific snippets of code that are syntax sensitive and indicate the use of the interoperate library or shell command. In total, we covered 173 libraries of the 19 programming languages: Python, C\#, Visual Basic.net, Java, Kotlin, Scala, Groovy, Go, Lua, Ruby, R, Perl, Haskell, JavaScript, Lisp, PowerShell, Bash\footnotemark[1], C and C++. \footnotetext[1]{Although bash is not a programming language in our definition, it is included as it is used to execute guest languages} We chose languages for which we have found \emph{interoperability} tools using search in Google and Stack Overflow. The list of code snippets is available in the supplements.

Besides searching for interoperability binding, we also search for shell binding. Similar to interoperability binding, we search for code snippets, but for shell binding, the code snippets are for executing another language interpreter or JIT. For instance, Python may execute Java using \texttt{os.system('java SomeApp')}. As code snippets for executing shell binding are much more complex than interoperability, we use regular expressions to search for the snippets. Each regular expression is a single snippet for a single language. To make sure the detected snippet executes a binary of a JIT or interpreter, we embed in the regular expression the name (and command line arguments if needed) of the target language binary to reduce false positives. As an example, to detect shell execution in C\#, we are using three different regular expression templates:
\begin{singlespace}
\begin{verbatimblock}
Process\.Start\([^\)]*['"`@]{}['"`@\s]
\\.FileName\s*=.*{}['"`@\s]*;
ProcessStartInfo\([^\)]+{}
\end{verbatimblock}
\end{singlespace}
For each regular expression template, we replace \{\} with the execution command to execute a program using the interpreter or JIT. For example, to detect C\# shell binding with Java, replace \{\} with \code{java}. To ensure we search only C\# source code, we use SourceGraph \cite{sourcegraph} filter for C\# source code: "lang:C\#". We search the regex and the filter with the case-insensitive flag turned on. Next, we iterate all the results of the regex search (using Python script) and ensure we count each project only once, regardless of how many matches were in the project.

An edge case of false positive would be a program with the same executable name of a popular language, for example, naming my application \emph{Python} or \emph{Java}. Also, we realize the regular expression approach can still miss shell bindings, but is does provide a lower bound.

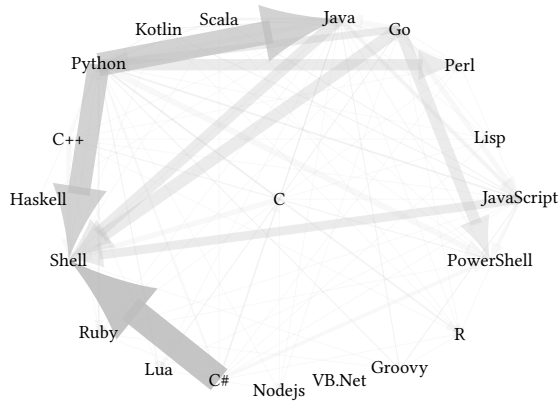
\begin{figure}[h]
\centering
    \begin{tikzpicture}[scale=0.8]
    \Vertices[size=0, style={inner sep=0pt}]{figures/shell-bindings_vertices.csv}
    \Edges[color=lightgray]{figures/shell-bindings_edges.csv}
    \end{tikzpicture}
    \caption{Shell Bindings}
    \label{fig:shell-bindings}
\end{figure}

As the data shows, shell bindings are rare compared to the interoperability libraries, but they are still used. Figure \ref{fig:shell-bindings} shows the detected shell bindings, where each edge represents a binding from the calling language to the called language. The percentages in this paragraph are of the total detected shell bindings (17,069 projects). The most notable shell binding is with the shells (e.g.,/bin/bash, bash.exe, cmd.exe, /bin/tcsh, /bin/zsh), where the most popular ones are detected from C\# (11.2\%). Also, other PLs execute the shell, for instance, Python (9.2\%), Go (5.3\%), and Java (5\%). The most popular shell binding between two PLs is Python $\rightarrow$ Java (9.7\%). Other notable shell bindings are Python $\rightarrow$ Perl (4.8\%), Go $\rightarrow$ Java (2.2\%), Go $\rightarrow$ Python (3.1\%), Java $\rightarrow$ Python (2\%), Go $\rightarrow$ JavaScript (2.1\%) and Java $\rightarrow$ JavaScript (2\%). 
Although shell binding was only found in 16,979 projects, it is still being used. There is no doubt that this methodology could be better, especially compared to a simple function or method call. We propose that developers defer to this methodology due to its simple implementation compared to other interoperate approaches. This finding leads us to conclude that although the alternative of \emph{shell binding} (i.e., \emph{interoperability binding}) is much more ideal, it is only used in some cases. The main benefit of \emph{shell binding} is its simplicity of use, while \emph{interoperability binding} can be complex.

\begin{figure*}[t]
\centering
  \begin{subfigure}[t]{0.4\textwidth}
    \begin{tikzpicture}[scale=0.6]
    \Vertices[size=0, style={inner sep=0pt}]{figures/interop-show-indirection-bindings_vertices.csv}
    \Edges{figures/interop-show-indirection-bindings_edges.csv}
    \end{tikzpicture}
    \caption{Interoperability binding including the intermediate}
    \label{fig:interop-show-indirection-bindings}
  \end{subfigure}
  \kern 13mm
  \begin{subfigure}[t]{0.4\textwidth}
    \begin{tikzpicture}[scale=0.6]
    \Vertices[size=0, style={inner sep=0pt}]{figures/interop-bindings_vertices.csv}
    \Edges{figures/interop-bindings_edges.csv}
    \end{tikzpicture}
    \caption{Interoperability binding excluding the intermediate}
    \label{fig:interop-bindings-no-indirection}
  \end{subfigure}
  \caption{Interoperability Bindings}
  \label{fig:interoperability-bindings}
\end{figure*}
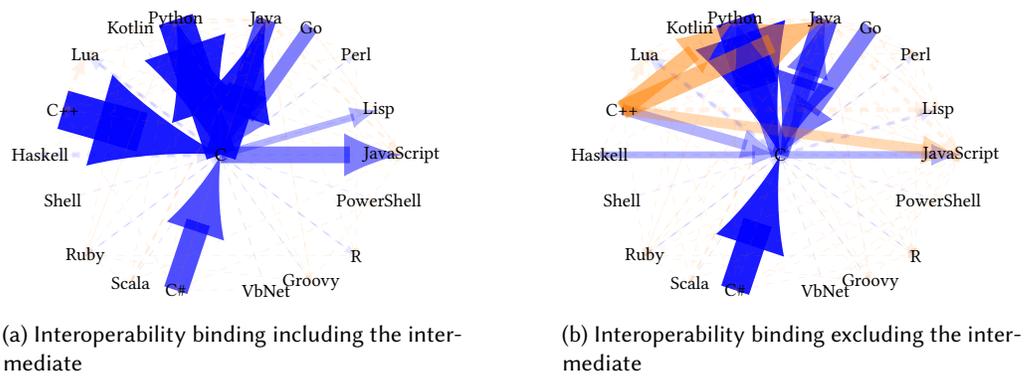

In order to analyze the interoperability binding for each of the 173 interoperability tools, we store the following:
\begin{itemize}
    \item snippets for detection
    \item the source languages
    \item the destination languages
    \item intermediate languages (if there are any)
    \item classification of the tool (detailed in $RQ_4$)
\end{itemize}
We also add snippets for interoperability that do not require $3^{rd}$ party tools (i.e., built-in interoperability tools). Upon a snippet detection, we mark in the graph the path detected where figure \ref{fig:interop-show-indirection-bindings} edges include the intermediate languages, and figure \ref{fig:interop-bindings-no-indirection} edges are from the source language directly to the destination language. Interoperability binding and the percentages presented in this paragraph refer to 147,655 projects detected with interoperability binding.

All edges pointing to or from C, the main intermediate language, are colored blue, and all other edges are colored orange. Rare bindings with usage less than 2.5\% of all detected interoperability bindings are dashed.

In figure \ref{fig:interop-show-indirection-bindings}, which shows the intermediate language, 74.9\% of the bindings are direct, where 67.7\% are to or from C. 25.1\% of the bindings are indirect and 16.4\% of all bindings are indirect via C (65.4\% of all indirect bindings). These findings show that C has a central role in language interoperability. In direct bindings, most of the binds are to/from C, and in indirect bindings, more of the indirect bindings go via C, such that it acts as a hub between languages. In section \ref{sec:discussion}, we discuss further the meaning of these findings.

Figure \ref{fig:interop-bindings-no-indirection} excludes the indirection and directly connects the source language to the destination languages. It is easy to see that C++ interoperate bindings to Python, Java, and JavaScript are the most common indirect binding. The top 5 interoperate bindings are $Python \rightarrow C$ (16\%), $C\# \rightarrow C$ (11.5\%), $C++ \rightarrow Java$ (7.6\%), $C \rightarrow Python$ (7.6\%), $C++ \rightarrow Python$ (6.8\%). The popularity of Python in its interoperability with C and C++ is interesting.

\subsection*{\bm{$RQ_4$}: What are the common interoperability tools?} \label{sec:what_are_the_common_interoperability_tools} 
\begin{figure}[h]
    \centering
    \includegraphics[scale=0.35,width=0.48\textwidth]{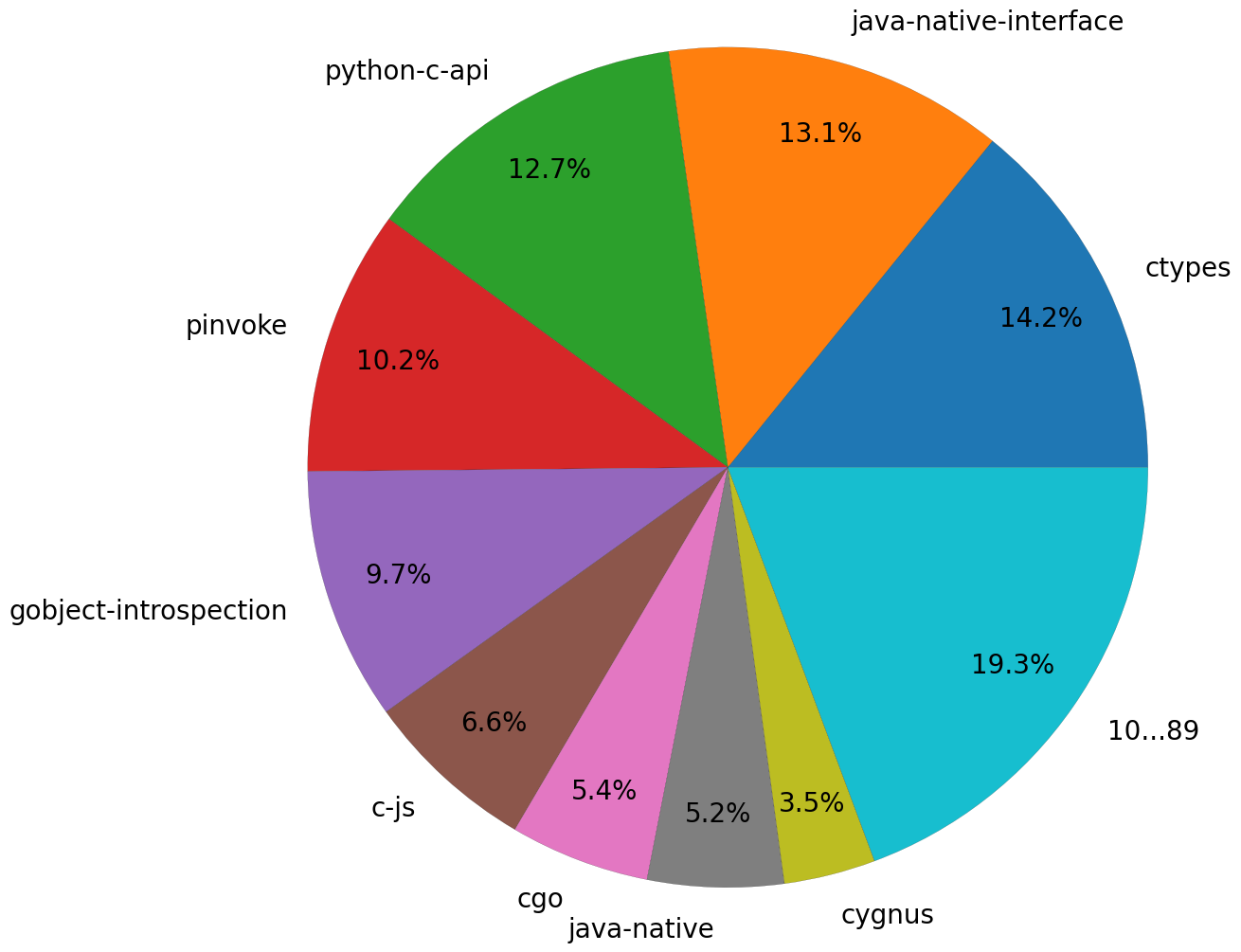}
    \caption{Most common interoperability tools}
    \label{fig:interop_libraries_count}
\end{figure}

Based on the findings presented in $RQ_3$, figure \ref{fig:interop_libraries_count} shows the most common interoperating tools. As the figure shows, most of the bindings are using C, again stressing the importance of C in interoperability. The top interoperate tools are CTypes \cite{python_ctypes} (14.2\%), JNI \cite{jni} (13.1\%), Python C-API \cite{cpython} (12.7\%) and PInvoke \cite{pinvoke} (10.2\%).
As we saw in $RQ_3$, interoperability between C/C++ and Python is common among interoperabilities, making up 26.9\% (CTypes and Python C-API).

By analyzing the interoperability tools detected in $RQ_3$, we define the following classification categories to interoperability tools mechanisms:
\begin{itemize}
\item API: Uses API library to achieve in-process interoperability (e.g., CPython \cite{cpython}, CTypes \cite{python_ctypes})\vspace{-2mm}
\item Compiler: A compiler that compiles the host language to the guest language or intermediate language (e.g., Lispyscript \cite{lispyscript}, Scala.JS \cite{scalajs})\vspace{-2mm}
\item Multi-Process: Executes the guest in a different process and communicates via command line arguments or message passing (e.g., RCaller \cite{rcaller}, RinRuby \cite{rinruby})\vspace{-2mm}
\item Language Port: Reimplemented guest language syntax in the host's runtime environment (e.g., Jython \cite{jython}, NLua \cite{nlua})
\end{itemize}

\begin{figure}[t!]
    \centering
    \includegraphics[scale=0.35,width=0.41\textwidth]{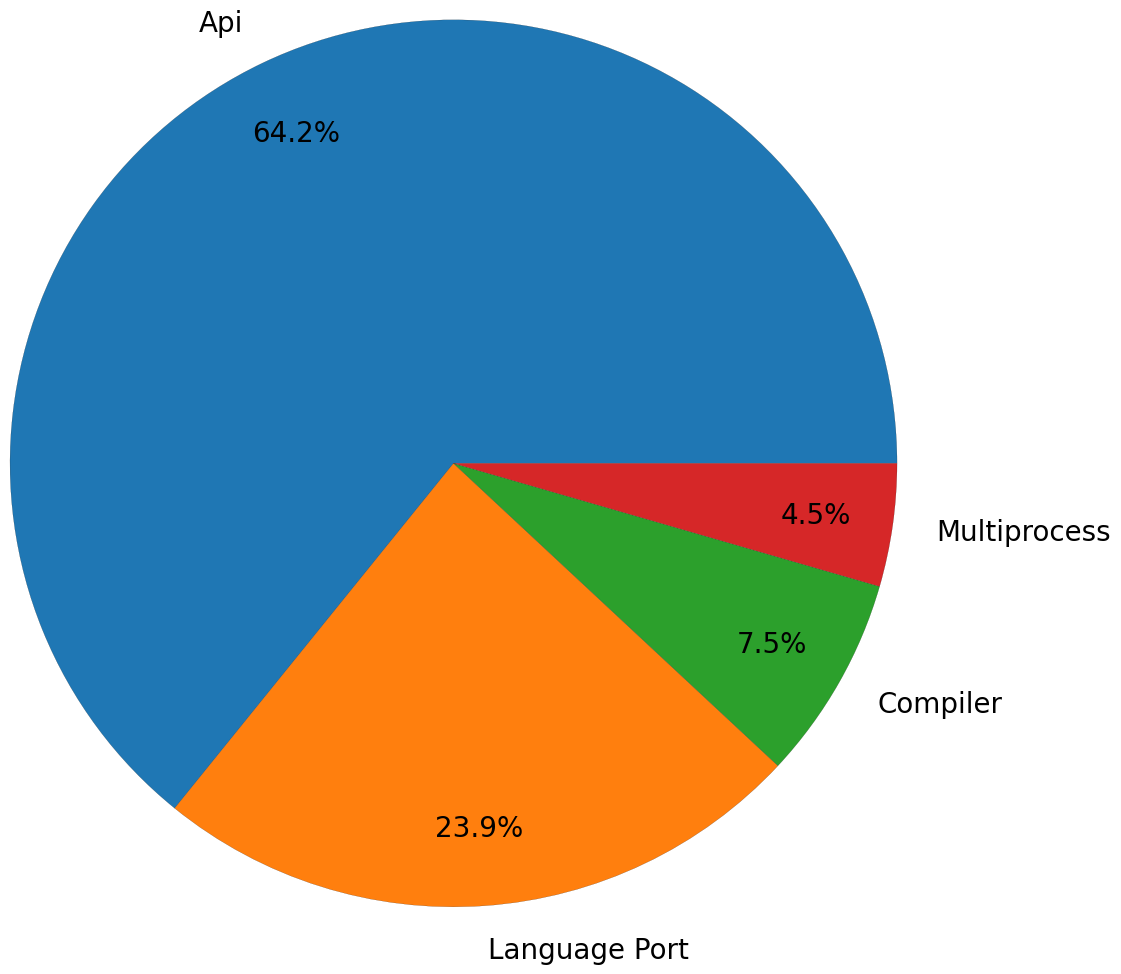}
    \caption{Direct Interoperability Tools Mechanisms}
    \label{fig:direct_interop_mechanisms}
\end{figure}

Figure \ref{fig:direct_interop_mechanisms} shows the interoperability mechanism categories of the analyzed tools.
The most common mechanism used for interoperability among the analyzed tools is API, with 64.2\% of the analyzed tools.

Language Port appears in 23.9\% of interoperability tools mechanisms and is popular in virtual-machine-based language. For example, JRuby \cite{jruby} is a reimplementation of Ruby for the JVM. It contains a compiler from Ruby to bytecode, simplifying interoperating with Java or other JVM-based languages. 

Compilers are 7.5\% of interoperability tools mechanisms. A compiler that compiles from one language to another, for example, gopherjs \cite{gopherjs} compiles Go to JavaScript. The analysis shows that compiler tools are popular among interoperability to JavaScript (88\% of the compiler interoperability tools).

\subsection*{\bm{$RQ_5$}: How many issues and discussions relate to multi-PL?} \label{sec:how_many_issues_discussions} 

In order to determine if an issue or discussion is relevant to multi-PL, we manually searched GitHub issues and discussions relating to multi-PL. We created a list of 48 sentences in English. We assume the text is in English, so non-English issues and comments are not matched. We added nine names of multi-PL tools to this list showing the sentence discusses multi-PL related subjects (e.g., swig, ctypes, JNI, libffi). 48 sentences should include a programming language name, for example \emph{Add support to [PL]}. In this case, we duplicate the sentence replacing \emph{[PL]} with different programming languages. We do not use all programming languages as some names might lead to false positives, especially if the name is a meaningful word (e.g., \emph{Go} programming language).

Each sentence is split into a set of words using and removing the stop words using NLTK Natural Language Toolkit (NLTK) \cite{nltk}. We call the created list of sets \emph{detection sets}.

Next, we search the detection set in the title, text, and comments for each project's issues and discussions. In order to perform the search, we split the text into sentences using NLTK Punkt Sentence Tokenizer \cite{punkt}. For each sentence, we split it into words using NLTK, and we seek a sentence containing a detection set. If a sentence is found, we determine that the issue or discussion is related to multi-PL, as it contains a set of words relating to multi-PL. We do not check sentences containing PL names that are used in the project. It is done to ensure the issue/discussion writer does not ask about a language used in the project but rather a language not in the project.

We have found that 13.3\% of the searched repositories (33,518 repositories) contain multi-PL-related issues and discussions. The findings show that multi-PL is a topic of interest. Moreover, it shows that developers seek programming language interoperability.

In some matched sentences, the sentence contains the programming languages or interoperability tool the user discusses. Using this information, we can detect which PLs the user discusses in a multi-PL context. In order to calculate the number of repositories containing an issue or discussion regarding interoperability with a PL, we count the number of repositories containing an issue or discussion containing a detectable PL.

\begin{table}[]
\centering
\begin{tabular}{|c|c|}
\hline
\multicolumn{1}{|c|}{PL in issues/discussions} & \multicolumn{1}{c|}{\%} \\ \hline
C                                              & 39.91                 \\
Python                                         & 31.3                 \\
Java                                           & 19.27                 \\
C\#                                            & 13.06                 \\
R                                              & 11.83                 \\
JavaScript                              & 10.6                    \\
C++                                         & 9.49 \\ \hline
\end{tabular}
\caption{Percentage of top 7 repositories containing issues and discussions by programming languages. Percentage out of repositories with detectable languages.}
\label{tbl:rq5_stats}
\end{table}

The analysis findings show that 98.8\% (33,124 repositories) contain an issue or discussion with detectable PL (table \ref{tbl:rq5_stats}). The most detected PLs are C and Python, which appear in 39.9\% and 31.3\% of the repositories, respectively.

An interesting point is the lack of issues and discussions regarding the interoperability between C and C++ compared to its popularity, as shown in $RQ_3$. This finding leads us to conclude that the interoperability between C and C++ is simple to use. Therefore, developers only have a few issues in this manner.

As shown in $RQ_4$, most interoperability tools provide interoperability with C, but as the analysis shows, support is needed in other programming languages.

\section{Threats To Validity} \label{sec:threats_to_validity}
A central \emph{external threat} is our reliance on GitHub\cite{github} source repository and Stack Overflow \cite{stackoverflow} Q\&A website. The size of the GitHub dataset and adding the Stack Overflow dataset is an improvement compared to previous work, where the reliance was only on GitHub with a smaller dataset. We also point out that the data is biased toward the type of users using these websites. Nonetheless, GitHub and Stack Overflow are still significant and popular enough to provide a good understanding of the language interoperability usage in the industry\cite{github_rank}\cite{stackoverflow_rank}. We filter out any GitHub project below 50 stars to exclude less popular and personal test projects.

An \emph{internal threat} exists in our programming language classification process, and PLs set, which is constructed manually by two researchers separately while minding the existing PL classifications in \cite{list_of_pl_wiki} and \cite{github_linguist} (detailed in section \ref{sec:prog_lang_classification}).

Another \emph{internal threat} is using the tags to detect multi-lingual questions in Stack Overflow. Since users manually apply the tags to their questions, some questions might miss tags resulting in missing them throughout the analysis. Some Stack Overflow questions might include multiple language tags while the question itself does not ask specifically about multi-lingual issues. The same threat appears in GitHub as we use the detected languages to detect multi-lingual projects. In these cases, we might miss multi-lingual projects as the language detection relies on file extensions. In cases of inline code embedding (e.g., CGo \cite{cgo_export_c}) where the foreign language source code is embedded within the host language source code, our analysis does not detect the embedded language. We also realize that having multiple languages in the same project does not mean the languages interact. Therefore the research and analysis split the multi-lingual projects analysis from the interoperability analysis.

In $RQ_3$ to find actual interoperability, we search source code snippets indexed by \emph{sourcegraph} \cite{sourcegraph} to detect snippets that match different binding types usage between the languages. The set of crafted snippets incurs an \emph{internal threat}. We know this methodology can miss many interoperability cases, but it does provide a lower bound. In order to minimize false positives as much as possible (even at the cost of having false negatives), we use very specific code snippets and search them in specific languages, even though they might appear in other languages as well. On top of that, we also made sure the project contained the expected languages (using GitHub API) and that the snippet was not in a single-line comment (multi-line comments might produce false positives). The search relies on the SourceGraph service, which incurs an \emph{external threat} as we use SourceGraph indexing and searching capabilities.

finally, another \emph{internal threat} may be due to the method we use in searching for multi-lingual related discussions and issues. Our methodology might miss a lot of discussions and issues since it is possible to ask about and discuss interoperability in many ways. In order to find as many discussions and issues as possible while minimizing false positives, we created a large list of partial sentences to search (as described in section \ref{sec:how_many_issues_discussions}). Also, we are looking for an exact match of the words in sentences. We have also manually checked about 100 matched sentences and made sure these languages discuss multi-lingual related discussions and issues. In case we find false positives in the manual validation, we have modified the list of partial sentences and started the whole process again until we had no false positives.

\section{Discussion}\label{sec:discussion}

The findings presented in section \ref{sec:multilang_result} show that multi-lingual and multi-PL development is a common practice in GitHub (66.4\% and 38\% respectively), at least from 2008, and C has a relatively major role in multi-lingual and multi-PL projects. We show that role in figure \ref{fig:interoperability-bindings} where C is used on its own and, in many cases, is used as an intermediate language to bind other languages.

As we have shown in $RQ_2$, C and C++ are common in multi-PL without web projects, and we can also see that C binding is quite popular, as Figure \ref{fig:interop_libraries_count} shows that the vast majority of interoperability tools provide interoperability with C. This finding raises the question: is C popular in multi-PL projects because of C, or is it because many tools provide interoperability just with C? 

As $RQ_4$ and $RQ_5$ show, there is a need to provide interoperability to languages other than C, while there needs to be more support for interoperability with other non-C languages. To provide interoperability with non-C languages with the available tools, a developer needs to implement the interoperability indirectly using C as an intermediate language. This task requires knowledge of C and makes the interoperability between languages more complex, requiring much more effort.

The extra effort might drive developers away from using $3^{rd}$ party libraries in different languages and look for alternatives. An alternative is to choose ports, which are different implementations of the original library. Ports might contain a different set of features than the original library and require maintenance. In a way, a port is code duplication in a different language. Therefore, we should strive to avoid ports. With that said, ports might also have a positive effect if the runtime of the ported language is more suitable for the task.

Another alternative is to look for a wrapper that implements the interoperability from $L$ to the library's language. Although wrappers provide a solution to use the original library, they require maintenance, updates, bug fixing, and mainly, a wrapper solves the problem for a specific library from a specific language.

Another approach is to use language ports that reimplement the language in the runtime we wish to interoperate. For example, to use Java Python code from Java or use $3^{rd}$ party library in Java, one can use Jython \cite{jython} as a JVM-based implementation of Python. The problem with language port is that we do not achieve interoperability between the languages we want (as Python $\neq$ Jython), but switch to another language with identical (or similar) syntax. Therefore, the original goal of interoperability between the two languages is not met.

Due to the lack of existing mechanisms or alternatives that provide interoperability between multiple non-C programming languages (shown in $RQ_4$), and as this is an issue of importance (as shown in $RQ_5$), we encourage to focus on providing interoperability between multiple programming languages and not just C. The mechanism should support the interoperability of source code in multiple programming languages and allow the use of $3^{rd}$ party libraries without implementing specific wrappers. This mechanism addresses the issues and discussions in $RQ_5$ where developers are looking to use libraries in different languages.

We also point out that multiple mechanisms where each provides access from the same $L_1$ to $L_2$ with a different API are weaker than a mechanism with a single API that provides access between multiple languages. It will be easier for a developer to learn one mechanism for multiple languages than multiple mechanisms. Such a mechanism will reduce the need and dependency on ports, wrappers, and language ports. Moreover, it will allow developers to easily fit the language to the task.

As we have pointed out in $RQ_5$, although C and C++ are the most popular interoperability (shown in $RQ_3$), there are a few issues and discussions regarding this interoperability compared to other interoperabilities. This can indicate that the interoperability between C and C++ is simple. We encourage future tools to replicate the principles of this interoperability. C and C++ have much in common regarding syntax and a common runtime (both compile to binary), which can be one of the main contributions to their simple interoperability. Also, it is important to point out that C cannot use all C++ features without the developers writing a wrapper in C++, like using the C++ STL library from within C.

\subsection{Simple Interoperability - step toward interoperability system }
As the empirical study shows, C plays a central role in programming language interoperability. This is an important finding, which provides a central pathway for a multi-PL interoperability system. In order to define such a system, we looked for central key features or guidelines that make an interoperability system "successful".

We define \emph{simple interoperability} as a set of features that an interoperability tool can implement to provide a simpler interoperability experience for developers, trying to mimic, as much as possible, the interoperability of C++ to C, which have been shown to be successful.

Let us assume a developer wishes to use code or library in $L_2$ from an existing project written in $L_1$. Both $L_1$ and $L_2$ are programming languages. We also assume the developer is knowledgeable in $L_1$ but only has some basic knowledge of $L_2$. Our assumption regarding $L_1$ and $L_2$ is that $L_1$ can access at least a callable entity (i.e. function, method) in $L_2$ using interoperability. It is not important (for this discussion) if $L_1$ access $L_2$ directly ($L_1 \rightarrow L_2$), or it is done indirectly using a third languages ($L_1 \rightarrow L_3 \rightarrow L_2$). Note that based on the findings of this paper, $L_3$ is usually C.

\subsubsection{Host-only coding}
In the case of C and C++, both languages share a similar syntax. As such, developers stay within their comfort zone (in terms of syntax) when performing interoperability tasks. In the case of arbitrary $L_1$ and $L_2$, this is not the case. As we assume the developer is knowledgeable in $L_1$, to allow the developers to stay within the comfort zone (like in C and C++), the interoperability tool must allow performing the interoperability, i.e., calling a callable entity, solely using $L_1$ syntax. We can see such behavior in Python CTypes \cite{python_ctypes}, where calling the C function does not require the developer to write in C code but stays within Python syntax. Notice that CTypes is a single-direction tool, initiating a call from Python to C, but not the other way around. Unlike CTypes, this is not the case with JNI \cite{jni}, a tool to interoperate between Java and C. JNI generates C code automatically to provide a call from Java to C, but the developer must implement code in C to complete the call. Therefore, \emph{Host-only coding}, states that the developer needs to write code only in the host (i.e., $L_1$) language.

\subsubsection{No manual Interface Definition Language}
C++ to C interoperability does not require any interface definition language (IDL) \cite{idl_wikipedia}, as C++ understands C definitions due to their similar syntax. This is not the case for arbitrary $L_1$ to $L_2$. For both languages to understand each other's signature, a tool needs to find a way to eliminate the need for an IDL, or at the least, generate the IDL automatically, releasing the burden of generating IDL. Writing IDL is a substantial and time-consuming task \cite{polispin} , especially for large libraries. Therefore, the interoperability tool should avoid IDLs or generate them automatically.

\subsubsection{Automatic runtime management}
C++ to C interoperability does not require runtime management. C++ does not need to explicitly load C runtime (nor the other way around) to call C code. In the case $L_2$ requires a runtime to be loaded into the process memory, the developer must explicitly load and manage the runtime. In order to mimic the C++ to C interoperability, a tool would need to manage the runtime automatically, or provide the capability of loading $L_2$ runtime from $L_1$ to maintain the \emph{Host-only coding} requirement.

\subsubsection{Common Data Type}
C++ is familiar with C's data types; therefore, C++ to C interoperability does not require any additional effort from the developer to use C data types, as they are well-known and supported by the C++ environment. This is true for both primitives and complex types defined in C. It is not true for arbitrary PLs $L_1$ and $L_2$. Therefore, an interoperability tool must provide a common data type that both $L_1$ and $L_2$ can understand and share. It is often impossible, especially if $L_1$ and $L_2$ use different runtimes. We point out that the common data type does not have to be implemented in either $L_1$ or $L_2$. However, it must be accessible to $L_1$ and $L_2$, using, for example, a shared intermediate language $L_i$, which both can interact with. In order not to break the \emph{Host-Only Coding} feature, a wrapper must be automatically generated from $L_1$ to $L_i$, or implement in $L_1$ directly. Notice that complex types are constructed of primitive types. Therefore, a common data type for primitives can provide a basis for supporting complex types. Built-in complex types (e.g., Python's dict, Go's map) might require special treatment, such as wrappers implemented in $L_2$. However, these wrappers are implemented in the interoperability tool and not by the end user.

To summarize, we define \emph{simple interoperability} as an interoperability tool that provides the following features:
\begin{itemize}
    \item Host-Only Coding\vspace{-2mm}
    \item No manual IDL\vspace{-2mm}
    \item Automatic runtime management\vspace{-2mm}
    \item Common Data Type
\end{itemize}

We have conducted an extensive research and have developed a plugin based cross-language interoperability system based on the simple interoperability concept and set of features named MetaFFI \cite{metaffi_paper}. The working proof of concept interoperates successfully between Python3 interpreter (CPython), Java (JVM) and Go programming languages providing access to code and library.

\section{Empirical Research Conclusions}\label{sec:conclusions}
The empirical study shows that multi-lingual and multi-programming language development are common practices regarding the amount of multi-lingual GitHub projects and the amount of Stack Overflow questions. While C is the most dominant language in multi-lingual projects and the most supported language in available interoperability tools, it is less popular in single-language projects. Also, C is used as a hub to interoperate between different languages. We showed a need for interoperability tools supporting programming languages other than C. Currently, the alternative is costly and requires developers skilled in C to make an extensive effort to implement code in other languages or $3^{rd}$ party languages. We have pointed out that C and C++ interoperability is the most common, with a very small number of issues and discussions. Therefore, we defined a set of features called \emph{Simple interoperability} inspired by the C++ to C interoperability features. A tool implementing these features would provide a more intuitive experience for developers to interoperate between two languages. Therefore, we research and develop MetaFFI \cite{metaffi_paper}, a POC which implements \emph{simple interoperability} to support multiple programming languages, specifically using a single API to interoperate between all the supported languages to provide a simple mechanism similar to C++ to C interoperability.

\section{Data Availability Statements} \label{sec:data_availability}
The authors declare that the data supporting the findings of this study, or a detailed explanation on how to acquire it, are available in the article and its Supplementary Information files.\\

The complete datasets and raw materials are available at\\\texttt{https://drive.google.com/file/d/1unGkN3cCZgu3qC6NxQK621Rnb82LNdqJ/view?usp=sharing}\\(60.75 GB zip)

For convenience, the GitHub metadata file in JSON is also available separately: \\\texttt{https://drive.google.com/file/d/1LJwi6jp6hXF6Ro3znm9Ix1f3-7Uv1oDC/view?usp=sharing}\\(47 MB 7z)

\bibliography{main}


\begin{thebibliography}{54}


\ifx \showCODEN    \undefined \def \showCODEN     #1{\unskip}     \fi
\ifx \showDOI      \undefined \def \showDOI       #1{#1}\fi
\ifx \showISBNx    \undefined \def \showISBNx     #1{\unskip}     \fi
\ifx \showISBNxiii \undefined \def \showISBNxiii  #1{\unskip}     \fi
\ifx \showISSN     \undefined \def \showISSN      #1{\unskip}     \fi
\ifx \showLCCN     \undefined \def \showLCCN      #1{\unskip}     \fi
\ifx \shownote     \undefined \def \shownote      #1{#1}          \fi
\ifx \showarticletitle \undefined \def \showarticletitle #1{#1}   \fi
\ifx \showURL      \undefined \def \showURL       {\relax}        \fi
\providecommand\bibfield[2]{#2}
\providecommand\bibinfo[2]{#2}
\providecommand\natexlab[1]{#1}
\providecommand\showeprint[2][]{arXiv:#2}

\bibitem[pun({[n.\,d.]})]%
        {punkt}
 \bibinfo{year}{[n.\,d.]}\natexlab{}.
\newblock \bibinfo{title}{{NLTK} :: nltk.tokenize.punkt module}.
\newblock
\newblock
\urldef\tempurl%
\url{https://www.nltk.org/api/nltk.tokenize.punkt.html}
\showURL{%
\tempurl}


\bibitem[doc(2023)]%
        {dockerfile}
 \bibinfo{year}{2023}\natexlab{}.
\newblock \bibinfo{title}{Dockerfile reference}.
\newblock
\newblock
\urldef\tempurl%
\url{https://docs.docker.com/engine/reference/builder/}
\showURL{%
\tempurl}


\bibitem[gop(2023)]%
        {gopherjs}
 \bibinfo{year}{2023}\natexlab{}.
\newblock \bibinfo{title}{gopherjs/gopherjs}.
\newblock
\newblock
\urldef\tempurl%
\url{https://github.com/gopherjs/gopherjs}
\showURL{%
\tempurl}
\newblock
\shownote{original-date: 2013-08-27T22:23:58Z}.


\bibitem[git(2023)]%
        {github_linguist}
 \bibinfo{year}{2023}\natexlab{}.
\newblock \bibinfo{title}{Linguist}.
\newblock
\newblock
\urldef\tempurl%
\url{https://github.com/github/linguist}
\showURL{%
\tempurl}
\newblock
\shownote{original-date: 2011-05-09T22:53:13Z}.


\bibitem[Abidi et~al\mbox{.}(2019)]%
        {behind_the_scenes}
\bibfield{author}{\bibinfo{person}{Mouna Abidi}, \bibinfo{person}{Manel Grichi}, {and} \bibinfo{person}{Foutse Khomh}.} \bibinfo{year}{2019}\natexlab{}.
\newblock \showarticletitle{Behind the scenes: developers' perception of multi-language practices}. In \bibinfo{booktitle}{\emph{Conference of the centre for advanced studies on collaborative research}}.
\newblock
\urldef\tempurl%
\url{https://api.semanticscholar.org/CorpusID:216589378}
\showURL{%
\tempurl}


\bibitem[Berners-Lee and Connolly(1995)]%
        {html_rfc}
\bibfield{author}{\bibinfo{person}{Tim Berners-Lee} {and} \bibinfo{person}{Daniel~W. Connolly}.} \bibinfo{year}{1995}\natexlab{}.
\newblock \bibinfo{title}{Hypertext markup language - 2.0}.
\newblock
\newblock
\urldef\tempurl%
\url{https://doi.org/10.17487/RFC1866}
\showDOI{\tempurl}
\newblock
\shownote{Number: 1866 Series: Request for comments tex.howpublished: RFC 1866 tex.pagetotal: 77}.


\bibitem[Bird and Klein(2009)]%
        {nltk}
\bibfield{author}{\bibinfo{person}{Edward~Loper Bird, Steven} {and} \bibinfo{person}{Ewan Klein}.} \bibinfo{year}{2009}\natexlab{}.
\newblock \bibinfo{booktitle}{\emph{Natural language processing with python}}.
\newblock \bibinfo{publisher}{O'Reilly Media Inc.}
\newblock


\bibitem[Bissyandé et~al\mbox{.}(2013)]%
        {100k_opensource}
\bibfield{author}{\bibinfo{person}{Tegawendé~F. Bissyandé}, \bibinfo{person}{Ferdian Thung}, \bibinfo{person}{David Lo}, \bibinfo{person}{Lingxiao Jiang}, {and} \bibinfo{person}{Laurent Réveillère}.} \bibinfo{year}{2013}\natexlab{}.
\newblock \showarticletitle{Popularity, interoperability, and impact of programming languages in 100,000 open source projects}. In \bibinfo{booktitle}{\emph{2013 {IEEE} 37th annual computer software and applications conference}}. \bibinfo{pages}{303--312}.
\newblock
\urldef\tempurl%
\url{https://doi.org/10.1109/COMPSAC.2013.55}
\showDOI{\tempurl}


\bibitem[Bray(2017)]%
        {json_rfc}
\bibfield{author}{\bibinfo{person}{Tim Bray}.} \bibinfo{year}{2017}\natexlab{}.
\newblock \bibinfo{title}{The {JavaScript} object notation ({JSON}) data interchange format}.
\newblock
\newblock
\urldef\tempurl%
\url{https://doi.org/10.17487/RFC8259}
\showDOI{\tempurl}
\newblock
\shownote{Number: 8259 Series: Request for comments tex.howpublished: RFC 8259 tex.pagetotal: 16}.


\bibitem[{Charles Oliver Nutter} et~al\mbox{.}(2022)]%
        {jruby}
\bibfield{author}{\bibinfo{person}{{Charles Oliver Nutter}}, \bibinfo{person}{{Thomas Enebo}}, \bibinfo{person}{{Ola Bini}}, {and} \bibinfo{person}{{Nick Sieger}}.} \bibinfo{year}{2022}\natexlab{}.
\newblock \bibinfo{title}{{JRuby} - an implementation of the {Ruby} language on the {JVM}}.
\newblock
\newblock
\urldef\tempurl%
\url{https://github.com/jruby/jruby}
\showURL{%
\tempurl}


\bibitem[Cherny-Shahar and Yehudai(2024)]%
        {metaffi_paper}
\bibfield{author}{\bibinfo{person}{Tsvi Cherny-Shahar} {and} \bibinfo{person}{Amiram Yehudai}.} \bibinfo{year}{2024}\natexlab{}.
\newblock \bibinfo{title}{{MetaFFI} – multilingual indirect interoperability system}.
\newblock
\newblock
\urldef\tempurl%
\url{https://arxiv.org/abs/2408.14175}
\showURL{%
\tempurl}
\newblock
\shownote{arXiv: 2408.14175 [cs.PL]}.


\bibitem[contributors(2022)]%
        {idl_wikipedia}
\bibfield{author}{\bibinfo{person}{Wikipedia contributors}.} \bibinfo{year}{2022}\natexlab{}.
\newblock \bibinfo{title}{Interface description language — {Wikipedia}, the free encyclopedia}.
\newblock
\newblock
\urldef\tempurl%
\url{https://en.wikipedia.org/w/index.php?title=Interface_description_language&oldid=1064057807}
\showURL{%
\tempurl}


\bibitem[Dahl and Crawford(2009)]%
        {rinruby}
\bibfield{author}{\bibinfo{person}{David~B. Dahl} {and} \bibinfo{person}{Scott Crawford}.} \bibinfo{year}{2009}\natexlab{}.
\newblock \showarticletitle{{RinRuby}: {Accessing} the {R} interpreter from pure ruby}.
\newblock \bibinfo{journal}{\emph{Journal of Statistical Software}} \bibinfo{volume}{29}, \bibinfo{number}{4} (\bibinfo{date}{Nov.} \bibinfo{year}{2009}), \bibinfo{pages}{1--18}.
\newblock
\showISSN{1548-7660}
\urldef\tempurl%
\url{http://www.jstatsoft.org/v29/i04}
\showURL{%
\tempurl}
\newblock
\shownote{tex.accepted: 2008-11-26 tex.bibdate: 2008-11-26 tex.coden: JSSOBK tex.submitted: 2008-07-03}.


\bibitem[Donaghy(2023)]%
        {hello_world_in_every_lang}
\bibfield{author}{\bibinfo{person}{Mike Donaghy}.} \bibinfo{year}{2023}\natexlab{}.
\newblock \bibinfo{title}{Hello {World}}.
\newblock
\newblock
\urldef\tempurl%
\url{https://github.com/leachim6/hello-world}
\showURL{%
\tempurl}
\newblock
\shownote{original-date: 2008-07-15T00:15:08Z}.


\bibitem[Flanagan(2016)]%
        {css_rfc}
\bibfield{author}{\bibinfo{person}{Heather Flanagan}.} \bibinfo{year}{2016}\natexlab{}.
\newblock \bibinfo{title}{Cascading style sheets ({CSS}) requirements for {RFCs}}.
\newblock
\newblock
\urldef\tempurl%
\url{https://doi.org/10.17487/RFC7993}
\showDOI{\tempurl}
\newblock
\shownote{Number: 7993 Series: Request for comments tex.howpublished: RFC 7993 tex.pagetotal: 14}.


\bibitem[Foundation(2020)]%
        {python_ctypes}
\bibfield{author}{\bibinfo{person}{Python~Software Foundation}.} \bibinfo{year}{2020}\natexlab{}.
\newblock \bibinfo{title}{ctypes — {A} foreign function library for {Python} — {Python} 3.8.3 documentation}.
\newblock
\newblock
\urldef\tempurl%
\url{https://docs.python.org/3/library/ctypes.html}
\showURL{%
\tempurl}


\bibitem[Foundation(2021a)]%
        {jython}
\bibfield{author}{\bibinfo{person}{Python~Software Foundation}.} \bibinfo{year}{2021}\natexlab{a}.
\newblock \bibinfo{title}{Jython}.
\newblock
\newblock
\urldef\tempurl%
\url{https://www.jython.org/}
\showURL{%
\tempurl}


\bibitem[Foundation(2021b)]%
        {cpython}
\bibfield{author}{\bibinfo{person}{Python~Software Foundation}.} \bibinfo{year}{2021}\natexlab{b}.
\newblock \bibinfo{title}{python/cpython: {The} {Python} programming language}.
\newblock
\newblock
\urldef\tempurl%
\url{https://github.com/python/cpython}
\showURL{%
\tempurl}


\bibitem[Grichi et~al\mbox{.}(2021)]%
        {impact_of_interlanguage_dependencies}
\bibfield{author}{\bibinfo{person}{Manel Grichi}, \bibinfo{person}{Mouna Abidi}, \bibinfo{person}{Fehmi Jaafar}, \bibinfo{person}{Ellis~E. Eghan}, {and} \bibinfo{person}{Bram Adams}.} \bibinfo{year}{2021}\natexlab{}.
\newblock \showarticletitle{On the impact of interlanguage dependencies in multilanguage systems empirical case study on java native interface applications ({JNI})}.
\newblock \bibinfo{journal}{\emph{IEEE Transactions on Reliability}} \bibinfo{volume}{70}, \bibinfo{number}{1} (\bibinfo{year}{2021}), \bibinfo{pages}{428--440}.
\newblock
\urldef\tempurl%
\url{https://doi.org/10.1109/TR.2020.3024873}
\showDOI{\tempurl}


\bibitem[Inc(2011)]%
        {github_repo_api}
\bibfield{author}{\bibinfo{person}{GitHub Inc}.} \bibinfo{year}{2011}\natexlab{}.
\newblock \bibinfo{title}{repositories - github docs}.
\newblock
\newblock
\urldef\tempurl%
\url{https://docs.github.com/en/rest/reference/repos}
\showURL{%
\tempurl}


\bibitem[Inc(2021a)]%
        {github}
\bibfield{author}{\bibinfo{person}{GitHub Inc}.} \bibinfo{year}{2021}\natexlab{a}.
\newblock \bibinfo{title}{build software better, together}.
\newblock
\newblock
\urldef\tempurl%
\url{https://github.com/}
\showURL{%
\tempurl}


\bibitem[Inc(2018)]%
        {stack_overflow_developer_survey_2018}
\bibfield{author}{\bibinfo{person}{Stack~Exchange Inc}.} \bibinfo{year}{2018}\natexlab{}.
\newblock \bibinfo{title}{Stack {Overflow} {Developers} {Survey} 2018}.
\newblock
\newblock
\urldef\tempurl%
\url{https://insights.stackoverflow.com/survey/2018#technology-_-programming-scripting-and-markup-languages}
\showURL{%
\tempurl}


\bibitem[Inc(2020)]%
        {stack_overflow_developer_survey_2020}
\bibfield{author}{\bibinfo{person}{Stack~Exchange Inc}.} \bibinfo{year}{2020}\natexlab{}.
\newblock \bibinfo{title}{Stack {Overflow} {Developers} {Survey} 2020}.
\newblock
\newblock
\urldef\tempurl%
\url{https://insights.stackoverflow.com/survey/2020#technology-programming-scripting-and-markup-languages}
\showURL{%
\tempurl}


\bibitem[Inc(2021b)]%
        {sede}
\bibfield{author}{\bibinfo{person}{Stack~Exchange Inc}.} \bibinfo{year}{2021}\natexlab{b}.
\newblock \bibinfo{title}{stack exchange data explorer}.
\newblock
\newblock
\urldef\tempurl%
\url{https://data.stackexchange.com/}
\showURL{%
\tempurl}


\bibitem[Inc(2021c)]%
        {stackoverflow}
\bibfield{author}{\bibinfo{person}{Stack~Exchange Inc}.} \bibinfo{year}{2021}\natexlab{c}.
\newblock \bibinfo{title}{stack overflow - where developers learn, share, \& build careers}.
\newblock
\newblock
\urldef\tempurl%
\url{https://stackoverflow.com/}
\showURL{%
\tempurl}


\bibitem[Kaplan et~al\mbox{.}(1998)]%
        {polispin}
\bibfield{author}{\bibinfo{person}{A. Kaplan}, \bibinfo{person}{J. Ridgway}, {and} \bibinfo{person}{J.~C. Wileden}.} \bibinfo{year}{1998}\natexlab{}.
\newblock \showarticletitle{Why {IDLs} are not ideal}. In \bibinfo{booktitle}{\emph{Proceedings of the 9th international workshop on software specification and design}} \emph{(\bibinfo{series}{{IWSSD} '98})}. \bibinfo{publisher}{IEEE Computer Society}, \bibinfo{address}{USA}, \bibinfo{pages}{2}.
\newblock
\showISBNx{0-8186-8439-9}


\bibitem[Li et~al\mbox{.}(2022a)]%
        {vulnerability_proneness_multilingual}
\bibfield{author}{\bibinfo{person}{Wen Li}, \bibinfo{person}{Li Li}, {and} \bibinfo{person}{Haipeng Cai}.} \bibinfo{year}{2022}\natexlab{a}.
\newblock \showarticletitle{On the vulnerability proneness of multilingual code}. In \bibinfo{booktitle}{\emph{Proceedings of the 30th {ACM} joint european software engineering conference and symposium on the foundations of software engineering}} \emph{(\bibinfo{series}{Esec/fse 2022})}. \bibinfo{publisher}{Association for Computing Machinery}, \bibinfo{address}{New York, NY, USA}, \bibinfo{pages}{847--859}.
\newblock
\showISBNx{978-1-4503-9413-0}
\urldef\tempurl%
\url{https://doi.org/10.1145/3540250.3549173}
\showDOI{\tempurl}
\newblock
\shownote{Number of pages: 13 Place: Singapore, Singapore}.


\bibitem[Li et~al\mbox{.}(2022b)]%
        {polyfax}
\bibfield{author}{\bibinfo{person}{Wen Li}, \bibinfo{person}{Li Li}, {and} \bibinfo{person}{Haipeng Cai}.} \bibinfo{year}{2022}\natexlab{b}.
\newblock \showarticletitle{{PolyFax}: a toolkit for characterizing multi-language software}. In \bibinfo{booktitle}{\emph{Proceedings of the 30th {ACM} joint european software engineering conference and symposium on the foundations of software engineering}} \emph{(\bibinfo{series}{Esec/fse 2022})}. \bibinfo{publisher}{Association for Computing Machinery}, \bibinfo{address}{New York, NY, USA}, \bibinfo{pages}{1662--1666}.
\newblock
\showISBNx{978-1-4503-9413-0}
\urldef\tempurl%
\url{https://doi.org/10.1145/3540250.3558925}
\showDOI{\tempurl}
\newblock
\shownote{Number of pages: 5 Place: Singapore, Singapore}.


\bibitem[Li et~al\mbox{.}(2024)]%
        {multilingual_systems_constructed}
\bibfield{author}{\bibinfo{person}{Wen Li}, \bibinfo{person}{Austin Marino}, \bibinfo{person}{Haoran Yang}, \bibinfo{person}{Na Meng}, \bibinfo{person}{Li Li}, {and} \bibinfo{person}{Haipeng Cai}.} \bibinfo{year}{2024}\natexlab{}.
\newblock \showarticletitle{How are multilingual systems constructed: {Characterizing} language use and selection in open-source multilingual software}.
\newblock \bibinfo{journal}{\emph{ACM Trans. Softw. Eng. Methodol.}} \bibinfo{volume}{33}, \bibinfo{number}{3} (\bibinfo{date}{March} \bibinfo{year}{2024}).
\newblock
\showISSN{1049-331X}
\urldef\tempurl%
\url{https://doi.org/10.1145/3631967}
\showDOI{\tempurl}
\newblock
\shownote{Number of pages: 46 Place: New York, NY, USA Publisher: Association for Computing Machinery tex.articleno: 63 tex.issue\_date: March 2024}.


\bibitem[Li et~al\mbox{.}(2021)]%
        {understanding_lang_selection}
\bibfield{author}{\bibinfo{person}{Wen Li}, \bibinfo{person}{Na Meng}, \bibinfo{person}{Li Li}, {and} \bibinfo{person}{Haipeng Cai}.} \bibinfo{year}{2021}\natexlab{}.
\newblock \showarticletitle{Understanding language selection in multi-language software projects on {GitHub}}. In \bibinfo{booktitle}{\emph{2021 {IEEE}/{ACM} 43rd international conference on software engineering: {Companion} proceedings ({ICSE}-companion)}}. \bibinfo{pages}{256--257}.
\newblock
\urldef\tempurl%
\url{https://doi.org/10.1109/ICSE-Companion52605.2021.00119}
\showDOI{\tempurl}


\bibitem[LLC({[n.\,d.]})]%
        {cgo_export_c}
\bibfield{author}{\bibinfo{person}{Google LLC}.} \bibinfo{year}{[n.\,d.]}\natexlab{}.
\newblock \bibinfo{title}{cgo - the go programming language 2020}.
\newblock
\newblock
\urldef\tempurl%
\url{https://golang.org/cmd/cgo/#hdr-C_references_to_Go}
\showURL{%
\tempurl}


\bibitem[Mayer et~al\mbox{.}(2017)]%
        {xlang_survey}
\bibfield{author}{\bibinfo{person}{Philip Mayer}, \bibinfo{person}{Michael Kirsch}, {and} \bibinfo{person}{Minh-Anh Le}.} \bibinfo{year}{2017}\natexlab{}.
\newblock \showarticletitle{On multi-language software development, cross-language links and accompanying tools: a survey of professional software developers}.
\newblock \bibinfo{journal}{\emph{Journal of Software Engineering Research and Development}}  \bibinfo{volume}{5} (\bibinfo{date}{Dec.} \bibinfo{year}{2017}).
\newblock
\urldef\tempurl%
\url{https://doi.org/10.1186/s40411-017-0035-z}
\showDOI{\tempurl}


\bibitem[{Microsoft Corporation}(2021)]%
        {pinvoke}
\bibfield{author}{\bibinfo{person}{{Microsoft Corporation}}.} \bibinfo{year}{2021}\natexlab{}.
\newblock \bibinfo{title}{Platform invoke ({P}/{Invoke})}.
\newblock
\newblock
\urldef\tempurl%
\url{https://docs.microsoft.com/en-us/dotnet/standard/native-interop/pinvoke}
\showURL{%
\tempurl}


\bibitem[{NLua}({[n.\,d.]})]%
        {nlua}
\bibfield{author}{\bibinfo{person}{{NLua}}.} \bibinfo{year}{[n.\,d.]}\natexlab{}.
\newblock \bibinfo{title}{Nlua}.
\newblock
\newblock
\urldef\tempurl%
\url{http://nlua.org/}
\showURL{%
\tempurl}


\bibitem[{Oracle}(2020)]%
        {jni}
\bibfield{author}{\bibinfo{person}{{Oracle}}.} \bibinfo{year}{2020}\natexlab{}.
\newblock \bibinfo{title}{{JNI} {APIs} and developer guides}.
\newblock
\newblock
\urldef\tempurl%
\url{https://docs.oracle.com/javase/8/docs/technotes/guides/jni/}
\showURL{%
\tempurl}


\bibitem[Ray et~al\mbox{.}(2017)]%
        {large_scale_study_github}
\bibfield{author}{\bibinfo{person}{Baishakhi Ray}, \bibinfo{person}{Daryl Posnett}, \bibinfo{person}{Premkumar Devanbu}, {and} \bibinfo{person}{Vladimir Filkov}.} \bibinfo{year}{2017}\natexlab{}.
\newblock \showarticletitle{A large-scale study of programming languages and code quality in {GitHub}}.
\newblock \bibinfo{journal}{\emph{Communications of The Acm}} \bibinfo{volume}{60}, \bibinfo{number}{10} (\bibinfo{date}{Sept.} \bibinfo{year}{2017}), \bibinfo{pages}{91--100}.
\newblock
\showISSN{0001-0782}
\urldef\tempurl%
\url{https://doi.org/10.1145/3126905}
\showDOI{\tempurl}
\newblock
\shownote{Number of pages: 10 Place: New York, NY, USA Publisher: Association for Computing Machinery tex.issue\_date: October 2017}.


\bibitem[{santoshrajan}(2020)]%
        {lispyscript}
\bibfield{author}{\bibinfo{person}{{santoshrajan}}.} \bibinfo{year}{2020}\natexlab{}.
\newblock \bibinfo{title}{santoshrajan/lispyscript: {A} javascript with {Lispy} syntax and macros}.
\newblock
\newblock
\urldef\tempurl%
\url{https://github.com/santoshrajan/lispyscript}
\showURL{%
\tempurl}


\bibitem[Satman(2014)]%
        {rcaller}
\bibfield{author}{\bibinfo{person}{M~Hakan Satman}.} \bibinfo{year}{2014}\natexlab{}.
\newblock \showarticletitle{{RCaller}: {A} software library for calling {R} from {Java}}.
\newblock \bibinfo{journal}{\emph{British Journal of Mathematics \& Computer Science}} \bibinfo{volume}{4}, \bibinfo{number}{15} (\bibinfo{year}{2014}), \bibinfo{pages}{2188}.
\newblock
\newblock
\shownote{Publisher: SCIENCEDOMAIN International}.


\bibitem[{Scala-JS}(2022)]%
        {scalajs}
\bibfield{author}{\bibinfo{person}{{Scala-JS}}.} \bibinfo{year}{2022}\natexlab{}.
\newblock \bibinfo{title}{Scala.js}.
\newblock
\newblock
\urldef\tempurl%
\url{https://www.scala-js.org/}
\showURL{%
\tempurl}


\bibitem[Similarweb({[n.\,d.]})]%
        {github_rank}
\bibfield{author}{\bibinfo{person}{Similarweb}.} \bibinfo{year}{[n.\,d.]}\natexlab{}.
\newblock \bibinfo{title}{github.com {Traffic} {Analytics} \& {Market} {Share}}.
\newblock
\newblock
\urldef\tempurl%
\url{https://www.similarweb.com/website/github.com/}
\showURL{%
\tempurl}


\bibitem[SimilarWeb({[n.\,d.]})]%
        {stackoverflow_rank}
\bibfield{author}{\bibinfo{person}{SimilarWeb}.} \bibinfo{year}{[n.\,d.]}\natexlab{}.
\newblock \bibinfo{title}{stackoverflow.com {Traffic} {Analytics} \& {Market} {Share}}.
\newblock
\newblock
\urldef\tempurl%
\url{https://www.similarweb.com/website/stackoverflow.com/}
\showURL{%
\tempurl}


\bibitem[Sorhus(2022)]%
        {awesome_list}
\bibfield{author}{\bibinfo{person}{Sindre Sorhus}.} \bibinfo{year}{2022}\natexlab{}.
\newblock \bibinfo{title}{Awesome lists about all kinds of interesting topics}.
\newblock
\newblock
\urldef\tempurl%
\url{https://github.com/sindresorhus/awesome}
\showURL{%
\tempurl}


\bibitem[Sourcegraph(2022)]%
        {sourcegraph}
\bibfield{author}{\bibinfo{person}{Sourcegraph}.} \bibinfo{year}{2022}\natexlab{}.
\newblock \bibinfo{title}{Sourcegraph}.
\newblock
\newblock
\urldef\tempurl%
\url{https://sourcegraph.com/search}
\showURL{%
\tempurl}


\bibitem[{Stack Exchange Community}(2021)]%
        {stackoverflow_data_dump}
\bibfield{author}{\bibinfo{person}{{Stack Exchange Community}}.} \bibinfo{year}{2021}\natexlab{}.
\newblock \bibinfo{title}{stack exchange data dump : stack exchange, inc. : free download, borrow, and streaming : internet archive}.
\newblock
\newblock
\urldef\tempurl%
\url{https://archive.org/details/stackexchange}
\showURL{%
\tempurl}


\bibitem[{Stack Exchange Inc}(2017)]%
        {stack_overflow_developer_survey_2017}
\bibfield{author}{\bibinfo{person}{{Stack Exchange Inc}}.} \bibinfo{year}{2017}\natexlab{}.
\newblock \bibinfo{title}{Stack {Overflow} {Developers} {Survey} 2017}.
\newblock
\newblock
\urldef\tempurl%
\url{https://insights.stackoverflow.com/survey/2017#technology-_-programming-scripting-and-markup-languages}
\showURL{%
\tempurl}


\bibitem[{Stack Exchange Inc}(2019)]%
        {stack_overflow_developer_survey_2019}
\bibfield{author}{\bibinfo{person}{{Stack Exchange Inc}}.} \bibinfo{year}{2019}\natexlab{}.
\newblock \bibinfo{title}{Stack {Overflow} {Developers} {Survey} 2019}.
\newblock
\newblock
\urldef\tempurl%
\url{https://insights.stackoverflow.com/survey/2019#technology-_-programming-scripting-and-markup-languages}
\showURL{%
\tempurl}


\bibitem[{Stack Exchange Inc}(2021)]%
        {stack_overflow_developer_survey_2021}
\bibfield{author}{\bibinfo{person}{{Stack Exchange Inc}}.} \bibinfo{year}{2021}\natexlab{}.
\newblock \bibinfo{title}{Stack {Overflow} {Developers} {Survey} 2021}.
\newblock
\newblock
\urldef\tempurl%
\url{https://insights.stackoverflow.com/survey/2021#most-popular-technologies-language-prof}
\showURL{%
\tempurl}


\bibitem[Stallman et~al\mbox{.}(2004)]%
        {makefile}
\bibfield{author}{\bibinfo{person}{Richard Stallman}, \bibinfo{person}{Roland McGrath}, {and} \bibinfo{person}{Paul~D. Smith}.} \bibinfo{year}{2004}\natexlab{}.
\newblock \bibinfo{booktitle}{\emph{{GNU} {Make}: a program for directing recompliation ; {GNU} make version 3.81}}.
\newblock \bibinfo{publisher}{Free Software Foundation}, \bibinfo{address}{Boston, Mass}.
\newblock
\showISBNx{978-1-882114-83-2}


\bibitem[Tomassetti and Torchiano(2014)]%
        {polyglot}
\bibfield{author}{\bibinfo{person}{Federico Tomassetti} {and} \bibinfo{person}{Marco Torchiano}.} \bibinfo{year}{2014}\natexlab{}.
\newblock \showarticletitle{An empirical assessment of polyglot-ism in {GitHub}}. In \bibinfo{booktitle}{\emph{Proceedings of the 18th {International} {Conference} on {Evaluation} and {Assessment} in {Software} {Engineering}}} \emph{(\bibinfo{series}{{EASE} '14})}. \bibinfo{publisher}{Association for Computing Machinery}, \bibinfo{address}{New York, NY, USA}, \bibinfo{pages}{1--4}.
\newblock
\showISBNx{978-1-4503-2476-2}
\urldef\tempurl%
\url{https://doi.org/10.1145/2601248.2601269}
\showDOI{\tempurl}


\bibitem[Tomassetti et~al\mbox{.}(2013)]%
        {lang_interaction}
\bibfield{author}{\bibinfo{person}{Federico Tomassetti}, \bibinfo{person}{Marco Torchiano}, {and} \bibinfo{person}{Antonio Vetro}.} \bibinfo{year}{2013}\natexlab{}.
\newblock \showarticletitle{Classification of {Language} {Interactions}}.
\newblock \bibinfo{journal}{\emph{2013 ACM / IEEE International Symposium on Empirical Software Engineering and Measurement}} (\bibinfo{date}{Oct.} \bibinfo{year}{2013}), \bibinfo{pages}{287--290}.
\newblock
\urldef\tempurl%
\url{https://doi.org/10.1109/ESEM.2013.34}
\showDOI{\tempurl}
\newblock
\shownote{Conference Name: 2013 ACM/IEEE International Symposium on Empirical Software Engineering and Measurement (ESEM) ISBN: 9780769550565 Place: Baltimore, Maryland Publisher: IEEE}.


\bibitem[{Wikipedia contributors}(2022)]%
        {markdown}
\bibfield{author}{\bibinfo{person}{{Wikipedia contributors}}.} \bibinfo{year}{2022}\natexlab{}.
\newblock \bibinfo{title}{Markdown — {Wikipedia}, the free encyclopedia}.
\newblock
\newblock
\urldef\tempurl%
\url{https://en.wikipedia.org/w/index.php?title=Markdown&oldid=1130043463}
\showURL{%
\tempurl}


\bibitem[{Wikipedia contributors}(2023)]%
        {list_of_pl_wiki}
\bibfield{author}{\bibinfo{person}{{Wikipedia contributors}}.} \bibinfo{year}{2023}\natexlab{}.
\newblock \bibinfo{title}{List of programming languages — {Wikipedia}, the free encyclopedia}.
\newblock
\newblock
\urldef\tempurl%
\url{https://en.wikipedia.org/w/index.php?title=List_of_programming_languages&oldid=1147985452}
\showURL{%
\tempurl}


\bibitem[Yang et~al\mbox{.}(2023)]%
        {demystifying_issues}
\bibfield{author}{\bibinfo{person}{Haoran Yang}, \bibinfo{person}{Weile Lian}, \bibinfo{person}{Shaowei Wang}, {and} \bibinfo{person}{Haipeng Cai}.} \bibinfo{year}{2023}\natexlab{}.
\newblock \showarticletitle{Demystifying issues, challenges, and solutions for multilingual software development}. In \bibinfo{booktitle}{\emph{2023 {IEEE}/{ACM} 45th international conference on software engineering ({ICSE})}}. \bibinfo{pages}{1840--1852}.
\newblock
\urldef\tempurl%
\url{https://doi.org/10.1109/ICSE48619.2023.00157}
\showDOI{\tempurl}


\bibitem[Yang et~al\mbox{.}(2024)]%
        {issue_challenges_solutions}
\bibfield{author}{\bibinfo{person}{Haoran Yang}, \bibinfo{person}{Yu Nong}, \bibinfo{person}{Shaowei Wang}, {and} \bibinfo{person}{Haipeng Cai}.} \bibinfo{year}{2024}\natexlab{}.
\newblock \showarticletitle{Multi-language software development: {Issues}, challenges, and solutions}.
\newblock \bibinfo{journal}{\emph{IEEE Transactions on Software Engineering}} \bibinfo{volume}{50}, \bibinfo{number}{3} (\bibinfo{year}{2024}), \bibinfo{pages}{512--533}.
\newblock
\urldef\tempurl%
\url{https://doi.org/10.1109/TSE.2024.3358258}
\showDOI{\tempurl}


\end{thebibliography}



\begin{thebibliography}{6}


\ifx \showCODEN    \undefined \def \showCODEN     #1{\unskip}     \fi
\ifx \showDOI      \undefined \def \showDOI       #1{#1}\fi
\ifx \showISBNx    \undefined \def \showISBNx     #1{\unskip}     \fi
\ifx \showISBNxiii \undefined \def \showISBNxiii  #1{\unskip}     \fi
\ifx \showISSN     \undefined \def \showISSN      #1{\unskip}     \fi
\ifx \showLCCN     \undefined \def \showLCCN      #1{\unskip}     \fi
\ifx \shownote     \undefined \def \shownote      #1{#1}          \fi
\ifx \showarticletitle \undefined \def \showarticletitle #1{#1}   \fi
\ifx \showURL      \undefined \def \showURL       {\relax}        \fi
\providecommand\bibfield[2]{#2}
\providecommand\bibinfo[2]{#2}
\providecommand\natexlab[1]{#1}
\providecommand\showeprint[2][]{arXiv:#2}

\bibitem[Bray(2017)]%
        {json_rfc}
\bibfield{author}{\bibinfo{person}{Tim Bray}.} \bibinfo{year}{2017}\natexlab{}.
\newblock \bibinfo{title}{The {JavaScript} object notation ({JSON}) data interchange format}.
\newblock
\newblock
\urldef\tempurl%
\url{https://doi.org/10.17487/RFC8259}
\showDOI{\tempurl}
\newblock
\shownote{Number: 8259 Series: Request for comments tex.howpublished: RFC 8259 tex.pagetotal: 16}.


\bibitem[Inc(2021)]%
        {sede}
\bibfield{author}{\bibinfo{person}{Stack~Exchange Inc}.} \bibinfo{year}{2021}\natexlab{}.
\newblock \bibinfo{title}{stack exchange data explorer}.
\newblock
\newblock
\urldef\tempurl%
\url{https://data.stackexchange.com/}
\showURL{%
\tempurl}


\bibitem[Sorhus(2022)]%
        {awesome_list}
\bibfield{author}{\bibinfo{person}{Sindre Sorhus}.} \bibinfo{year}{2022}\natexlab{}.
\newblock \bibinfo{title}{Awesome lists about all kinds of interesting topics}.
\newblock
\newblock
\urldef\tempurl%
\url{https://github.com/sindresorhus/awesome}
\showURL{%
\tempurl}


\bibitem[Sourcegraph(2022)]%
        {sourcegraph}
\bibfield{author}{\bibinfo{person}{Sourcegraph}.} \bibinfo{year}{2022}\natexlab{}.
\newblock \bibinfo{title}{Sourcegraph}.
\newblock
\newblock
\urldef\tempurl%
\url{https://sourcegraph.com/search}
\showURL{%
\tempurl}


\bibitem[{Stack Exchange Community}(2021)]%
        {stackoverflow_data_dump}
\bibfield{author}{\bibinfo{person}{{Stack Exchange Community}}.} \bibinfo{year}{2021}\natexlab{}.
\newblock \bibinfo{title}{stack exchange data dump : stack exchange, inc. : free download, borrow, and streaming : internet archive}.
\newblock
\newblock
\urldef\tempurl%
\url{https://archive.org/details/stackexchange}
\showURL{%
\tempurl}


\bibitem[{Wikipedia contributors}(2022)]%
        {markdown}
\bibfield{author}{\bibinfo{person}{{Wikipedia contributors}}.} \bibinfo{year}{2022}\natexlab{}.
\newblock \bibinfo{title}{Markdown — {Wikipedia}, the free encyclopedia}.
\newblock
\newblock
\urldef\tempurl%
\url{https://en.wikipedia.org/w/index.php?title=Markdown&oldid=1130043463}
\showURL{%
\tempurl}


\end{thebibliography}

\end{document}


\title{Multi-Lingual Development \& Programming Languages Interoperability: An Empirical Study - Supplementary Information
}


\newcommand*\wrapletters[1]{\wr@pletters#1\@nil}
\def\wr@pletters#1#2\@nil{#1\allowbreak\if&#2&\else\wr@pletters#2\@nil\fi}
\newcommand{\code}[1]{\texttt{\wrapletters{#1}}}
\newcommand{\numberofissuesprojects}{9,495}

\maketitle

The supplementary information details the technical steps taken in this research, in order to allow other researchers to repeat the experiments with the same dataset, or in the future, to redo the experiment with a larger dataset.

\section{StackOverflow Data Collection \& Data Preparation}
StackOverflow posts were acquired from the Stack Overflow data dump stored at archive.org \cite{stackoverflow_data_dump}. The data dump begins at 2014, and although data dumps since 2008 do circle around the web, we prefered to gather the data from "official" sources.\\
In order to acquire data from 2008 to 2014, we used Stack Exchange Data Explorer (SEDE) \cite{sede}, a tool provides us to execute SQL commands on Stack Exchange databases. SEDE has a limit returning 50,000 rows per query. To overcome the limitation, we split the search into multiple searches. Overall we acquired 22,156,001 questions ranging from 2008 to end of December 2021.
The query executed in SEDE to retrieve the data:
\begin{singlespace}
\begin{verbatimblock}
-- @offset hops intervals of 50000 starting at 0
select *
from Posts
order by Id asc
OFFSET @offset ROWS
FETCH FIRST 50000 ROWS ONLY
\end{verbatimblock}
\end{singlespace}
The acquired data contains only the \texttt{Posts} table, which contains the questions and the answer. The data is stored in SQLite database.

As part of the data preparation, we split the "Tags" column, containing a list of text tags in a single varchar column. As Stack Overflow question can contains between 1 to 5 tags, we split the tags into 5 different columns, \texttt{tag1, tag2, tag3, tag4, tag5}.
The splitting is done by the SQL command:
\begin{singlespace}
\begin{verbatimblock}
update StackOverflowPosts
set tag1=REPLACE(substr(tags, 0, instr(tags,',')),'"', ''),
tag2=REPLACE(substr( substr(tags, instr(tags,',')+1) ,
 0,
 instr(substr(tags, instr(tags,',')+1),',')),'"',''),
tag3=REPLACE(substr(substr(substr(tags, instr(tags,',')+1),
 instr(substr(tags, instr(tags,',')+1),',')+1),
 0,
 instr(substr(substr(tags, instr(tags,',')+1),
 instr(substr(tags, instr(tags,',')+1),',')+1),',')), '"', ''),
tag4=REPLACE(substr( substr(substr(substr(tags, instr(tags,',')+1),
 instr(substr(tags, instr(tags,',')+1),',')+1),
 instr(substr(substr(tags, instr(tags,',')+1),
 instr(substr(tags, instr(tags,',')+1),',')+1),',')+1) ,
 0,
 instr(substr(substr(substr(tags, instr(tags,',')+1),
 instr(substr(tags, instr(tags,',')+1),',')+1),
 instr(substr(substr(tags, instr(tags,',')+1),
 instr(substr(tags, instr(tags,',')+1),',')+1),',')+1),',')),'"',''),
tag5=REPLACE(substr(substr(substr(substr(tags,instr(tags,',')+1),
 instr(substr(tags, instr(tags,',')+1),',')+1),
 instr(substr(substr(tags, instr(tags,',')+1),
 instr(substr(tags, 
 instr(tags,',')+1),',')+1),',')+1),
 instr(substr(substr(substr(tags, instr(tags,',')+1),
 instr(substr(tags, instr(tags,',')+1),',')+1),
 instr(substr(substr(tags, instr(tags,',')+1),
 instr(substr(tags,instr(tags,',')+1),',')+1),',')+1),',')+1),'"','')
where post_type_id = 1; -- post_type_id = 1 returns only QUESTION posts

update StackOverflowPosts set tag1=NULL where tag1='';
update StackOverflowPosts set tag2=NULL where tag2='';
update StackOverflowPosts set tag3=NULL where tag3='';
update StackOverflowPosts set tag4=NULL where tag4='';
update StackOverflowPosts set tag5=NULL where tag5='';
\end{verbatimblock}
\end{singlespace}

As part of the initial analysis, we have gathered four groups of tags relating to multi-lingual development: Languages (section \ref{sec:so_lang_set}), Wrappers \& Language ports (list detailed in section \ref{langport}), Cross-Language (section \ref{xlang}) and Interoperability Tools (section \ref{sec:interop_tools_tags}). For each group we have created a table holding their relevant tags. For the Language table, we also set if the language is a \emph{programming language} (list detailed in section \ref{sec:craft_prog_lang_sets}).\\
Stack Overflow also contain specific language-version tags, which we see as aliases (e.g. \texttt{c++11} is an alias to \texttt{c++}). Therefore we also created a language alias table holding the alias and its language (detailed in section \ref{sec:so_lang_set}).

Next, we have created a linking tables between Posts to each of the tables described above and added a link between questions to the tags they are containing:
\begin{singlespace}
\begin{verbatimblock}
-- fills PostsToLanguages table
insert into PostsToLanguages(PostID, Language)
SELECT StackOverflowPosts.id, Languages.Name
from StackOverflowPosts, Languages
where
    (tag1=Languages.Name OR tag1 in 
      (select Alias from LanguagesAliases where 
        LanguagesAliases.Language=Languages.Name)) OR
    (tag2=Languages.Name OR tag2 in
      (select Alias from LanguagesAliases where 
        LanguagesAliases.Language=Languages.Name)) OR
    (tag3=Languages.Name OR tag3 in 
      (select Alias from LanguagesAliases where 
        LanguagesAliases.Language=Languages.Name)) OR
    (tag4=Languages.Name OR tag4 in 
      (select Alias from LanguagesAliases where 
        LanguagesAliases.Language=Languages.Name)) OR
    (tag5=Languages.Name OR tag5 in 
      (select Alias from LanguagesAliases where 
        LanguagesAliases.Language=Languages.Name));


-- fills PostsToCrossLanguage table
insert into PostsToCrossLanguage(PostID, Tag)
SELECT id, tag1 from StackOverflowPosts where
    tag1 in (select name from CrossLanguage);
insert into PostsToCrossLanguage(PostID, Tag)
SELECT id, tag2 from StackOverflowPosts where
    tag2 in (select name from CrossLanguage);
insert into PostsToCrossLanguage(PostID, Tag)
SELECT id, tag3 from StackOverflowPosts where
    tag3 in (select name from CrossLanguage);
insert into PostsToCrossLanguage(PostID, Tag)
SELECT id, tag4 from StackOverflowPosts where
    tag4 in (select name from CrossLanguage);
insert into PostsToCrossLanguage(PostID, Tag)
SELECT id, tag5 from StackOverflowPosts where
    tag5 in (select name from CrossLanguage);

-- fills PostsToInteroperabilityTools table
insert into PostsToInteroperabilityTools(PostID, Tag) 
SELECT id, tag1 from StackOverflowPosts where
    tag1 in (select name from InteroperabilityTools);
insert into PostsToInteroperabilityTools(PostID, Tag) 
SELECT id, tag2 from StackOverflowPosts where
    tag2 in (select name from InteroperabilityTools);
insert into PostsToInteroperabilityTools(PostID, Tag) 
SELECT id, tag3 from StackOverflowPosts where
    tag3 in (select name from InteroperabilityTools);
insert into PostsToInteroperabilityTools(PostID, Tag) 
SELECT id, tag4 from StackOverflowPosts where
    tag4 in (select name from InteroperabilityTools);
insert into PostsToInteroperabilityTools(PostID, Tag)
SELECT id, tag5 from StackOverflowPosts where
    tag5 in (select name from InteroperabilityTools);

-- fills PostsToWrappersAndLanguagePorts table
insert into PostsToWrappersAndLanguagePorts(PostID, Tag)
SELECT id, tag1 from StackOverflowPosts where tag1 in
        (select name from WrappersAndLanguagePorts);
insert into PostsToWrappersAndLanguagePorts(PostID, Tag)
SELECT id, tag2 from StackOverflowPosts where tag2 in
        (select name from WrappersAndLanguagePorts);
insert into PostsToWrappersAndLanguagePorts(PostID, Tag) 
SELECT id, tag3 from StackOverflowPosts where tag3 in
        (select name from WrappersAndLanguagePorts);
insert into PostsToWrappersAndLanguagePorts(PostID, Tag)
SELECT id, tag4 from StackOverflowPosts where tag4 in
        (select name from WrappersAndLanguagePorts);
insert into PostsToWrappersAndLanguagePorts(PostID, Tag) 
SELECT id, tag5 from StackOverflowPosts where tag5
    in (select name from WrappersAndLanguagePorts);
\end{verbatimblock}
\end{singlespace}

One last calculation, is to calculate when a question has been resolved (if question was resolved). If \texttt{accepted\_answer\_id} is set, it contains the ID of the answer (stored in Posts table). We set the creation date of the accepted answer under a new column \texttt{answer\_date}.

\section{GitHub Data Collection}
\subsection{Projects MetaData}\label{sec:github_projects_metadata}
We have used GitHub API in order to acquire the metadata of all public GitHub projects (or repositories, used interchangeably). To be able to filter only for projects with at least 50 stars, we have used GitHub Search API (\texttt{https://api.github.com/search/repositories}).\\
The search API allows us to filter by the number of stars that we receive by applying the stars filter in the search (\texttt{stars:>=50}).\\
The GitHub search API has a limit of returning 1,000 projects per search. In order to overcome this limitation, we have used the creation date filter to minimise the search\\(for example, \texttt{created:"2010-02-15..2010-02-25"}). In order to download all the projects (with at least 50 stars), we have searched the time frame we required, and if exactly 1,000 results returned, we split the request into two requests, splitting the time frame in half. The \code{created} filter does also support time, but in our case (due to the 50 stars filter), we did not need to use it. Also, GitHub has a rate limit of requests, therefore to speed up the acquiring process we are using several GitHub accounts to have a higher rate limit.\\
In order to retrieve the language statistics of each project returned from the search API, we perform additional REST API request. The URL returning the language statistics is set under the \texttt{languages\_url} field within the data returned from the search API for each project.\\
The data returned for each project, including the language statistics, is stored for the research analysis in 360Mb file in JSON \cite{json_rfc} format. The list of fields is detailed in section \ref{sec:github_project_metadata_fields}.

\subsection{Interoperability Tools}
We have tracked and analysed 173 interoperability tools. For each interoperability tool we detailed the calling language, the target language, intermediate languages (if there are any) and the type/s of interoperability Tool (Language port, compiler, library and multi-process).\\
Tools we have found to have a Stack Overflow tag, are part of the Interoperability Tools tags, which are taken into account when calculated "multi-lingual related" posts.

The Interoperability tools are as follows:
\begin{longtblr}[caption = {Interoperability Tools}, label = {tab:interoperabilitytools},]{
  colspec = {|XXXXX|},
  rowhead = 1,
  hlines,
  row{even} = {gray9},
  row{1} = {olive9},
}
\hline
\textbf{Tool} & \textbf{From} & \textbf{To} & \textbf{Intermediate} & \textbf{Mechanism} \\ \hline
lua-api & c++,c & lua & - & api \\ \hline
luabind & lua & c & - & api \\ \hline
rnetlogo & r & netlogo & c,java & api \\ \hline
pynetlogo & python & netlogo & c,java & api \\ \hline
bridging-header & \begin{tabular}[c]{@{}l@{}}objective-c++,c++,\\ objective-c,swift,c\end{tabular} & \begin{tabular}[c]{@{}l@{}}objective-c++,c++,\\ objective-c,swift,c\end{tabular} & - & api \\ \hline
objc-bridging-header & \begin{tabular}[c]{@{}l@{}}objective-c,\\ objective-c++\end{tabular} & swift,c++,c & - & api \\ \hline
c2hs & haskell & c & - & api \\ \hline
hsc2hs & c & haskell & - & api \\ \hline
c2hsc & c & haskell & - & api \\ \hline
cgo & go & c & - & api \\ \hline
gccgo & go & c & - & api \\ \hline
ctypes & python & c & - & api \\ \hline
jsctypes & javascript & c & - & api \\ \hline
python-ctypes & python & c & - & api \\ \hline
cython & python & c & - & api \\ \hline
cythonize & python & c & - & api \\ \hline
derelict3 & d & c & - & api \\ \hline
f2c & fortran & c & - & api \\ \hline
fortran-iso-c-binding & fortran,c & c & - & api \\ \hline
gobject-introspection & \begin{tabular}[c]{@{}l@{}}rust,smalltalk,rex,\\ guile,julia,java,\\ javascript,j,gambas,\\ lua,objective-c,fortran\\ ,c,groovy,jython,\\ scala,r,c++,pascal,\\ crystal-lang,ocaml,\\ c\#,ruby,vala,prolog\\ ,freebasic,\\ ada,d,perl,kotlin,\\ erlang,go,\\ tcl,ml,genie,haskell,\\ php,python\end{tabular} & c & - & api \\ \hline
inline-assembly & c++,c & assembly & - & api \\ \hline
javah & java & c & - & api \\ \hline
java-native-interface & java,c & java,c & - & api \\ \hline
jpl & java & prolog & c & api \\ \hline
lua-userdata & lua & c & - & api \\ \hline
pyobjc & python & objective-c & c & api \\ \hline
python-c-api & c & python & - & api \\ \hline
python-cffi & python & c & - & api \\ \hline
ruby-c-extension & ruby & c & - & api \\ \hline
swig & \begin{tabular}[c]{@{}l@{}}modula-3,scilab,\\ ocaml,lisp,ruby,\\ c\#,clisp,guile,java,\\ javascript,d,\\ scheme,lua,r,perl,\\ go,octave,tcl,\\ pike,php,python\end{tabular} & c++,c & - & api \\ \hline
lbffi & \begin{tabular}[c]{@{}l@{}}perl,lisp,ruby,guile,\\ haskell,java,javascript,\\ f-script,racket,python,d\end{tabular} & c & - & api \\ \hline
vb.net-to-c\# & vb.net & c\# & - & api \\ \hline
c\#-to-vb.net & c\# & vb.net & - & api \\ \hline
python4delphi & delphi & python & c & api \\ \hline
luainterface & vb.net,c\#,c++-cli,lua & \begin{tabular}[c]{@{}l@{}}vb.net,c\#,\\ c++-cli,lua\end{tabular} & - & language\_port     \\ \hline
luabridge & lua & c++ & c & api \\ \hline
nlua & c\# & lua & - & language\_port     \\ \hline
jnlua & java & lua & c & api \\ \hline
luajava & java,lua & java,lua & c & api \\ \hline
luaj & java & lua & - & api \\ \hline
fable-f\# & f\# & javascript & - & compiler           \\ \hline
f\#-giraffe & f\# & asp.net-core & - & api \\ \hline
js-of-ocaml & ocaml & javascript & - & api \\ \hline
erlang-nif & erlang & c & - & api \\ \hline
jni4net & vb.net,c\#,java,c++-cli & \begin{tabular}[c]{@{}l@{}}vb.net,c\#,java,\\ c++-cli\end{tabular} & c & api \\ \hline
py4j & java,python & java,python & - & multiprocess       \\ \hline
interprolog & java,prolog & java,prolog & - & multiprocess       \\ \hline
boost-python & c++ & python & - & api \\ \hline
f2py & fortran & python & c & api \\ \hline
rpy2 & python & r & c & api \\ \hline
python.net & python & vb.net,c\#,c++-cli & c & api \\ \hline
lispyscript & lisp & javascript & - & compiler           \\ \hline
com-interop & \begin{tabular}[c]{@{}l@{}}vb.net,c\#,c++,\\ delphi,powershell,\\ python,c\end{tabular} & \begin{tabular}[c]{@{}l@{}}vb.net,c\#,c++,\\ delphi,c\end{tabular} & - & api \\ \hline
interopservices & vb.net,c\# & c & - & api \\ \hline
kotlin-js-interop & kotlin & javascript & - & language\_port     \\ \hline
jruby-win32ole & ruby & c & - & api \\ \hline
gwt-jsinterop & java & javascript & - & compiler           \\ \hline
scala-java-interop & java,scala & java,scala & - & api \\ \hline
clojure-java-interop & java,clojure & java,clojure & - & api \\ \hline
jruby-java-interop & java,ruby & java,ruby & - & language\_port     \\ \hline
kotlin-java-interop & kotlin & java & - & api \\ \hline
clojurescript-javascript-interop & clojure & javascript & - & compiler           \\ \hline
ballerina-java-interop & ballerina & java & - & api \\ \hline
pybind11 & c++ & python & c & api \\ \hline
jna & java & c++ & c & api \\ \hline
jnativehook & java & c & - & api \\ \hline
jnaerator & java & objective-c,c++,c & c & api \\ \hline
djnativeswing & java & c & - & api \\ \hline
autoit-c\#-wrapper & autoit & c\# & c & api \\ \hline
android-jsinterface & java & javascript & c & api \\ \hline
pycall & julia,ruby & python & c & api \\ \hline
pythoninterpreter & \begin{tabular}[c]{@{}l@{}}kotlin,java,\\ groovy,scala\end{tabular} & python & - & language\_port     \\ \hline
ironpython & vb.net,c\#,python & python & - & language\_port     \\ \hline
scalapy & scala & python & c & api \\ \hline
jpype & python & java & c & api \\ \hline
pyjnius & python & java & c & api \\ \hline
scalajs & scala & javascript & - & compiler           \\ \hline
gopherjs & javascript & go & - & compiler           \\ \hline
renjin & java,r & java,r & - & language\_port     \\ \hline
perlapi & c & perl & - & api \\ \hline
rcaller & java & r & - & multiprocess       \\ \hline
haskell-ffi & haskell & c & - & api \\ \hline
quickjs & c & javascript & - & api \\ \hline
pinvoke & vb.net,c\# & c & - & api \\ \hline
fiddle & ruby & c & - & api \\ \hline
rjava & r & java & c & api \\ \hline
rhino & java & javascript & - & language\_port     \\ \hline
llvm & \begin{tabular}[c]{@{}l@{}}rust,c++,hydra,esl,\\ crack,ruby,ada,\\ scheme,d,vuo,lua,\\ emscripten,fortran,\\ c,pure,pony\end{tabular} & \begin{tabular}[c]{@{}l@{}}rust,c++,hydra,esl,\\ crack,ruby,ada,\\ scheme,d,vuo,lua,\\ emscripten,fortran,\\ c,pure,pony\end{tabular} & llvm-ir & compiler           \\ \hline
smgo & go & javascript & c & api \\ \hline
javabridge & python & java & c & api \\ \hline
pythonkit & switft & python & c & api \\ \hline
rubypython & ruby & python & c & api \\ \hline
go-python & go & python & c & api \\ \hline
python-script-engine & \begin{tabular}[c]{@{}l@{}}kotlin,java,\\ groovy,scala\end{tabular} & python & - & language\_port     \\ \hline
java-script-engine & \begin{tabular}[c]{@{}l@{}}kotlin,java,\\ groovy,scala\end{tabular} & java & - & language\_port     \\ \hline
groovy-script-engine & \begin{tabular}[c]{@{}l@{}}kotlin,java,\\ groovy,scala\end{tabular} & groovy & - & language\_port     \\ \hline
scala-script-engine & \begin{tabular}[c]{@{}l@{}}kotlin,java,\\ groovy,scala\end{tabular} & scala & - & language\_port     \\ \hline
kotlin-script-engine & \begin{tabular}[c]{@{}l@{}}kotlin,java,\\ groovy,scala\end{tabular} & kotlin & - & language\_port     \\ \hline
lua-script-engine & \begin{tabular}[c]{@{}l@{}}kotlin,java,\\ groovy,scala\end{tabular} & lua & - & language\_port     \\ \hline
ruby-script-engine & \begin{tabular}[c]{@{}l@{}}kotlin,java,\\ groovy,scala\end{tabular} & ruby & - & language\_port     \\ \hline
perl-script-engine & \begin{tabular}[c]{@{}l@{}}kotlin,java,\\ groovy,scala\end{tabular} & perl & - & language\_port     \\ \hline
haskell-script-engine & \begin{tabular}[c]{@{}l@{}}kotlin,java,\\ groovy,scala\end{tabular} & haskell & - & language\_port     \\ \hline
javascript-script-engine & \begin{tabular}[c]{@{}l@{}}kotlin,java,\\ groovy,scala\end{tabular} & javascript & - & language\_port     \\ \hline
lisp-script-engine & \begin{tabular}[c]{@{}l@{}}kotlin,java,\\ groovy,scala\end{tabular} & lisp & - & language\_port     \\ \hline
.net-load & c & vb.net,c\# & - & api \\ \hline
mono-load & c & vb.net,c\# & - & api \\ \hline
csml & ocaml & vb.net,c\# & c & api \\ \hline
scala-java-conversion & scala & java & - & api \\ \hline
lua.vm.js & javascript & lua & - & language\_port     \\ \hline
dynamic-lua & vb.net,c\# & lua & - & language\_port     \\ \hline
gopher-lua & go & lua & - & language\_port     \\ \hline
go-lua & go & lua & - & language\_port     \\ \hline
goluajit & go & lua & c & api \\ \hline
rustpython & rust & python & - & language\_port     \\ \hline
m2cgen-ruby & python & ruby & - & compiler           \\ \hline
ruby-api & c & ruby & - & api \\ \hline
rufus-lua & ruby & lua & c & api \\ \hline
coldruby & javascript & ruby & - & compiler           \\ \hline
go-mruby & go & ruby & c & api \\ \hline
embedded-r & c & r & - & api \\ \hline
statistics::r & perl & r & - & multiprocess       \\ \hline
rinruby & ruby & r & - & multiprocess       \\ \hline
ocaml-r & ocaml & r & c & api \\ \hline
campher & go & perl & c & api \\ \hline
hapy & haskell,python & haskell,python & c & api \\ \hline
jaskell & java & haskell & - & language\_port     \\ \hline
c-js & c & javascript & - & api \\ \hline
clearscript & c\# & javascript & - & language\_port     \\ \hline
v8.net & c\# & javascript & c & api \\ \hline
js2py & python & javascript & - & language\_port     \\ \hline
pyminirace & python & javascript & c & api \\ \hline
mini\_racer & ruby & javascript & c & api \\ \hline
therubyracer & ruby & javascript & c & api \\ \hline
pyv8 & python & javascript & c & api \\ \hline
otto & go & javascript & - & language\_port     \\ \hline
gov8 & go & javascript & c & api \\ \hline
v8go & go & javascript & c & api \\ \hline
elsa & go & javascript & - & language\_port     \\ \hline
lisp-c & c & lisp & - & api \\ \hline
hy & python & lisp & - & language\_port     \\ \hline
cslisp & c\# & lisp & - & language\_port     \\ \hline
cs-powershell & c\# & powershell & - & api \\ \hline
rust-ffi & rust & c & - & api \\ \hline
ruby-ffi & ruby & c & - & api \\ \hline
perl-ffi & perl & c & - & api \\ \hline
lua-ffi & lua & c & - & api \\ \hline
lua-cffi & lua & c & - & api \\ \hline
luaffi & lua & c & - & api \\ \hline
luaffifb & lua & c & - & api \\ \hline
ffi-platypus & perl & c & - & api \\ \hline
sbffi & javascript & c & - & api \\ \hline
nodeffi & javascript & c & - & api \\ \hline
lisp-ffi & lisp & c & - & api \\ \hline
matlabffi & matlab & c & - & api \\ \hline
cygnus & c++ & java & c & api \\ \hline
dart:ffi & dart & c & - & api \\ \hline
php-ffi & php & c & - & api \\ \hline
racket-ffi & racket & c & - & api \\ \hline
julia-ffi & julia & c & - & api \\ \hline
java-native & java & c & - & api \\ \hline
groovy-native & groovy & c & - & api \\ \hline
kotlin-native & kotlin & c & - & api \\ \hline
scala-native & scala & c & - & api \\ \hline
jsm & javascript & c & - & api \\ \hline
jscocoa & javascript & objective-c,javascript & c & api \\ \hline
rubycocoa & ruby & objective-c & c & api \\ \hline
macruby & ruby & objective-c & c & api \\ \hline
luacore & lua & objective-c & c & api \\ \hline
r-ffi & r & c & - & api \\ \hline
c-haskell-ffi & c & haskell & - & api \\ \hline
execjs & ruby & javascript & c & api \\ \hline
therubyrhino & ruby & javascript & - & language\_port     \\ \hline
duktape & ruby & javascript & c & api \\ \hline
javascriptcore & objective-c,swift,c & javascript & - & api \\ \hline
\end{longtblr}

\subsection{Searching Within Source Code}
Acquiring all the source code is a time consuming effort with steep storage requirements. Therefore, we are using sourcegraph.com \cite{sourcegraph}, which indexes all GitHub public projects. SourceGraph allow us to search using text and regular expressions on all GitHub projects by filtering to a specific language (based on file extension).\\
In order to search, we are using SourceGraph REST API, using the REST function \texttt{https://sourcegraph.com/.api/search/stream}.

\subsection{GitHub Issues \& Discussions}
Similar to acquiring source code, acquiring all the issues and discussions is a time consuming effort with steep storage requirements. Therefore, we have acquired issues and discussion groups of \numberofissuesprojects{} selected public projects in different languages using awesome lists provided by \emph{sindresorhus' Awesome List} \cite{awesome_list}. \\
In order to retrieve all the projects in all the Awesome lists, we acquire the awesome lists and parse them (Awesome lists are written in Markdown \cite{markdown}). For example, \emph{sindresorhus' Awesome List} links to C++ awesome list in the URL \texttt{https://github.com/fffaraz/awesome-cpp/}, therefore we download the awesome list raw file: \code{https://raw.githubusercontent.com/fffaraz/awesome-cpp/master/README.md}. For each awesome list, we are acquiring only projects within GitHub.\\
Acquiring issues and discussions of GitHub projects is only available via GitHub GraphQL API.\\
To retrieve information regarding the project we perform the following query:
\begin{singlespace}
\begin{verbatimblock}
query {
  repository(owner: "[OWNER]", name: "[NAME]") {
    name
    url
    homepageUrl
    languages(first: 100) {
      nodes {
        name
      }
    }
  }
}
\end{verbatimblock}
\end{singlespace}

To retrieve project discussions we perform the following query:
\begin{singlespace}
\begin{verbatimblock}
query {
  repository(owner: "[OWNER]", name: "[NAME]") {
    # "after" is used only if pagination is required
    discussions(first: 100, after: [AFTER]) {
      totalCount

      pageInfo {
        endCursor
        hasNextPage
      }

      nodes {
        # type: Discussion
        id
        title
        bodyText
        answer {
          isAnswer
          id
          bodyText
        }

        comments(first: 100) {
          totalCount
          pageInfo {
            endCursor
            hasNextPage
          }
          nodes {
            id
            bodyText
          }
        }
      }
    }
  }
}
\end{verbatimblock}
\end{singlespace}

To retrieve project issues we perform the following query:
\begin{singlespace}
\begin{verbatimblock}
query {
  repository(owner: "[OWNER]", name: "[NAME]") {
    # "after" is used only if pagination is required
    issues(first: 100, after: [AFTER]) {
      totalCount

      pageInfo {
        endCursor
        hasNextPage
      }

      nodes {
        id
        title
        bodyText
        state
        comments(first: 100) {
          nodes {
            bodyText
          }
        }

        labels(first: 100) {
          nodes {
            name
            description
          }
        }
      }
    }
  }
}
\end{verbatimblock}
\end{singlespace}

The discussions and issues are stored in JSON files.

\section{Language \& Programming Language Sets}

\subsection{Crafting GitHub Language Set}
By unifying all the languages of all the projects returned by GitHub API (as described in section \ref{sec:github_projects_metadata}), we create a single set of all the languages found in GitHub. Notice, GitHub detects languages using the file extension, therefore it might not find languages exist in Stack Overflow. For example, C++ and C++.Net are not the same language, but use the same file extension (\texttt{CPP}).

The GitHub language set is as follows:\\
\begin{multicols}{4}
    \begin{itemize}
        \item	1c enterprise
        \item	abap
        \item	abap cds
        \item	actionscript
        \item	ada
        \item	agda
        \item	ags script
        \item	aidl
        \item	al
        \item	alloy
        \item	ampl
        \item	angelscript
        \item	antlr
        \item	apacheconf
        \item	apex
        \item	api blueprint
        \item	apl
        \item	applescript
        \item	arc
        \item	arduino
        \item	asciidoc
        \item	asl
        \item	asp
        \item	asp.net
        \item	aspectj
        \item	assembly
        \item	astro
        \item	asymptote
        \item	ats
        \item	augeas
        \item	autohotkey
        \item	autoit
        \item	awk
        \item	ballerina
        \item	basic
        \item	batchfile
        \item	beef
        \item	befunge
        \item	bicep
        \item	bison
        \item	bitbake
        \item	blade
        \item	blitzbasic
        \item	blitzmax
        \item	bluespec
        \item	boo
        \item	boogie
        \item	brainfuck
        \item	brightscript
        \item	bro
        \item	c
        \item	c\#
        \item	c++
        \item	capn proto
        \item	cartocss
        \item	ceylon
        \item	chapel
        \item	charity
        \item	chuck
        \item	cirru
        \item	clarion
        \item	classic asp
        \item	clean
        \item	click
        \item	clips
        \item	clojure
        \item	closure templates
        \item	cmake
        \item	cobol
        \item	codeql
        \item	coffeescript
        \item	coldfusion
        \item	common lisp
        \item	common workflow language
        \item	component pascal
        \item	cool
        \item	coq
        \item	crystal
        \item	csound
        \item	csound document
        \item	csound score
        \item	css
        \item	cuda
        \item	cue
        \item	cweb
        \item	cycript
        \item	cython
        \item	d
        \item	dafny
        \item	dart
        \item	dataweave
        \item	dcpu-16 asm
        \item	dhall
        \item	diff
        \item	digital command language
        \item	dm
        \item	dockerfile
        \item	dogescript
        \item	dtrace
        \item	dylan
        \item	e
        \item	eagle
        \item	ec
        \item	ecl
        \item	eiffel
        \item	ejs
        \item	elixir
        \item	elm
        \item	emacs lisp
        \item	emberscript
        \item	eq
        \item	erlang
        \item	f\#
        \item	f*
        \item	factor
        \item	fancy
        \item	fantom
        \item	faust
        \item	fennel
        \item	filebench wml
        \item	fluent
        \item	flux
        \item	forth
        \item	fortran
        \item	freebasic
        \item	freemarker
        \item	frege
        \item	futhark
        \item	g-code
        \item	game maker language
        \item	gaml
        \item	gams
        \item	gap
        \item	gcc machine description
        \item	gdb
        \item	gdscript
        \item	genie
        \item	genshi
        \item	gettext catalog
        \item	gherkin
        \item	glsl
        \item	glyph
        \item	gnuplot
        \item	go
        \item	golo
        \item	gosu
        \item	grace
        \item	gradle
        \item	grammatical framework
        \item	graphviz (dot)
        \item	groff
        \item	groovy
        \item	hack
        \item	haml
        \item	handlebars
        \item	haproxy
        \item	harbour
        \item	haskell
        \item	haxe
        \item	hcl
        \item	hiveql
        \item	hlsl
        \item	holyc
        \item	html
        \item	hy
        \item	hyphy
        \item	idl
        \item	idris
        \item	igor pro
        \item	imagej macro
        \item	inform 7
        \item	ini
        \item	inno setup
        \item	io
        \item	ioke
        \item	isabelle
        \item	j
        \item	jasmin
        \item	java
        \item	javascript
        \item	jflex
        \item	jinja
        \item	jolie
        \item	jq
        \item	json
        \item	jsoniq
        \item	jsonnet
        \item	julia
        \item	kaitai struct
        \item	kakounescript
        \item	kicad layout
        \item	kicad schematic
        \item	kit
        \item	kotlin
        \item	krl
        \item	labview
        \item	lasso
        \item	latte
        \item	lean
        \item	less
        \item	lex
        \item	lfe
        \item	lilypond
        \item	limbo
        \item	liquid
        \item	livescript
        \item	llvm
        \item	logos
        \item	logtalk
        \item	lolcode
        \item	lookml
        \item	loomscript
        \item	lsl
        \item	lua
        \item	m
        \item	m4
        \item	macaulay2
        \item	makefile
        \item	mako
        \item	markdown
        \item	marko
        \item	mask
        \item	mathematica
        \item	matlab
        \item	max
        \item	maxscript
        \item	mcfunction
        \item	mercury
        \item	meson
        \item	metal
        \item	mirah
        \item	mirc script
        \item	mlir
        \item	modelica
        \item	modula-2
        \item	modula-3
        \item	module management system
        \item	monkey
        \item	moocode
        \item	moonscript
        \item	mql4
        \item	mql5
        \item	mtml
        \item	mupad
        \item	mustache
        \item	myghty
        \item	nasl
        \item	ncl
        \item	nearley
        \item	nemerle
        \item	nesc
        \item	netlinx
        \item	netlinx+erb
        \item	netlogo
        \item	newlisp
        \item	nextflow
        \item	nginx
        \item	nim
        \item	nit
        \item	nix
        \item	nodejs
        \item	nsis
        \item	nu
        \item	nunjucks
        \item	nwscript
        \item	objective-c
        \item	objective-c++
        \item	objective-j
        \item	objectscript
        \item	ocaml
        \item	odin
        \item	omgrofl
        \item	ooc
        \item	opa
        \item	opal
        \item	open policy agent
        \item	openedge abl
        \item	openqasm
        \item	openscad
        \item	org
        \item	ox
        \item	oxygene
        \item	oz
        \item	p4
        \item	pan
        \item	papyrus
        \item	parrot
        \item	pascal
        \item	pawn
        \item	peg.js
        \item	pep8
        \item	perl
        \item	perl 6
        \item	php
        \item	picolisp
        \item	piglatin
        \item	pike
        \item	plantuml
        \item	plpgsql
        \item	plsql
        \item	pogoscript
        \item	pony
        \item	postscript
        \item	pov-ray sdl
        \item	powerbuilder
        \item	powershell
        \item	processing
        \item	prolog
        \item	propeller spin
        \item	protocol buffer
        \item	pug
        \item	puppet
        \item	pure data
        \item	purebasic
        \item	purescript
        \item	python
        \item	q
        \item	q\#
        \item	qmake
        \item	qml
        \item	qt script
        \item	quake
        \item	r
        \item	racket
        \item	ragel
        \item	ragel in ruby host
        \item	raku
        \item	raml
        \item	rascal
        \item	realbasic
        \item	reason
        \item	rebol
        \item	red
        \item	redcode
        \item	renpy
        \item	renderscript
        \item	rescript
        \item	restructuredtext
        \item	rexx
        \item	rich text format
        \item	ring
        \item	riot
        \item	rmarkdown
        \item	robotframework
        \item	roff
        \item	rouge
        \item	rpc
        \item	ruby
        \item	runoff
        \item	rust
        \item	sage
        \item	saltstack
        \item	sas
        \item	sass
        \item	scala
        \item	scaml
        \item	scheme
        \item	scilab
        \item	scss
        \item	sed
        \item	self
        \item	shaderlab
        \item	shell
        \item	shellsession
        \item	shen
        \item	sieve
        \item	singularity
        \item	slash
        \item	slice
        \item	slim
        \item	smali
        \item	smalltalk
        \item	smarty
        \item	smpl
        \item	smt
        \item	solidity
        \item	sourcepawn
        \item	sqf
        \item	sql
        \item	sqlpl
        \item	squirrel
        \item	srecode template
        \item	stan
        \item	standard ml
        \item	starlark
        \item	stata
        \item	stringtemplate
        \item	stylus
        \item	supercollider
        \item	svelte
        \item	svg
        \item	swift
        \item	swig
        \item	systemverilog
        \item	tcl
        \item	tea
        \item	terra
        \item	tex
        \item	thrift
        \item	ti program
        \item	tla
        \item	tsql
        \item	turing
        \item	twig
        \item	txl
        \item	typescript
        \item	uno
        \item	unrealscript
        \item	urweb
        \item	v
        \item	vala
        \item	vba
        \item	vbscript
        \item	vcl
        \item	verilog
        \item	vhdl
        \item	vim script
        \item	vim snippet
        \item	visual basic
        \item	visual basic .net
        \item	volt
        \item	vue
        \item	wdl
        \item	web ontology language
        \item	webassembly
        \item	webidl
        \item	wisp
        \item	wollok
        \item	x10
        \item	xbase
        \item	xc
        \item	xml
        \item	xojo
        \item	xonsh
        \item	xpages
        \item	xproc
        \item	xquery
        \item	xs
        \item	xslt
        \item	xtend
        \item	yacc
        \item	yaml
        \item	yara
        \item	yasnippet
        \item	zap
        \item	zeek
        \item	zenscript
        \item	zephir
        \item	zig
        \item	zil
        \item	zimpl
    \end{itemize}
\end{multicols}

\subsection{Stack Overflow Language Set}\label{sec:so_lang_set}
Languages in Stack Overflow are expressed by tags. Each Stack Overflow question have must between one to five tags to the questions, where some of the tags are represent languages. In order to find language tags (and other types of tags as presented in section \ref{sec:so_tags}), we have used Stack Exchange Data Explorer (SEDE) \cite{sede} to get a list of all the tags and their description. Once we have a list of all the tags, we went through them manually to find relevant tags.\\
Due to the large number of tags that have been used at least once (45,175 tags), we went through tags that have been used at least 400 times (i.e. in 400 different questions, which is 0.001\% of questions), which drops the amount of tags to 9,712.\\
To get the tags and their description, we execute the following SQL command in Stack Overflow SEDE (https://data.stackexchange.com/stackoverflow/):
\begin{singlespace}
\begin{verbatimblock}
select TagName, Body As Description
from Tags, Posts
where Tags.count >= 400 and Tags.ExcerptPostId=Posts.Id
\end{verbatimblock}
\end{singlespace}

As stated earlier, some language tags are version specific, therefore we treat them as aliases. The list of aliases is as follows:\\
\begin{multicols}{3}
    \begin{itemize}
        \item f\#-3.0 $\rightarrow$ f\#
        \item f\#-4.0 $\rightarrow$ f\#
        \item f\#-3.1 $\rightarrow$ f\#
        \item f\#-4.1 $\rightarrow$ f\#
        \item silverlight-3.0 $\rightarrow$ silverlight
        \item silverlight-5.0 $\rightarrow$ silverlight
        \item silverlight-2.0 $\rightarrow$ silverlight
        \item silverlight-plugin $\rightarrow$ silverlight
        \item silverlight-4.0 $\rightarrow$ silverlight
        \item silverlight-embedded $\rightarrow$ silverlight-oob
        \item jabaco $\rightarrow$ vb6
        \item nativescript-vue $\rightarrow$ nativescript
        \item angular2-nativescript $\rightarrow$ nativescript
        \item nativescript-angular $\rightarrow$ nativescript
        \item kawa $\rightarrow$ scheme
        \item sisc $\rightarrow$ scheme
        \item guile $\rightarrow$ scheme
        \item bigloo $\rightarrow$ scheme
        \item jython-2.7 $\rightarrow$ python
        \item jython-2.5 $\rightarrow$ python
        \item python-2.x $\rightarrow$ python
        \item activepython $\rightarrow$ python
        \item python2.7 $\rightarrow$ python
        \item python-3.6 $\rightarrow$ python
        \item jython $\rightarrow$ python
        \item jython-2.2 $\rightarrow$ python
        \item python-3.x $\rightarrow$ python
        \item python-3.7 $\rightarrow$ python
        \item python-3.4 $\rightarrow$ python
        \item graalpython $\rightarrow$ python
        \item python-3.8 $\rightarrow$ python
        \item python-3.9 $\rightarrow$ python
        \item genie.jl $\rightarrow$ julia
        \item oorex $\rightarrow$ rexx
        \item netrexx $\rightarrow$ rexx
        \item iso-prolog $\rightarrow$ prolog
        \item b-prolog $\rightarrow$ prolog
        \item turbo-prolog $\rightarrow$ prolog
        \item yap-prolog $\rightarrow$ prolog
        \item prologscript $\rightarrow$ prolog
        \item jiprolog $\rightarrow$ prolog
        \item sicstus-prolog $\rightarrow$ prolog
        \item gnu-prolog $\rightarrow$ prolog
        \item tau-prolog $\rightarrow$ prolog
        \item lambda-prolog $\rightarrow$ prolog
        \item ruby-prolog $\rightarrow$ prolog
        \item swi-prolog $\rightarrow$ prolog
        \item visual-prolog $\rightarrow$ prolog
        \item swi-prolog-for-sharing $\rightarrow$ prolog
        \item llvm-3.1 $\rightarrow$ llvm
        \item llvm5.1 $\rightarrow$ llvm
        \item llvm-3.2 $\rightarrow$ llvm
        \item llvm-3.0 $\rightarrow$ llvm
        \item llvm-4.0 $\rightarrow$ llvm
        \item genie $\rightarrow$ vala
        \item c++11 $\rightarrow$ c++
        \item visual-c++ $\rightarrow$ c++
        \item c++17 $\rightarrow$ c++
        \item c++14 $\rightarrow$ c++
        \item c++20 $\rightarrow$ c++
        \item mirah $\rightarrow$ ruby
        \item jruby $\rightarrow$ ruby
        \item truffleruby $\rightarrow$ ruby
        \item eta $\rightarrow$ haskell
        \item frege $\rightarrow$ haskell
        \item php-7 $\rightarrow$ php
        \item clojure-1.3 $\rightarrow$ clojure
        \item clojurescript $\rightarrow$ clojure
        \item clojureclr $\rightarrow$ clojure
        \item bash4 $\rightarrow$ bash
        \item swift2 $\rightarrow$ swift
        \item swift5 $\rightarrow$ swift
        \item vb.net-2010 $\rightarrow$ vb.net
        \item react $\rightarrow$ javascript
        \item vue.js $\rightarrow$ javascript
        \item oxygene $\rightarrow$ pascal
        \item freepascal $\rightarrow$ pascal
        \item pascalscript $\rightarrow$ pascal
        \item omnipascal $\rightarrow$ pascal
        \item component-pascal $\rightarrow$ pascal
        \item turbo-pascal $\rightarrow$ pascal
        \item asp.net-mvc-4 $\rightarrow$ asp.net
        \item asp.net-mvc-5 $\rightarrow$ asp.net
        \item asp.net-mvc $\rightarrow$ asp.net
        \item vba7 $\rightarrow$ vba
        \item vba6 $\rightarrow$ vba
        \item masm64 $\rightarrow$ masm
        \item armasm $\rightarrow$ masm
        \item masm32 $\rightarrow$ masm
        \item fastr $\rightarrow$ r
        \item renjin $\rightarrow$ r
        \item asp.net-core-mvc $\rightarrow$ asp.net-core
        \item asp.net-core-2 $\rightarrow$ asp.net-core
        \item inline-assembly $\rightarrow$ assembly
        \item cypher-shell $\rightarrow$ neo4j
        \item anormcypher $\rightarrow$ neo4j
        \item cypher $\rightarrow$ neo4j
        \item cypher-3.1 $\rightarrow$ neo4j
        \item game-maker-language $\rightarrow$ gml
        \item ruby-on-rails-3.2 $\rightarrow$ ruby-on-rails
        \item jrubyonrails $\rightarrow$ ruby-on-rails
        \item ruby-on-rails-5 $\rightarrow$ ruby-on-rails
        \item typescript2.0 $\rightarrow$ typescript
        \item typescript1.8 $\rightarrow$ typescript
        \item typescript1.6 $\rightarrow$ typescript
        \item typescript1.7 $\rightarrow$ typescript
        \item typescript-2.5 $\rightarrow$ typescript
        \item typescript2.8 $\rightarrow$ typescript
        \item typescript2.7 $\rightarrow$ typescript
        \item typescript1.5 $\rightarrow$ typescript
        \item typescript1.4 $\rightarrow$ typescript
        \item typescript4.0 $\rightarrow$ typescript
        \item typescript2.3 $\rightarrow$ typescript
        \item typescript2.2 $\rightarrow$ typescript
        \item typescript3.8 $\rightarrow$ typescript
        \item typescript-3.6 $\rightarrow$ typescript
        \item cobol85 $\rightarrow$ cobol
        \item gnu-smalltalk $\rightarrow$ smalltalk
        \item dolphin-smalltalk $\rightarrow$ smalltalk
        \item visual-foxpro $\rightarrow$ foxpro
        \item delphi4php $\rightarrow$ radphp
        \item delphi-xe6 $\rightarrow$ delphi
        \item delphi-4 $\rightarrow$ delphi
        \item delphi-3 $\rightarrow$ delphi
        \item delphi-xe $\rightarrow$ delphi
        \item delphi-xe5 $\rightarrow$ delphi
        \item delphi-2007 $\rightarrow$ delphi
        \item delphi-2006 $\rightarrow$ delphi
        \item delphi-6 $\rightarrow$ delphi
        \item delphi-xe2 $\rightarrow$ delphi
        \item delphi7 $\rightarrow$ delphi
        \item delphi-xe7 $\rightarrow$ delphi
        \item delphi-10.4.2 $\rightarrow$ delphi
        \item delphi-11-alexandria $\rightarrow$ delphi
        \item delphi-10.2-tokyo $\rightarrow$ delphi
        \item delphi-10.3-rio $\rightarrow$ delphi
        \item delphi-xe8 $\rightarrow$ delphi
        \item delphi-5 $\rightarrow$ delphi
        \item delphi-10.4-sydney $\rightarrow$ delphi
        \item delphi-2010 $\rightarrow$ delphi
        \item delphi-2009 $\rightarrow$ delphi
        \item delphi-10-seattle $\rightarrow$ delphi
        \item delphi-2005 $\rightarrow$ delphi
        \item delphi-xe3 $\rightarrow$ delphi
        \item delphi-10.1-berlin $\rightarrow$ delphi
        \item delphi-xe4 $\rightarrow$ delphi
        \item fortran95 $\rightarrow$ fortran
        \item fortran77 $\rightarrow$ fortran
        \item fortran90 $\rightarrow$ fortran
        \item fortran2003 $\rightarrow$ fortran
        \item fortran2008 $\rightarrow$ fortran
        \item smlnj $\rightarrow$ sml
        \item mosml $\rightarrow$ sml
        \item visualj\# $\rightarrow$ j\#
        \item modula-2 $\rightarrow$ modula
        \item modula-3 $\rightarrow$ modula
        \item java-9 $\rightarrow$ java
        \item java-11 $\rightarrow$ java
        \item java-8 $\rightarrow$ java
        \item java-7 $\rightarrow$ java
        \item java-10 $\rightarrow$ java
        \item delphi.net $\rightarrow$ delphi-prism
        \item delphi-prism-2010 $\rightarrow$ delphi-prism
        \item msbuild-4.0 $\rightarrow$ msbuild
        \item msbuild-15 $\rightarrow$ msbuild
        \item msbuild-14.0 $\rightarrow$ msbuild
        \item typed-racket $\rightarrow$ racket
        \item powershell-5.1 $\rightarrow$ powershell
        \item powershell-6.0 $\rightarrow$ powershell
        \item powershell-2.0 $\rightarrow$ powershell
        \item powershell-3.0 $\rightarrow$ powershell
        \item powershell-4.0 $\rightarrow$ powershell
        \item powershell-5.0 $\rightarrow$ powershell
        \item powershell-7.0 $\rightarrow$ powershell
        \item powershell-1.0 $\rightarrow$ powershell
        \item actionscript-1 $\rightarrow$ actionscript
        \item actionscript-2 $\rightarrow$ actionscript
        \item actionscript-3 $\rightarrow$ actionscript
        \item c\#-3.0 $\rightarrow$ c\#
        \item c\#-4.0 $\rightarrow$ c\#
        \item perl5 $\rightarrow$ perl
        \item perl6 $\rightarrow$ perl
        \item raku $\rightarrow$ perl
        \item rakudo $\rightarrow$ perl
        \item jacl $\rightarrow$ tcl
        \item lua-5.3 $\rightarrow$ lua
        \item luaj $\rightarrow$ lua
        \item luajit $\rightarrow$ lua
        \item lua-5.2 $\rightarrow$ lua
        \item lua-5.1 $\rightarrow$ lua
        \item common-lisp $\rightarrow$ lisp
        \item elisp $\rightarrow$ lisp
        \item clisp $\rightarrow$ lisp
        \item autolisp $\rightarrow$ lisp
        \item lispworks $\rightarrow$ lisp
        \item lisp-2 $\rightarrow$ lisp
        \item maclisp $\rightarrow$ lisp
        \item gambas $\rightarrow$ basic
        \item freebasic $\rightarrow$ basic
        \item coldfusion-8 $\rightarrow$ coldfusion
        \item coldfusion-9 $\rightarrow$ coldfusion
        \item coldfusion-11 $\rightarrow$ coldfusion
        \item coldfusion-2016 $\rightarrow$ coldfusion
        \item coldfusion-2018 $\rightarrow$ coldfusion
        \item coldfusion-6 $\rightarrow$ coldfusion
        \item coldfusion-2021 $\rightarrow$ coldfusion
        \item cfml $\rightarrow$ coldfusion
        \item railo $\rightarrow$ coldfusion
        \item coldfusion-10 $\rightarrow$ coldfusion
        \item coldfusion-7 $\rightarrow$ coldfusion
        \item bluedragon $\rightarrow$ coldfusion
        \item c99 $\rightarrow$ c
        \item c11 $\rightarrow$ c
        \item c17 $\rightarrow$ c
    \end{itemize}
\end{multicols}

The Stack Overflow language set is as follows:\\
\begin{multicols}{4}
    \begin{itemize}
    \item	actionscript
    \item	ada
    \item	applescript
    \item	arden-syntax
    \item	asp.net
    \item	asp.net-core
    \item	assembly
    \item	ballerina
    \item	bash
    \item	batch-file
    \item	beanshell
    \item	c
    \item	c\#
    \item	c++
    \item	c++-cli
    \item	ceylon
    \item	clojure
    \item	cmake
    \item	cobol
    \item	coffeescript
    \item	coldfusion
    \item	crystal-lang
    \item	css
    \item	d
    \item	dart
    \item	delphi
    \item	delphi-prism
    \item	e
    \item	ecmascript-6
    \item	eiffel
    \item	el
    \item	erlang
    \item	f\#
    \item	fortran
    \item	foxpro
    \item	fscript
    \item	f-script
    \item	gml
    \item	go
    \item	gosu
    \item	groovy
    \item	hacklang
    \item	haskell
    \item	haxe
    \item	html
    \item	ioke
    \item	j
    \item	j\#
    \item	java
    \item	java-me
    \item	javascript
    \item	jelly
    \item	jquery
    \item	jscript
    \item	json
    \item	julia
    \item	kivy-language
    \item	kotlin
    \item	ksh
    \item	lisp
    \item	llvm
    \item	lua
    \item	lucee
    \item	makefile
    \item	masm
    \item	matlab
    \item	maxscript
    \item	mel
    \item	ml
    \item	modula
    \item	msbuild
    \item	nasm
    \item	nativescript
    \item	neo4j
    \item	netlogo
    \item	newlisp
    \item	node.js
    \item	objective-c
    \item	objective-c++
    \item	objective-j
    \item	ocaml
    \item	octave
    \item	pascal
    \item	perl
    \item	php
    \item	pike
    \item	powershell
    \item	processing
    \item	prolog
    \item	purescript
    \item	python
    \item	r
    \item	racket
    \item	radphp
    \item	rascal
    \item	react-native-android
    \item	rexx
    \item	ruby
    \item	ruby-on-rails
    \item	rust
    \item	scala
    \item	scheme
    \item	scilab
    \item	silverlight
    \item	silverlight-oob
    \item	smalltalk
    \item	sml
    \item	swift
    \item	tasm
    \item	tcl
    \item	tcsh
    \item	typescript
    \item	vala
    \item	vb.net
    \item	vba
    \item	vbscript
    \item	vhdl
    \item	wolfram-language
    \item	wolfram-mathematica
    \item	x10-language
    \item	xtend
    \item	yasm
    \item	yeti
    \item	zsh
    \end{itemize}
\end{multicols}

\subsection{Crafting Programming Language Sets} \label{sec:craft_prog_lang_sets}
We found 451 languages in GitHub. Both researchers classified which of them are Programming languages. There were 238 languages that both agreed were PL (Programming Language), and 93 that both agreed are non-PL. 25 languages were classified as PL by the first researcher and non-PL by the second, whereas 81 languages were classified as non-PL by the first researcher and PL by the second. In addition, the second researcher left 14 languages as undecided.\\
To reach an agreed upon list of PLs, the second researcher first reclassified the 14 undecided the way they were classified by the first. Then it was agreed that only languages that both researchers classified as PL would accepted as PL. We took this conservative approach to make sure we do not overestimate e.g. the number of multi-PL projects. We repeated the process for Stack Overflow languages. Out of the 124 languages identified in Stack Overflow, 74 had appeared in GitHub, so their status had already been decided. Of the other 50 there were 21 languages that both agreed were PL, and they were accepted as PL, whereas all the others remain non-PL.

\textbf{GitHub Programming Language Set:}\\
\begin{multicols}{4}
    \begin{itemize}
        \item	actionscript
        \item	ada
        \item	agda
        \item	al
        \item	angelscript
        \item	apex
        \item	apl
        \item	applescript
        \item	arc
        \item	arduino
        \item	asl
        \item	aspectj
        \item	assembly
        \item	astro
        \item	ats
        \item	autohotkey
        \item	autoit
        \item	ballerina
        \item	basic
        \item	beef
        \item	befunge
        \item	blitzbasic
        \item	blitzmax
        \item	bluespec
        \item	boo
        \item	boogie
        \item	brainfuck
        \item	brightscript
        \item	bro
        \item	c
        \item	c\#
        \item	c++
        \item	ceylon
        \item	chapel
        \item	cirru
        \item	clarion
        \item	clean
        \item	click
        \item	clips
        \item	clojure
        \item	cobol
        \item	coffeescript
        \item	coldfusion
        \item	common lisp
        \item	cool
        \item	crystal
        \item	cuda
        \item	cycript
        \item	cython
        \item	d
        \item	dart
        \item	dataweave
        \item	digital command language
        \item	dm
        \item	dogescript
        \item	dylan
        \item	e
        \item	ec
        \item	eiffel
        \item	elixir
        \item	elm
        \item	emacs lisp
        \item	emberscript
        \item	erlang
        \item	f\#
        \item	f*
        \item	factor
        \item	fancy
        \item	fantom
        \item	faust
        \item	fennel
        \item	flux
        \item	forth
        \item	fortran
        \item	freebasic
        \item	frege
        \item	futhark
        \item	game maker language
        \item	gaml
        \item	gams
        \item	gap
        \item	gdscript
        \item	genie
        \item	glsl
        \item	go
        \item	golo
        \item	gosu
        \item	groovy
        \item	hack
        \item	haskell
        \item	haxe
        \item	holyc
        \item	hy
        \item	idris
        \item	igor pro
        \item	io
        \item	ioke
        \item	isabelle
        \item	j
        \item	jasmin
        \item	java
        \item	javascript
        \item	jolie
        \item	julia
        \item	kit
        \item	kotlin
        \item	lasso
        \item	latte
        \item	lfe
        \item	limbo
        \item	livescript
        \item	logos
        \item	logtalk
        \item	lolcode
        \item	lua
        \item	m
        \item	macaulay2
        \item	mako
        \item	mathematica
        \item	matlab
        \item	maxscript
        \item	mercury
        \item	mirah
        \item	mlir
        \item	modelica
        \item	modula-3
        \item	moonscript
        \item	mql4
        \item	mql5
        \item	mupad
        \item	myghty
        \item	ncl
        \item	nemerle
        \item	nesc
        \item	netlogo
        \item	newlisp
        \item	nim
        \item	nit
        \item	nix
        \item	nu
        \item	objective-c
        \item	objective-c++
        \item	objective-j
        \item	objectscript
        \item	ocaml
        \item	odin
        \item	ooc
        \item	opa
        \item	opal
        \item	openscad
        \item	ox
        \item	oxygene
        \item	oz
        \item	parrot
        \item	pascal
        \item	pawn
        \item	perl
        \item	perl 6
        \item	php
        \item	picolisp
        \item	pike
        \item	plsql
        \item	pogoscript
        \item	pony
        \item	pov-ray sdl
        \item	powerbuilder
        \item	processing
        \item	prolog
        \item	propeller spin
        \item	purebasic
        \item	purescript
        \item	python
        \item	q
        \item	q\#
        \item	qt script
        \item	r
        \item	racket
        \item	ragel
        \item	raku
        \item	realbasic
        \item	reason
        \item	red
        \item	renderscript
        \item	renpy
        \item	rescript
        \item	rexx
        \item	ring
        \item	ruby
        \item	rust
        \item	sage
        \item	saltstack
        \item	sas
        \item	scala
        \item	scaml
        \item	scheme
        \item	scilab
        \item	shen
        \item	singularity
        \item	slash
        \item	smali
        \item	smalltalk
        \item	smpl
        \item	sourcepawn
        \item	sqlpl
        \item	squirrel
        \item	stan
        \item	standard ml
        \item	starlark
        \item	stata
        \item	supercollider
        \item	swift
        \item	systemverilog
        \item	tcl
        \item	tea
        \item	terra
        \item	turing
        \item	typescript
        \item	uno
        \item	unrealscript
        \item	urweb
        \item	v
        \item	vala
        \item	vba
        \item	vbscript
        \item	vcl
        \item	verilog
        \item	vhdl
        \item	visual basic
        \item	visual basic .net
        \item	webassembly
        \item	wisp
        \item	x10
        \item	xbase
        \item	xc
        \item	xojo
        \item	xonsh
        \item	xtend
        \item	zap
        \item	zeek
        \item	zenscript
        \item	zephir
        \item	zig
        \item	zil
    \end{itemize}
\end{multicols}

\textbf{Stack Overflow Programming Language Set:}\\
\begin{multicols}{4}
    \begin{itemize}
        \item	actionscript
        \item	ada
        \item	applescript
        \item	assembly
        \item	ballerina
        \item	beanshell
        \item	c
        \item	c\#
        \item	c++
        \item	c++-cli
        \item	ceylon
        \item	clojure
        \item	cobol
        \item	coffeescript
        \item	coldfusion
        \item	crystal-lang
        \item	d
        \item	dart
        \item	delphi
        \item	e
        \item	ecmascript-6
        \item	eiffel
        \item	el
        \item	erlang
        \item	f\#
        \item	fortran
        \item	fscript
        \item	go
        \item	gosu
        \item	groovy
        \item	hacklang
        \item	haskell
        \item	haxe
        \item	ioke
        \item	j
        \item	j\#
        \item	java
        \item	javascript
        \item	jscript
        \item	julia
        \item	kotlin
        \item	lisp
        \item	lua
        \item	lucee
        \item	masm
        \item	matlab
        \item	maxscript
        \item	mel
        \item	ml
        \item	modula
        \item	nasm
        \item	netlogo
        \item	newlisp
        \item	objective-c
        \item	objective-c++
        \item	objective-j
        \item	ocaml
        \item	pascal
        \item	perl
        \item	php
        \item	pike
        \item	processing
        \item	prolog
        \item	purescript
        \item	python
        \item	r
        \item	racket
        \item	rexx
        \item	ruby
        \item	rust
        \item	scala
        \item	scheme
        \item	scilab
        \item	smalltalk
        \item	sml
        \item	swift
        \item	tasm
        \item	tcl
        \item	typescript
        \item	vala
        \item	vb.net
        \item	vba
        \item	vbscript
        \item	vhdl
        \item	x10-language
        \item	xtend
    \end{itemize}
\end{multicols}

\section{Research Analysis}
\subsection{RQ1 - How common is multi-lingual development?}
In order to count the number of languages in each project, we group the projects by their language numbers. In order to count the number of \textbf{programming} languages in each project, we simply count the projects grouping them by their \textbf{programming language} numbers, discarding completely non-PL languages.

In Stack Overflow, to count the total number of questions and sum of all views is simply query:
\begin{singlespace}
\begin{verbatimblock}select count(*) As counts, sum(view_count) As views
from StackOverflowPosts
where post_type_id = 1;
\end{verbatimblock}
\end{singlespace}

multi-lingual related question is a question linked to at least one of the tables CrossLanguage, InteroperabilityTools, WrappersAndLanguagePorts, or contains at least 2 languages. We find them (by year) using the following query:
\begin{singlespace}
\begin{verbatimblock}
select strftime('%Y', creation_date) As year, 
    count(*) As counts, sum(view_count) As views
from StackOverflowPosts
where post_type_id = 1 and
(id in (select PostID from PostsToLanguages
    group by PostID having count(*) > 1) or
id in (select PostID from PostsToCrossLanguage) or
id in (select PostID from PostsToInteroperabilityTools) or
id in (select PostID from PostsToWrappersAndLanguagePorts))
group by strftime('%Y', creation_date);
\end{verbatimblock}
\end{singlespace}

\subsection{RQ2 - What are the dominant languages?}
The calculation in GitHub is very straight forward. Counting the existence of languages in projects, for different types of projects (single-lingual, multi-lingual and multi-PL projects).

For Stack Overflow, to count and calculate the views we execute the following:
\begin{singlespace}
\begin{verbatimblock}
-- count of languages, for questions that ask for
-- EXACTLY 1 language - single_lang_statistics_count
select Language, count(*) As counts, sum(view_count) As views
from PostsToLanguages, StackOverflowPosts
where PostID in (select PostID from PostsToLanguages
    group by PostID having count(*) = 1) AND PostID=id
group by Language;

-- count of languages, for questions that ask for
-- MORE than 1 language - multi_lang_statistics_count
select Language, count(*) As counts, sum(view_count) As views
from PostsToLanguages, StackOverflowPosts
where PostID in (select PostID from PostsToLanguages
    group by PostID having count(*) > 1) AND PostID=id
group by Language;

-- count of languages, for questions that ask for
-- MORE than 1 PROGRAMMING language -
-- prog_multi_lang_statistics_count / prog_multi_lang_statistics_view
select Language, count(*) As counts, sum(view_count) As views
from PostsToLanguages, StackOverflowPosts
where PostID in (select PostID from PostsToLanguages
    group by PostID having count(*) > 1) AND
    Language in (select Name from
    Languages where IsProgrammingLanguage = 1) AND PostID=id
group by Language;
\end{verbatimblock}
\end{singlespace}

\subsection{RQ3 - Which programming languages are mostly used together?}
Friendship binding can be calculated from GitHub project Metadata. By calculating if the size of the code in one language is at least 10\% of that in another language in the same project:
\begin{singlespace}
\begin{verbatimblock}
def __calc_project_friendship(self):
  def __cb_project_binding(projid, langs_size, created_at):
    if langs_size is None:
      return

    for l1 in langs_size:
      for l2 in langs_size:
        if l1 == l2:
          continue

        if (not is_programmable_language(l1)) or (not \
        is_programmable_language(l2)):
          continue

        
        if langs_size[l1] * 0.1 <= langs_size[l2] or \
          langs_size[l2] * 0.1 <= langs_size[l1]:
          key = sorted([l1, l2])
          key = (key[0], key[1])
          inc_entry(self.project_bindings, key)

  self.for_each_repo(__cb_project_binding)
\end{verbatimblock}
\end{singlespace}

Unlike project friendship binding, to calculate shell binding and interoperate binding, we must dive into the source code. To search and analyse the source code, we're using sourcegraph \cite{sourcegraph}, on the set of projects from \emph{sindresorhus' Awesome List} \cite{awesome_list}. In order to find the bindings we search using text, regular expressions and filters to find these bindings.

To detect \emph{shell binding} for each language we execute the following queries, where for each language, we're searching for their interpreter or JIT command. The \emph{lang} filter searches only files of the specified language:
\begin{singlespace}
\begin{verbatimblock}
[^a-z_]system( ['"`]{[COMMAND]}["'`\s] lang:C
[^a-z_]system( ['"`]{[COMMAND]}["'`\s] lang:C++
[^a-z_]system( ['"`]{[COMMAND]}["'`\s] lang:Objective-C
[^a-z_]system( ['"`]{[COMMAND]}["'`\s] lang:Objective-C++
os\.system( ['"`]{[COMMAND]}["'`\s] lang:Python
Process\.Start\([^\)]*['"`@]{[COMMAND]}['"`@\textbackslash{}s] lang:C#
Process\.Start\([^\)]*['"`@]{[COMMAND]}['"`@\textbackslash{}s]
                                            lang:"Visual Basic .NET" 
\\.FileName\textbackslash{}s*=.*{[COMMAND]}['"`@\textbackslash{}s]*;
                                                            lang:C#
(new ProcessBuilder)\([^\)]*{[COMMAND]} lang:Java
(ProcessBuilder)\([^\)]*{[COMMAND]} lang:Kotlin
(new ProcessBuilder)\([^\)]*{[COMMAND]} lang:Scala
(new ProcessBuilder)\([^\)]*{[COMMAND]} lang:Groovy
CommandLine\.parse\([ ]*[@]?"{[COMMAND]} lang:Java
CommandLine\.parse\([ ]*[@]?"{[COMMAND]} lang:Kotlin
CommandLine\.parse\([ ]*[@]?"{[COMMAND]} lang:Scala
CommandLine\.parse\([ ]*[@]?"{[COMMAND]} lang:Groovy
CreateProcess[AW]?\([^\)]+{[COMMAND]} lang:C
CreateProcess[AW]?\([^\)]+{[COMMAND]} lang:C++
exec.Command\([^\)]*["|`]{[COMMAND]} lang:Go
{[COMMAND]}[^"'`\)]*["'`]\.execute\(\) lang:Groovy # Groovy
(Unix\.open_process_in \"{[COMMAND]}) lang:OCaml
pexpect\.spawn\([^"'\)]+["']{[COMMAND]} lang:Python
spawn\([^"'\)]+["']{[COMMAND]} lang:JavaScript
spawn\([^"'\)]+["']{[COMMAND]} lang:TypeScript
thread::spawn\([^"'\)]+["']{[COMMAND]} lang:Rust
process\.spawn\([^"'\)]+["']{[COMMAND]} lang:Lua
check_output\([^"'\)]+["'][^\)]*['"]{[COMMAND]}['"\s] lang:Python
check_call\([^\)]?['"\s]{[COMMAND]}['"\s] lang:Python
run_cmd![^\\n]+{[COMMAND]}[^\\n\)]*\) lang:Rust
run_cmd! [\{{\(]{{[^\}}\)]* {[COMMAND]} lang:Rust
Popen\([^\)"']+["']{[COMMAND]}['"\s] lang:Python
[^a-z]exec\([^\)"']+['"]{[COMMAND]}['"\s] lang:Python
\.exec\([^\)"']*['"]{[COMMAND]}\s?[^'"]*['"] lang:Java
[^a-z]exec\([^\)"']+['"]{[COMMAND]}['"\s] lang:Go
[^a-z]exec\([^\)"']+['"]{[COMMAND]}['"\s] lang:C++
[^a-z]exec\([^\)"']+['"]{[COMMAND]}['"\s] lang:Rust
ProcessStartInfo\([^\)]+{[COMMAND]} lang:C#
ProcessStartInfo\([^\)]+{[COMMAND]} lang:"Visual Basic .NET" 
processcall\([^\)]+'{[COMMAND]}[^a-z] lang:Dart
subprocess\.call\([^\)]+'{[COMMAND]}[^a-z] lang:Python 
runtime\.exec\([^\)]+{[COMMAND]} lang:Java
runtime\.exec\([^\)]+{[COMMAND]} lang:Scala
runtime\.exec\([^\)]+{[COMMAND]} lang:Groovy
runtime\.exec\([^\)]+{[COMMAND]} lang:Kotlin
runtime\.getruntime\(\)\.exec\([^\)]+{[COMMAND]} lang:Java
runtime\.getruntime\(\)\.exec\([^\)]+{[COMMAND]} lang:Scala
runtime\.getruntime\(\)\.exec\([^\)]+{[COMMAND]} lang:Groovy
runtime\.getruntime\(\)\.exec\([^\)]+{[COMMAND]} lang:Kotlin
ProcessExecutor \([^\)]+{[COMMAND]} lang:Java
ProcessExecutorParams\.ofcommand\([^\)]+{[COMMAND]} lang:Java
ProcessExecutorParams\.ofcommand\([^\)]+{[COMMAND]} lang:Scala
ProcessExecutorParams\.ofcommand\([^\)]+{[COMMAND]} lang:Groovy
ProcessExecutorParams\.ofcommand\([^\)]+{[COMMAND]} lang:Kotlin
ProcessExecutorParams\.builder\(\)\.addcommand {[COMMAND]} lang:Java
ProcessExecutorParams\.builder\(\)\.addcommand {[COMMAND]} lang:Scala
ProcessExecutorParams\.builder\(\)\.addcommand {[COMMAND]} lang:Groovy
ProcessExecutorParams\.builder\(\)\.addcommand {[COMMAND]} lang:Kotlin
execSync\([^'"`]*['`"]{[COMMAND]}[^'`"]*['`"] lang:JavaScript
execSync\([^'"`]*['`"]{[COMMAND]}[^'`"]*['`"] lang:TypeScript
command::new\([^\)]*{[COMMAND]} lang:Rust
\end{verbatimblock}
\end{singlespace}

For each results, we're checking if it is not a single-line comment, which is removed. Each result signifies a shell binding from the specified \emph{lang} filter to the language the [COMMAND] belongs to. The list of COMMANDs by language is as follows:
\begin{itemize}
    \item	python - python, python3, jython, cython, bpython, numba
    \item	.net - mono, csharp
    \item	java - java
    \item	groovy - groovy
    \item	scala - scala
    \item	kotlin - kotlin
    \item	go - [\string^a-z]go\textbackslash srun[\string^a-z]
    \item	lua - lua, luajit
    \item	rust - cargo\textbackslash smiri (run\textbar test)
    \item	ruby - ruby, jruby
    \item	r - Rscript
    \item	perl - perl, perl5, perl5i, perl6, rakudo
    \item	haskell - runhaskell
    \item	javascript - js, v8, k8, jjs
    \item	nodejs - nodejs, mx js
    \item	lisp - clisp, scheme, lisp, tclsh
    \item	powershell - powershell, powershell\.exe
    \item	shell - /bin/bash, bash\.exe, cmd\.exe, /bin/tcsh, /bin/zsh
\end{itemize}

In order to analyse interoperability binding, we analysed 173 interoperability tools. For each language, we search source code using the libraries to call the current language. Also, we search for \emph{"extern"}-ing functions to other languages, based on our research. For example, if we find code snippets of externing functions to "C", we understand this code can be called from a different language (e.g. C++).\\
The list of patterns we're looking for interoperate binding:

\begin{longtblr}[
  caption = {Interoperability Tools Detection Patterns},
]{
  width = \linewidth,
  colspec = {Q[9]Q[12]Q[23]},
  cell{2}{1} = {r=14}{},
  cell{5}{2} = {r=4}{},
  cell{11}{2} = {r=2}{},
  cell{16}{1} = {r=3}{},
  cell{16}{2} = {r=3}{},
  cell{19}{1} = {r=9}{},
  cell{24}{2} = {r=4}{},
  cell{28}{1} = {r=12}{},
  cell{30}{2} = {r=2}{},
  cell{35}{2} = {r=3}{},
  cell{40}{1} = {r=5}{},
  cell{40}{2} = {r=4}{},
  cell{45}{1} = {r=4}{},
  cell{45}{2} = {r=3}{},
  cell{49}{1} = {r=4}{},
  cell{49}{2} = {r=3}{},
  cell{55}{1} = {r=12}{},
  cell{55}{2} = {r=3}{},
  cell{58}{2} = {r=3}{},
  cell{68}{1} = {r=9}{},
  cell{77}{1} = {r=9}{},
  cell{86}{1} = {r=4}{},
  cell{86}{2} = {r=3}{},
  cell{91}{1} = {r=10}{},
  cell{92}{2} = {r=3}{},
  cell{95}{2} = {r=2}{},
  cell{101}{1} = {r=27}{},
  cell{101}{2} = {r=10}{},
  cell{111}{2} = {r=2}{},
  cell{113}{2} = {r=4}{},
  cell{117}{2} = {r=2}{},
  cell{128}{1} = {r=4}{},
  cell{130}{2} = {r=2}{},
  cell{132}{1} = {r=11}{},
  cell{132}{2} = {r=5}{},
  cell{139}{2} = {r=3}{},
  cell{143}{1} = {r=3}{},
  cell{143}{2} = {r=3}{},
  cell{146}{1} = {r=16}{},
  cell{147}{2} = {r=2}{},
  cell{152}{2} = {r=2}{},
  cell{154}{2} = {r=2}{},
  cell{157}{2} = {r=2}{},
  cell{162}{1} = {r=21}{},
  cell{180}{2} = {r=2}{},
  vlines,
  hline{1-2,16,19,28,40,45,49,53-55,67-68,77,86,90-91,101,128,132,143,146,162,183} = {-}{},
  hline{3-5,9-11,13-15,20-24,29-30,32-35,38-39,44,48,52,58,61-66,69-76,78-85,89,92,95,97-100,111,113,117,119-127,129-130,137-139,142,147,149-152,154,156-157,159-161,163-180,182} = {2-3}{},
  hline{6-8,12,17-18,25-27,31,36-37,41-43,46-47,50-51,56-57,59-60,87-88,93-94,96,102-110,112,114-116,118,131,133-136,140-141,144-145,148,153,155,158,181} = {3}{},
}
\textbf{Target Language} & \textbf{Interoperability Tool} & \textbf{Regular Expression}                                                                                                                                  \\
Python                   & go-python                      & github\textbackslash{}.com/sbinet/go-python                                                                                                                  \\
                         & ironpython                     & Python\textbackslash{}.CreateRuntime                                                                                                                         \\
                         & pycall                         & (require 'pycall)                                                                                                                                            \\
                         & python-c-api                   & Py\_InitializeEx\textbackslash{}(                                                                                                                            \\
                         &                                & Py\_Initialize\textbackslash{}(                                                                                                                              \\
                         &                                & python\textbackslash{}.hpp                                                                                                                                   \\
                         &                                & python\textbackslash{}.h                                                                                                                                     \\
                         & pythoninterpreter              & {new (PythonInterpreter\textbar{}\\JythonInterpreter\textbar{}CPythonInterpreter)}                                                                           \\
                         & pythonkit                      & import PythonKit lang:Swift                                                                                                                                  \\
                         & python-script-engine           & \textbackslash{}.getEngineByName\textbackslash{}(\textbackslash{}s*["'\`{}]+python["'\`{}]+\textbackslash{}s*\textbackslash{})                               \\
                         &                                & \textbackslash{}.getEngineByExtension\textbackslash{}(\textbackslash{}s*["'\`{}]+py["'\`{}]+\textbackslash{}s*\textbackslash{})                              \\
                         & rubypython                     & require 'rubypython                                                                                                                                          \\
                         & rustpython                     & use rustpython\_vm                                                                                                                                           \\
                         & scalapy                        & ScalaPy                                                                                                                                                      \\
.NET                     & .net-load                      & CLRCreateInstance\textbackslash{}(\textbackslash{})                                                                                                          \\
                         &                                & CorBindToRuntimeEx\textbackslash{}(\textbackslash{})                                                                                                         \\
                         &                                & hostfxr\_initialize\_for\_runtime\_config                                                                                                                    \\
.NET                     & mono-load                      & mono\_jit\_init\textbackslash{}(                                                                                                                             \\
                         & .net-load                      & \#using mscorlib\textbackslash{}.dll                                                                                                                         \\
                         & csml                           & csstub csml                                                                                                                                                  \\
                         & ironpython                     & clr.AddReference                                                                                                                                             \\
                         & jni4net                        & using net\textbackslash{}.sf\textbackslash{}.jni4net                                                                                                         \\
                         & Extern                         & \textbackslash{}{[}DllExport                                                                                                                                 \\
                         &                                & dllexport                                                                                                                                                    \\
                         &                                & \textbackslash{}{[}ComVisible\textbackslash{}(true\textbackslash{})\textbackslash{}]                                                                         \\
                         &                                & \textless{}ComClass\textbackslash{}([\^\textgreater{}]*\textgreater{}                                                                                        \\
Java                     & pyjnius                        & jnius.*autoclass                                                                                                                                             \\
                         & scala-java-conversion          & scala.collection.JavaConversions                                                                                                                             \\
                         & jni4net                        & imports net\textbackslash{}.sf\textbackslash{}.jni4net                                                                                                       \\
                         &                                & Bridge\textbackslash{}.CreateJVM                                                                                                                             \\
                         & py4j                           & py4j JavaGateway                                                                                                                                             \\
                         & javabridge                     & javabridge.start\_vm\textbackslash{}(                                                                                                                        \\
                         & jpype                          & jpype startJVM                                                                                                                                               \\
                         & java-native-interface          & JNI\_CreateJavaVM                                                                                                                                            \\
                         &                                & \#include jni\textbackslash{}.h                                                                                                                              \\
                         &                                & JNIEXPORT                                                                                                                                                    \\
                         & cygnus                         & \#include\textbackslash{}sgcj/cni.h                                                                                                                          \\
                         & Extern                         & import com\textbackslash{}.sun\textbackslash{}.jna lang:Java                                                                                                 \\
Groovy                   & groovy-script-engine           & new GroovyEngine\textbackslash{}(\textbackslash{})                                                                                                           \\
                         &                                & groovy\textbackslash{}.lang\textbackslash{}.GroovyShell                                                                                                      \\
                         &                                & \textbackslash{}.getEngineByName\textbackslash{}(\textbackslash{}s*["'\`{}]+groovy["'\`{}]+\textbackslash{}s*\textbackslash{})                               \\
                         &                                & {\textbackslash{}.getEngineByExtension\textbackslash{}\\(\textbackslash{}s*["'\`{}]+groovy["'\`{}]+\textbackslash{}s*\textbackslash{})}                      \\
                         & Extern                         & import com\textbackslash{}.sun\textbackslash{}.jna lang:Groovy                                                                                               \\
Scala                    & scala-script-engine            & \textbackslash{}.getEngineByExtension\textbackslash{}(\textbackslash{}s*["'\`{}]+sc["'\`{}]+\textbackslash{}s*\textbackslash{})                              \\
                         &                                & \textbackslash{}.getEngineByExtension\textbackslash{}(\textbackslash{}s*["'\`{}]+scala["'\`{}]+\textbackslash{}s*\textbackslash{})                           \\
                         &                                & \textbackslash{}.getEngineByName\textbackslash{}(\textbackslash{}s*["'\`{}]+scala["'\`{}]+\textbackslash{}s*\textbackslash{})                                \\
                         & Extern                         & import com\textbackslash{}.sun\textbackslash{}.jna lang:Scala                                                                                                \\
Kotlin                   & kotlin-script-engine           & \textbackslash{}.getEngineByName\textbackslash{}(\textbackslash{}s*["'\`{}]+kotlin["'\`{}]+\textbackslash{}s*\textbackslash{})                               \\
                         &                                & \textbackslash{}.getEngineByExtension\textbackslash{}(\textbackslash{}s*["'\`{}]+kt["'\`{}]+\textbackslash{}s*\textbackslash{})                              \\
                         &                                & \textbackslash{}.getEngineByExtension\textbackslash{}(\textbackslash{}s*["'\`{}]+kts["'\`{}]+\textbackslash{}s*\textbackslash{})                             \\
                         & Extern                         & import com\textbackslash{}.sun\textbackslash{}.jna lang:Kotlin                                                                                               \\
Go                       & gopherjs                       & syscall/js                                                                                                                                                   \\
Go                       & Extern                         & \^{}//export.*\textbackslash{}nfunc lang:Go                                                                                                                  \\
Lua                      & lua-script-engine              & \textbackslash{}.getEngineByName\textbackslash{}(\textbackslash{}s*["'\`{}]+lua["'\`{}]+\textbackslash{}s*\textbackslash{})                                  \\
                         &                                & \textbackslash{}.getEngineByName\textbackslash{}(\textbackslash{}s*["'\`{}]+luaj["'\`{}]+\textbackslash{}s*\textbackslash{})                                 \\
                         &                                & \textbackslash{}.getEngineByExtension\textbackslash{}(\textbackslash{}s*["'\`{}]+lua["'\`{}]+\textbackslash{}s*\textbackslash{})                             \\
                         & lua-api                        & luaL\_newstate\textbackslash{}(\textbackslash{})                                                                                                             \\
                         &                                & lua\_open\textbackslash{}(\textbackslash{})                                                                                                                  \\
                         &                                & new Lua\textbackslash{}(\textbackslash{})                                                                                                                    \\
                         & luaj                           & import org\textbackslash{}.luaj                                                                                                                              \\
                         & lua.vm.js                      & require\textbackslash{}('lua\textbackslash{}.vm\textbackslash{}.js'\textbackslash{})                                                                         \\
                         & dynamic-lua                    & DynamicLua\textbackslash{}(\textbackslash{})                                                                                                                 \\
                         & gopher-lua                     & github\textbackslash{}.com/yuin/gopher-lua                                                                                                                   \\
                         & go-lua                         & github\textbackslash{}.com/Shopify/go-lua                                                                                                                    \\
                         & goluajit                       & github\textbackslash{}.com/antonvolkoff/goluajit                                                                                                             \\
Rust                     & Extern                         & \#\textbackslash{}{[}no\_mangle\textbackslash{}] lang:Rust                                                                                                   \\
Ruby                     & ruby-script-engine             & \textbackslash{}.getEngineByName\textbackslash{}(\textbackslash{}s*["'\`{}]+ruby["'\`{}]+\textbackslash{}s*\textbackslash{})                                 \\
                         & ruby-script-engine             & \textbackslash{}.getEngineByName\textbackslash{}(\textbackslash{}s*["'\`{}]+jruby["'\`{}]+\textbackslash{}s*\textbackslash{})                                \\
                         & ruby-script-engine             & \textbackslash{}.getEngineByExtension\textbackslash{}(\textbackslash{}s*["'\`{}]+rb["'\`{}]+\textbackslash{}s*\textbackslash{})                              \\
                         & m2cgen-ruby                    & RubyInterpreter\textbackslash{}(\textbackslash{})                                                                                                            \\
                         & ruby-api                       & ruby\_init\textbackslash{}(\textbackslash{})                                                                                                                 \\
                         & ruby-api                       & execute\_rb                                                                                                                                                  \\
                         & jruby-java-interop             & org\textbackslash{}.jruby                                                                                                                                    \\
                         & coldruby                       & require\textbackslash{}('ruby'\textbackslash{})                                                                                                              \\
                         & go-mruby                       & github\textbackslash{}.com/mitchellh/go-mruby                                                                                                                \\
R                        & embedded-r                     & Rf\_initEmbeddedR\textbackslash{}(                                                                                                                           \\
                         & embedded-r                     & \#include Rembedded\textbackslash{}.h                                                                                                                        \\
                         & embedded-r                     & \#include RInside\textbackslash{}.h                                                                                                                          \\
                         & rcaller                        & RCaller\textbackslash{}.create\textbackslash{}(                                                                                                              \\
                         & renjin                         & new\textbackslash{}sRenjinScriptEngine\textbackslash{}(                                                                                                      \\
                         & rpy2                           & (from\textbar{}import)\textbackslash{}srpy[2]?                                                                                                               \\
                         & statistics-r                   & use\textbackslash{}sStatistics::R                                                                                                                            \\
                         & rinruby                        & require\textbackslash{}s"rinruby"                                                                                                                            \\
                         & ocaml-r                        & ocaml-r                                                                                                                                                      \\
Perl                     & perl-script-engine             & \textbackslash{}.getEngineByName\textbackslash{}(\textbackslash{}s*["'\`{}]+perl["'\`{}]+\textbackslash{}s*\textbackslash{})                                 \\
                         &                                & \textbackslash{}.getEngineByName\textbackslash{}(\textbackslash{}s*["'\`{}]+perl5["'\`{}]+\textbackslash{}s*\textbackslash{})                                \\
                         &                                & \textbackslash{}.getEngineByExtension\textbackslash{}(\textbackslash{}s*["'\`{}]+pl["'\`{}]+\textbackslash{}s*\textbackslash{})                              \\
                         & campher                        & github\textbackslash{}.com/bradfitz/campher/perl                                                                                                             \\
Perl                     & perlapi                        & \#include perl\textbackslash{}.h                                                                                                                             \\
Haskell                  & Extern                         & (foreign import ccall)                                                                                                                                       \\
                         & haskell-script-engine          & \textbackslash{}.getEngineByName\textbackslash{}(\textbackslash{}s*["'\`{}]+haskell["'\`{}]+\textbackslash{}s*\textbackslash{})                              \\
                         &                                & \textbackslash{}.getEngineByName\textbackslash{}(\textbackslash{}s*["'\`{}]+jaskell["'\`{}]+\textbackslash{}s*\textbackslash{})                              \\
                         &                                & \textbackslash{}.getEngineByExtension\textbackslash{}(\textbackslash{}s*["'\`{}]+hs["'\`{}]+\textbackslash{}s*\textbackslash{})                              \\
                         & c-haskell-ffi                  & \#include HsFFI.h                                                                                                                                            \\
                         &                                & void hs\_init\textbackslash{}(IntPtr                                                                                                                         \\
                         & hapy                           & (import\textbar{}from)[\^a-zA-Z]*HaPy[ ]                                                                                                                     \\
                         & jaskell                        & import Jaskell                                                                                                                                               \\
                         &                                &                                                                                                                                                              \\
                         &                                &                                                                                                                                                              \\
JavaScript               & javascript-script-engine       & {\textbackslash{}.getEngineByName\textbackslash{}\\(\textbackslash{}s*["'\`{}]+javascript["'\`{}]+\textbackslash{}s*\textbackslash{})}                       \\
                         &                                & \textbackslash{}.getEngineByName\textbackslash{}(\textbackslash{}s*["'\`{}]+nashorn["'\`{}]+\textbackslash{}s*\textbackslash{})                              \\
                         &                                & \textbackslash{}.getEngineByName\textbackslash{}(\textbackslash{}s*["'\`{}]+rhino["'\`{}]+\textbackslash{}s*\textbackslash{})                                \\
                         &                                & {\textbackslash{}.getEngineByName\textbackslash{}\\(\textbackslash{}s*["'\`{}]+ecmascript["'\`{}]+\textbackslash{}s*\textbackslash{})}                       \\
                         &                                & {\textbackslash{}.getEngineByName\textbackslash{}\\(\textbackslash{}s*["'\`{}]+typescript["'\`{}]+\textbackslash{}s*\textbackslash{})}                       \\
                         &                                & \textbackslash{}.getEngineByName\textbackslash{}(\textbackslash{}s*["'\`{}]+graal.js["'\`{}]+\textbackslash{}s*\textbackslash{})                             \\
                         &                                & \textbackslash{}.getEngineByName\textbackslash{}(\textbackslash{}s*["'\`{}]+js["'\`{}]+\textbackslash{}s*\textbackslash{})                                   \\
                         &                                & \textbackslash{}.getEngineByExtension\textbackslash{}(\textbackslash{}s*["'\`{}]+js["'\`{}]+\textbackslash{}s*\textbackslash{})                              \\
                         &                                & (new V8Engine) lang:Java                                                                                                                                     \\
                         &                                & getEngineByName\textbackslash{}("javascript"\textbackslash{})                                                                                                \\
                         & v8.net                         & new V8ScriptEngine                                                                                                                                           \\
                         &                                & (new V8Engine) lang:C\#                                                                                                                                      \\
                         & c-js                           & ExecuteJavaScript                                                                                                                                            \\
                         &                                & \#include \textless{}emscripten[\^\textgreater{}]*\textgreater{}                                                                                             \\
                         &                                & \#include[\^\textless{}]*\textless{}v8[\^\textgreater{}]*\textgreater{}                                                                                      \\
                         &                                & \#include jsapi.h                                                                                                                                            \\
                         & quickjs                        & quickjs\textbackslash{}.h                                                                                                                                    \\
                         &                                & (from\textbar{}import) quickjs                                                                                                                               \\
                         & js2py                          & (from\textbar{}import) js2py                                                                                                                                 \\
                         & pyminirace                     & (from\textbar{}import) py\_mini\_racer                                                                                                                       \\
                         & pyv8                           & (from\textbar{}import) PyV8                                                                                                                                  \\
                         & smgo                           & github\textbackslash{}.com/realint/monkey                                                                                                                    \\
                         & gov8                           & github\textbackslash{}.com/idada/go-v8                                                                                                                       \\
                         & elsa                           & github.com/elsaland/elsa                                                                                                                                     \\
                         & scalajs                        & enablePlugins\textbackslash{}(ScalaJSPlugin\textbackslash{})                                                                                                 \\
                         & rhino                          & import org\textbackslash{}.mozilla\textbackslash{}.javascript                                                                                                \\
                         & c-js                           & createV8Runtime\textbackslash{}(\textbackslash{})                                                                                                            \\
JavaScript               & execjs                         & require ["']execjs["'] lang:Ruby                                                                                                                             \\
                         & cpp-js                         & JavaScript::Runtime-new\textbackslash{}(\textbackslash{}))                                                                                                   \\
                         & jscocoa                        & import JSCocoa/JSCocoa\textbackslash{}.h                                                                                                                     \\
                         &                                & import "JSCocoa\textbackslash{}.h"                                                                                                                           \\
Lisp                     & lisp-script-engine             & \textbackslash{}.getEngineByName\textbackslash{}(\textbackslash{}s*["'\`{}]+lisp["'\`{}]+\textbackslash{}s*\textbackslash{})                                 \\
                         &                                & \textbackslash{}.getEngineByName\textbackslash{}(\textbackslash{}s*["'\`{}]+tcl["'\`{}]+\textbackslash{}s*\textbackslash{})                                  \\
                         &                                & \textbackslash{}.getEngineByName\textbackslash{}(\textbackslash{}s*["'\`{}]+scheme["'\`{}]+\textbackslash{}s*\textbackslash{})                               \\
                         &                                & \textbackslash{}.getEngineByExtension\textbackslash{}(\textbackslash{}s*["'\`{}]+lisp["'\`{}]+\textbackslash{}s*\textbackslash{})                            \\
                         &                                & \textbackslash{}.getEngineByExtension\textbackslash{}(\textbackslash{}s*["'\`{}]+lsp["'\`{}]+\textbackslash{}s*\textbackslash{})                             \\
                         & lisp-c                         & \#include ecl/ech\textbackslash{}.h                                                                                                                          \\
                         & hy                             & import hy[\^a-zA-Z] lang:Python                                                                                                                              \\
                         & lisp-c                         & \#include \textless{}emscripten[\^\textgreater{}]*\textgreater{}                                                                                             \\
                         &                                & Lisp\textbackslash{}.new lang:Go                                                                                                                             \\
                         &                                & Lisp\textbackslash{}.new lang:Perl                                                                                                                           \\
                         & cslisp                         & using CSLisp                                                                                                                                                 \\
PowerShell               & cs-powershell                  & PowerShell\textbackslash{}.Create\textbackslash{}(\textbackslash{})                                                                                          \\
                         &                                & PowerShell\textbackslash{}.openSession\textbackslash{}(\textbackslash{})                                                                                     \\
                         &                                & RunspaceFactory.CreateRunspace                                                                                                                               \\
C                        & ctypes                         & (from ctypes)\textbar{}(import ctypes)                                                                                                                       \\
                         & pinvoke                        & \textbackslash{}{[}dllimport[\^\textbackslash{}]]*]                                                                                                          \\
                         &                                & \textless{}dllimport[\^\textgreater{}]*\textgreater{}                                                                                                        \\
                         & cgo                            & import\textbackslash{}s"C" lang:Go                                                                                                                           \\
                         & rust-ffi                       & \#\textbackslash{}{[}link extern[\^\{]*\{{[}\^\}]*fn lang:Rust                                                                                               \\
                         & java-native-interface          & JNIIMPORT                                                                                                                                                    \\
                         & perl-ffi                       & require[\^'"]*['"]ffi lang:Perl                                                                                                                              \\
                         &                                & require[\^\textbackslash{}(\textbackslash{}n]*([\^"'\textbackslash{}n]*['"]ffi[\^'"\textbackslash{}n]*["'][\^\textbackslash{})\textbackslash{}n]*) lang:Perl \\
                         & ruby-ffi                       & require[\^\textbackslash{}(\textbackslash{}n]*([\^"'\textbackslash{}n]*['"]ffi[\^'"\textbackslash{}n]*["'][\^\textbackslash{})\textbackslash{}n]*) lang:Ruby \\
                         &                                & require[\^'"]*['"]ffi lang:Ruby                                                                                                                              \\
                         & fiddle                         & require[\^'"]*['"]fiddle                                                                                                                                     \\
                         & lua-ffi                        & require[\^'"]*['"]ffi lang:Lua                                                                                                                               \\
                         &                                & require[\^\textbackslash{}(\textbackslash{}n]*([\^"'\textbackslash{}n]*['"]ffi[\^'"\textbackslash{}n]*["'][\^\textbackslash{})\textbackslash{}n]*) lang:Lua  \\
                         & luaffi                         & require[\^\textbackslash{}(\textbackslash{}n]*([\^"\textbackslash{}n]*"luaffi[\^"\textbackslash{}n]*"[\^\textbackslash{})\textbackslash{}n]*)                \\
                         & lua-cffi                       & require[\^\textbackslash{}(\textbackslash{}n]*([\^"\textbackslash{}n]*"cffi[\^"\textbackslash{}n]*"[\^\textbackslash{})\textbackslash{}n]*) lang:Lua         \\
                         & luaffifb                       & require[\^\textbackslash{}(\textbackslash{}n]*([\^"\textbackslash{}n]*"luaffifb[\^"\textbackslash{}n]*"[\^\textbackslash{})\textbackslash{}n]*)              \\
C                        & ffi-platypus                   & use FFI::Platypus lang:Perl                                                                                                                                  \\
                         & sbffi                          & require[\^\textbackslash{}(\textbackslash{}n]*([\^"'\textbackslash{}n]*['"]sbffi[\^'"\textbackslash{}n]*["'][\^\textbackslash{})\textbackslash{}n]*)         \\
                         & nodeffi                        & require[\^\textbackslash{}(\textbackslash{}n]*([\^"'\textbackslash{}n]*['"]node-ffi[\^'"\textbackslash{}n]*["'][\^\textbackslash{})\textbackslash{}n]*)      \\
                         & lisp-ffi                       & ffi:def-c-call-out lang:"common lisp"                                                                                                                        \\
                         & matlabffi                      & calllib lang:MATLAB                                                                                                                                          \\
                         & swig                           & \%module lang:SWIG                                                                                                                                           \\
                         & dart:ffi                       & import 'dart:ffi'                                                                                                                                            \\
                         & php-ffi                        & FFI::load lang:PHP                                                                                                                                           \\
                         & racket-ffi                     & (require ffi/unsafe) lang:Racket                                                                                                                             \\
                         & julia-ffi                      & ccall\textbackslash{}( lang:Julia                                                                                                                            \\
                         & java-native                    & (public\textbackslash{}sstatic\textbackslash{}snative\textbackslash{}s)\textbar{}(public\textbackslash{}snative\textbackslash{}s) lang:Java                  \\
                         & scala-native                   & (public\textbackslash{}sstatic\textbackslash{}snative\textbackslash{}s)\textbar{}(public\textbackslash{}snative\textbackslash{}s) lang:Scala                 \\
                         & kotlin-native                  & (public\textbackslash{}sstatic\textbackslash{}snative\textbackslash{}s)\textbar{}(public\textbackslash{}snative\textbackslash{}s) lang:Kotlin                \\
                         & groovy-native                  & (public\textbackslash{}sstatic\textbackslash{}snative\textbackslash{}s)\textbar{}(public\textbackslash{}snative\textbackslash{}s) lang:Groovy                \\
                         & jsm                            & {Components\textbackslash{}.utils\textbackslash{}.import\textbackslash{}\\("resource://gre/modules/ctypes\textbackslash{}.jsm lang:JavaScript}               \\
                         & r-ffi                          & \textbackslash{}scfunction\textbackslash{}(c\textbackslash{}( lang:R                                                                                         \\
                         & rust-ffi                       & (use libc::) lang:Rust                                                                                                                                       \\
                         & haskell-ffi                    & (foreign import ccall) lang:Haskell                                                                                                                          \\
                         & com-iunknown                   & (public IUnknown\textbackslash{}s)                                                                                                                           \\
                         &                                & (HRESULT STDMETHODCALLTYPE QueryInterface)                                                                                                                   \\
                         & corba                          & \#include[\^\textbackslash{}\textbackslash{}n]+CORBA\textbackslash{}.h                                                                                       
\end{longtblr}

Once we have found all the bindings and their libraries, we plot two types of graphs, "Interoperability bindings \textbf{including} the intermediary" and "Interoperability bindings \textbf{excluding} the intermediary". As the analysis of the tools show, some tools are using intermediate languages (In most cases, C, for one tool, C and Java). The "including" graph plots the indirection, such that if a tool is used from $L_1$ to $L_2$ using indirection with language $L_i$, the graph shows $L_1 \rightarrow L_i \rightarrow L_2$. In the "excluding" graph we skip the intermediate language.

To plot the graphs we're going over the detected bindings, and create two CSVs for each graph, one for vertices and one for edges. Last, we're using Tikz to plot the graphs from the CSVs.

\subsection{RQ4 - What are the common interoperability tools}
To detect the popularity of interoperability tools and popular interoperability tool type, we count the usage of each tool and its type.

\subsection{RQ5 - How difficult is code interoperability?}
In order to find relevant multi-lingual discussions and issues, we're searching the text for keywords and regular expressions. To minimize the chance of false positives, for each match of the keywords and patterns we search for a programming language that \textbf{does not} exist in the project's list of languages. To avoid matching containing text, we are using to following regular expression to ensure the keywords are made of complete text:
\begin{verbatim}
(?=(?:^|[^a-zA-Z0-9_/])({})(?:[^a-zA-Z0-9_/]|$))
\end{verbatim}
The keywords that invoke searching for another programming language not existing in the project are:
\begin{itemize}
    \item wrapper
    \item (port (to|it)|a port|porting)
    \item interop|interoperate|interoperating
    \item bind|binding
    \item jni|jna (project must not have JVM languages to satisfy detection)
\end{itemize}

\section{Acquired GitHub Repository Fields}
\subsection{Project}\label{sec:github_project_metadata_fields}
\begin{table}[H]
\centering
\begin{tabular}{|l|l|l|}
\hline
\multicolumn{1}{|l|}{Name} & \multicolumn{1}{l|}{Type} & \multicolumn{1}{l|}{Description}                         \\ \hline
id                         & int                       & Repository ID                                            \\
name                       & string                    & Name of repository                                       \\
full\_name                 & string                    & Full name of repository                                  \\
url                        & string                    & URL of repository                                        \\
languages\_url             & string                    & URL to acquire list of languages                         \\
created\_at                & datetime                  & Time of creation                                         \\
size                       & int                       & Size of project (in bytes)                               \\
forks\_count               & int                       & Number of forks of the project                           \\
watchers                   & int                       & Number of watchers to receive updates                    \\
stargazers\_count          & int                       & Number of stars                                          \\
description                & string                    & Description of the project                               \\
read\_me                   & string                    & URL to the main readme.md (if there is one)              \\
languages\_sizes           & map                       & List of languages in the project and their size in bytes \\ \hline
\end{tabular}
\end{table}

\subsection{Issues \& Discussions}
\begin{table}[H]
\centering
\begin{tabular}{|l|l|l|}
\hline
\multicolumn{1}{|l|}{Name} & \multicolumn{1}{l|}{Type} & \multicolumn{1}{l|}{Description} \\ \hline
id                         & int                       & Issue ID                         \\
creation\_date             & datetime                  & Time of creation                 \\
is\_resolved               & boolean                   & Is issue resolved                \\
tags                       & string{[}{]}              & Tags                             \\
title                      & string                    & Title of issue                   \\ \hline
\end{tabular}
\end{table}

\begin{table}[H]
\centering
\begin{tabular}{|l|l|l|}
\hline
\multicolumn{1}{|l|}{Name} & \multicolumn{1}{l|}{Type} & \multicolumn{1}{l|}{Description}         \\ \hline
id                         & string                    & Discussion ID                            \\
title                      & string                    & Title of discussion                      \\
body\_text                 & string                    & Body of discussion                       \\
state                      & string                    & State of discussion (e.g. Open, Close)   \\
comments                   & string{[}{]}              & Comments (threads) of discussions        \\
labels                     & string{[}{]}              & Labels (tags) attached to the discussion \\ \hline
\end{tabular}
\end{table}

\section{Acquired Stack Overflow Post Fields}\label{sec:so_posts_fields}
\begin{table}[H]
\centering
\begin{tabular}{|l|l|l|}
\hline
\multicolumn{1}{|l|}{Name} & \multicolumn{1}{l|}{Type} & \multicolumn{1}{l|}{Description}        \\ \hline
id                         & int                       & Repository ID                           \\
post\_type\_id             & int                       & Type of Post (question or answer)       \\
accepted\_answer\_id       & int                       & Answer - is the post an accepted answer \\
parent\_id                 & int                       & Answer - the ID of the question         \\
creation\_date             & datetime                  & Date of creation                        \\
deletion\_date             & datetime                  & Date of deletion                        \\
view\_count                & int                       & Number of question views                \\
body                       & string                    & Body of post                            \\
title                      & string                    & Question - title                        \\
tags                       & string{[}{]}              & Tags delimited by comma                 \\
answer\_count              & int                       & Question - how many answers posts       \\
favorite\_count            & int                       & Question - is marked as "favoriate"     \\
closed\_date               & datetime                  & Question - is question closed and when  \\ \hline
\end{tabular}
\end{table}

\section{Stack Overflow Tags}\label{sec:so_tags}
\subsection{Wrappers \& Language ports} \label{langport}
\begin{multicols}{4}
    \begin{itemize}
        \item lua-api
        \item luabind
        \item pyopengl
        \item qtopengl
        \item pyqt
        \item ruby-llvm
        \item llvm-fs
        \item pcap4j
        \item llvm-py
        \item pyqt4
        \item jslint4java
        \item im4java
        \item ghost4j
        \item syslog4j
        \item mpi4py
        \item qt4dotnet
        \item winscp-net
        \item pycurl
        \item pylucene
        \item gitpython
        \item pyffmpeg
        \item pyopencl
        \item pyportmidi
        \item celerity
        \item wxpython
        \item pywin32
        \item pyhook
        \item m2crypto
        \item pcap.net
        \item cocoalibspotify-2.0
        \item jnetpcap
        \item pyfftw
        \item ghostscript.net
        \item curlpp
        \item python-fu
        \item pcre.net
        \item fgsl
        \item ffmpeg-python
        \item jrubyfx
        \item jruby-openssl
        \item openssl-net
        \item pyopenssl
        \item php-openssl
        \item pyspark
        \item discord.py
        \item pyqt5
        \item pymongo
        \item openpyxl
        \item hadoop-yarn
        \item rhadoop
        \item pyqtgraph
        \item wxhaskell
        \item wxperl
        \item wxruby
        \item wxerlang
        \item wxlua
        \item wxphp
        \item wxmpl
        \item wxgo
        \item tensorflow.js
        \item opencv-python
        \item opencvsharp
        \item pyspark-dataframes
        \item sparkr
        \item sparklyr
        \item spark-java
        \item rselenium
        \item selenium-ruby
        \item soap4r
        \item p4python
        \item log4r
        \item log4cxx
        \item log4cpp
        \item log4perl
        \item log4cplus
        \item log4javascript
        \item log4php
        \item log4net
        \item linq4j
        \item ice4j
        \item log4qt
        \item docx4j.net
        \item porting
        \item log4jna
    \end{itemize}
\end{multicols}

\subsection{Cross-Language} \label{xlang}
\begin{multicols}{4}
    \begin{itemize}
        \item rnetlogo
        \item bridging-header
        \item objc-bridging-header
        \item c2hs
        \item hsc2hs
        \item c2hsc
        \item cgo
        \item gccgo
        \item ctypes
        \item jsctypes
        \item python-ctypes
        \item cython
        \item cythonize
        \item derelict3
        \item f2c
        \item fortran-iso-c-binding
        \item gobject-introspection
        \item inline-assembly
        \item javah
        \item java-native-interface
        \item jpl
        \item lua-api
        \item lua-userdata
        \item pyobjc
        \item python-c-api
        \item python-cffi
        \item ruby-c-extension
        \item swig
        \item libffi
        \item vb.net-to-c\#
        \item c\#-to-vb.net
        \item python4delphi
        \item luainterface
        \item luabridge
        \item nlua
        \item jnlua
        \item luajava
        \item lua-c++-connection
        \item c\#-to-f\#
        \item fable-f\#
        \item f\#-giraffe
        \item cmake-js
        \item js-of-ocaml
        \item erlang-nif
        \item jni4net
        \item py4j
        \item interprolog
        \item boost-python
        \item f2py
        \item rpy2
        \item python.net
        \item lispyscript
        \item interop
        \item com-interop
        \item language-interoperability
        \item office-interop
        \item dart-js-interop
        \item kotlin-interop
        \item pkcs11interop
        \item interopservices
        \item kotlin-js-interop
        \item jruby-rack
        \item jruby-win32ole
        \item blazor-jsinterop
        \item java-interop
        \item gwt-jsinterop
        \item j-interop
        \item scala-java-interop
        \item clojure-java-interop
        \item jruby-java-interop
        \item kotlin-java-interop
        \item cinterop
        \item primary-interop-assembly
        \item javascript-interop
        \item clojurescript-javascript-interop
        \item ballerina-java-interop
        \item pybind11
        \item jna
        \item jnativehook
        \item jnaerator
        \item djnativeswing
        \item autoit-c\#-wrapper
        \item android-jsinterface
    \end{itemize}
\end{multicols}

\subsection{Interoperability Tools}\label{sec:interop_tools_tags}
\begin{multicols}{4}
    \begin{itemize}
        \item lua-api
        \item luabind
        \item rnetlogo
        \item pynetlogo
        \item bridging-header
        \item objc-bridging-header
        \item c2hs
        \item hsc2hs
        \item c2hsc
        \item cgo
        \item gccgo
        \item ctypes
        \item jsctypes
        \item python-ctypes
        \item cython
        \item cythonize
        \item derelict3
        \item f2c
        \item fortran-iso-c-binding
        \item gobject-introspection
        \item inline-assembly
        \item javah
        \item java-native-interface
        \item jpl
        \item lua-userdata
        \item pyobjc
        \item python-c-api
        \item python-cffi
        \item ruby-c-extension
        \item swig
        \item lbffi
        \item vb.net-to-c\#
        \item c\#-to-vb.net
        \item python4delphi
        \item luainterface
        \item luabridge
        \item nlua
        \item jnlua
        \item luajava
        \item luaj
        \item fable-f\#
        \item f\#-giraffe
        \item js-of-ocaml
        \item erlang-nif
        \item jni4net
        \item py4j
        \item interprolog
        \item boost-python
        \item f2py
        \item rpy2
        \item python.net
        \item lispyscript
        \item com-interop
        \item interopservices
        \item kotlin-js-interop
        \item jruby-win32ole
        \item gwt-jsinterop
        \item scala-java-interop
        \item clojure-java-interop
        \item jruby-java-interop
        \item kotlin-java-interop
        \item clojurescript-javascript-interop
        \item ballerina-java-interop
        \item pybind11
        \item jna
        \item jnativehook
        \item jnaerator
        \item djnativeswing
        \item autoit-c\#-wrapper
        \item android-jsinterface
        \item pycall
        \item pythoninterpreter
        \item ironpython
        \item scalapy
        \item jpype
        \item pyjnius
        \item scalajs
        \item gopherjs
        \item renjin
        \item perlapi
        \item rcaller
        \item haskell-ffi
        \item quickjs
        \item pinvoke
        \item fiddle
        \item rjava
        \item rhino
        \item llvm
    \end{itemize}
\end{multicols}

\subsection{Other interoperability tools analysed (\textbf{Not tags!})}
\begin{multicols}{4}
    \begin{itemize}
        \item pythonkit
        \item rubypython
        \item go-python
        \item python-script-engine
        \item java-script-engine
        \item groovy-script-engine
        \item scala-script-engine
        \item kotlin-script-engine
        \item lua-script-engine
        \item ruby-script-engine
        \item perl-script-engine
        \item haskell-script-engine
        \item javascript-script-engine
        \item lisp-script-engine
        \item .net-load
        \item mono-load
        \item csml
        \item scala-java-conversion
        \item lua.vm.js
        \item dynamic-lua
        \item gopher-lua
        \item go-lua
        \item goluajit
        \item rustpython
        \item m2cgen-ruby
        \item ruby-api
        \item rufus-lua
        \item coldruby
        \item go-mruby
        \item embedded-r
        \item statistics::r
        \item rinruby
        \item ocaml-r
        \item campher
        \item hapy
        \item jaskell
        \item c-js
        \item clearscript
        \item v8.net
        \item js2py
        \item pyminirace
        \item mini\_racer
        \item therubyracer
        \item pyv8
        \item otto
        \item gov8
        \item v8go
        \item elsa
        \item lisp-c
        \item hy
        \item cslisp
        \item cs-powershell
        \item rust-ffi
        \item ruby-ffi
        \item perl-ffi
        \item lua-ffi
        \item lua-cffi
        \item luaffi
        \item luaffifb
        \item ffi-platypus
        \item sbffi
        \item nodeffi
        \item lisp-ffi
        \item matlabffi
        \item cygnus
        \item dart:ffi
        \item php-ffi
        \item racket-ffi
        \item julia-ffi
        \item java-native
        \item groovy-native
        \item kotlin-native
        \item scala-native
        \item jsm
        \item jscocoa
        \item rubycocoa
        \item macruby
        \item luacore
        \item r-ffi
        \item c-haskell-ffi
        \item execjs
        \item therubyrhino
        \item duktape
        \item javascriptcore
    \end{itemize}
\end{multicols}

\section{Extra Figures}
The following figures were removed from the main paper due to space limitation.

\subsection{GitHub}
\begin{figure}[H]
  \centering
  \captionsetup[subfigure]{width=\linewidth}
  \begin{subfigure}[h]{0.27\textwidth}
    \includegraphics[scale=0.27,width=\textwidth]{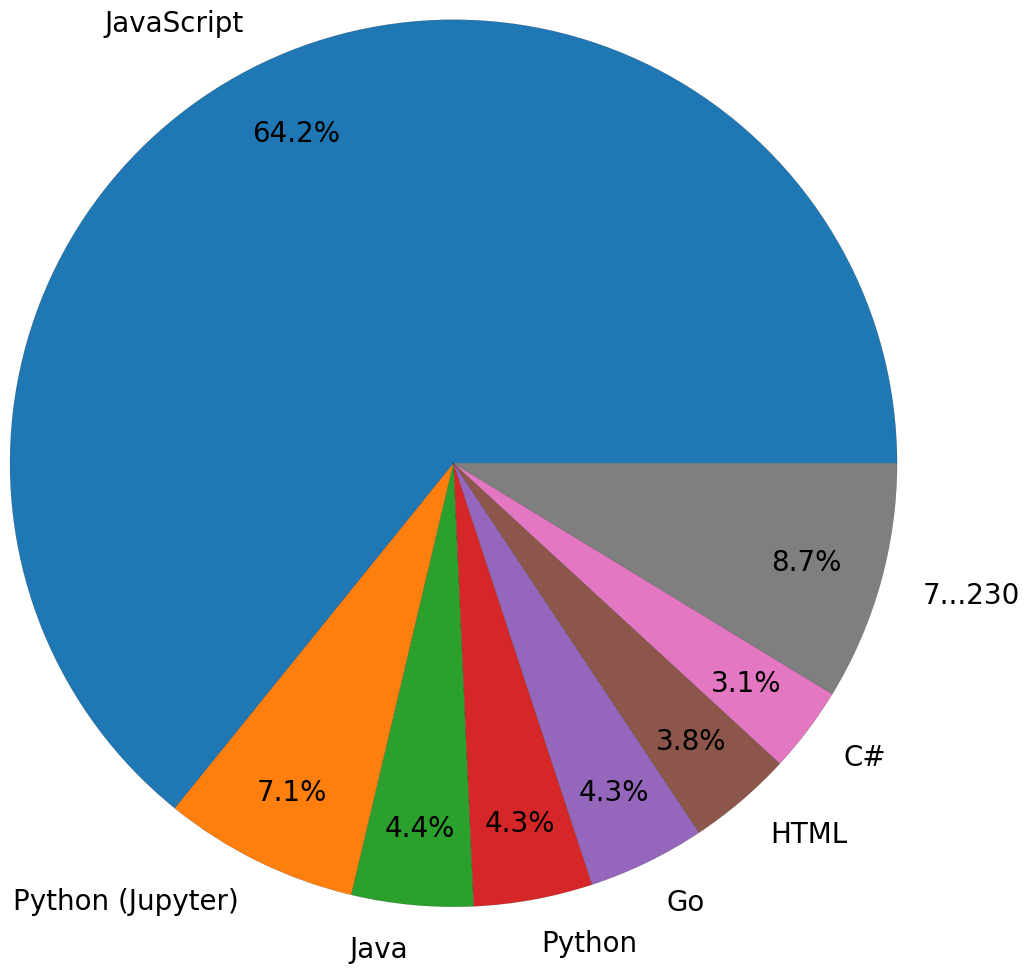}
    \caption{GitHub languages in single-language projects by size}
    \label{fig:github-single-language-language-bytes-percentage}
  \end{subfigure}
  \begin{subfigure}[h]{0.3\textwidth}
    \includegraphics[scale=0.3,width=\textwidth]{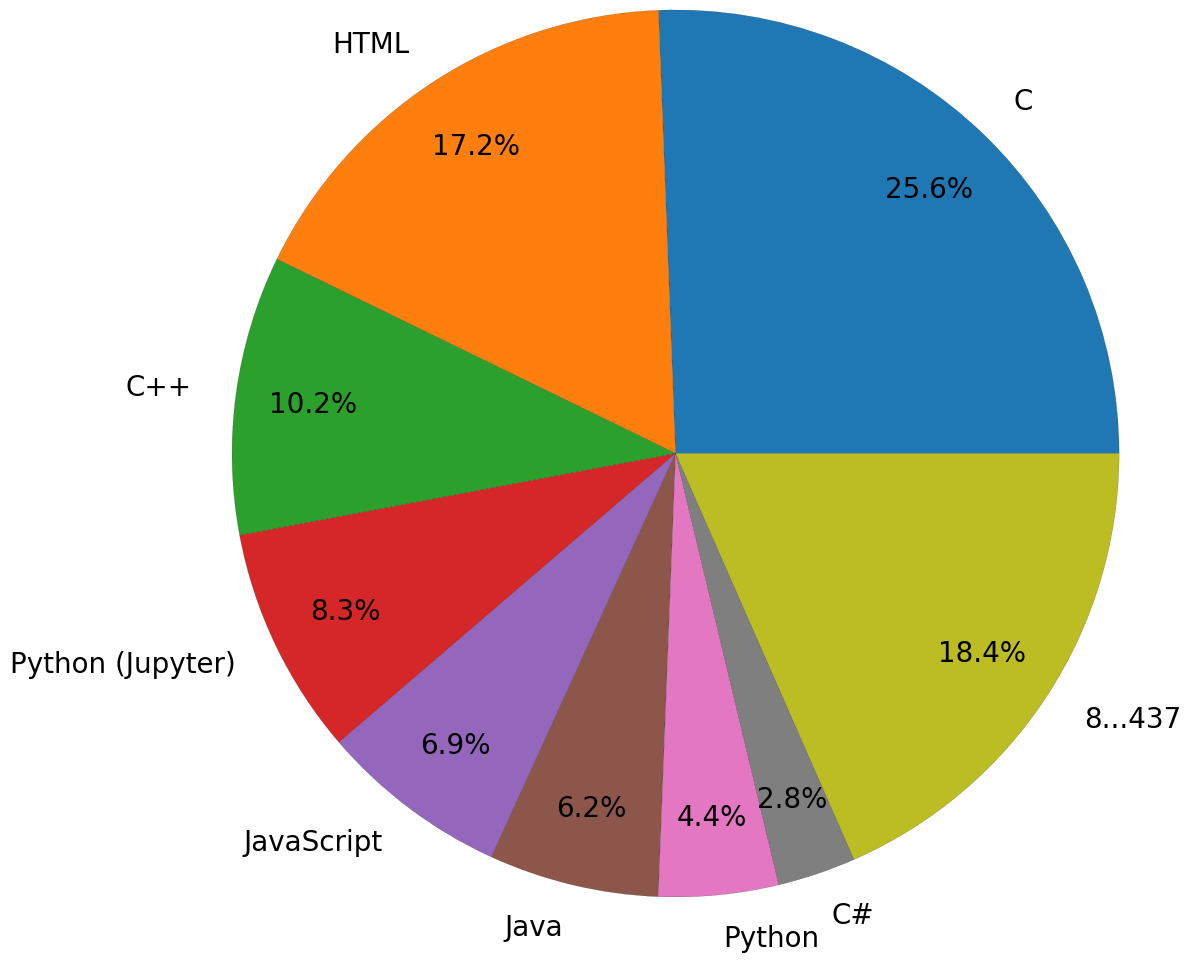}
    \caption{GitHub languages in multi-lingual projects by size}
  \label{fig:github-multi-lingual-language-bytes-percentage}
  \end{subfigure}
  \begin{subfigure}[h]{0.31\textwidth}
    \includegraphics[scale=0.31,width=\textwidth]{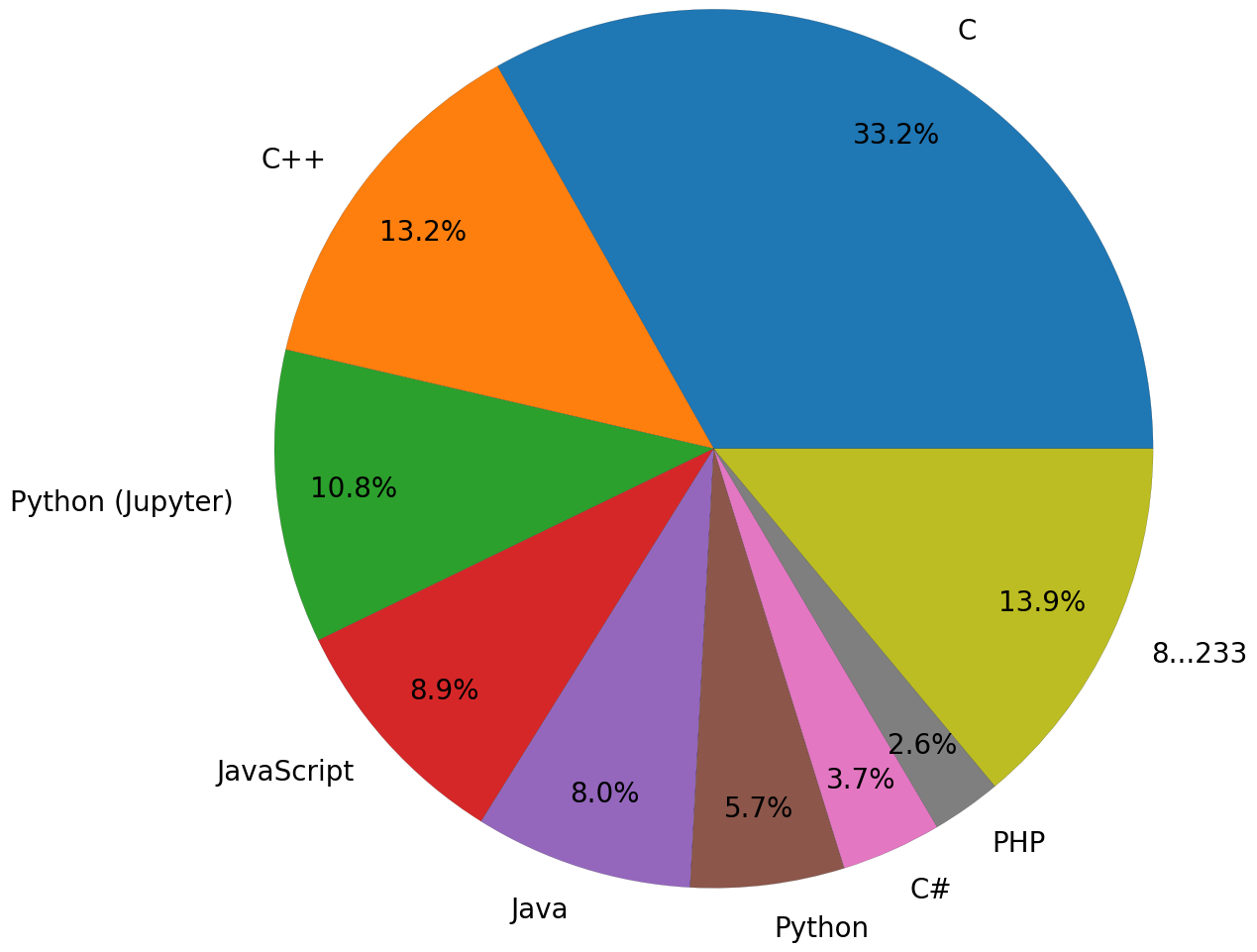}
    \caption{GitHub languages in multi-PL projects by size}
    \label{fig:github-prog-multi-lingual-language-bytes-percentage}
  \end{subfigure}
  \caption{GitHub Languages by size and count}
  \begin{subfigure}[h]{\textwidth}
    \includegraphics[scale=1,width=\textwidth]{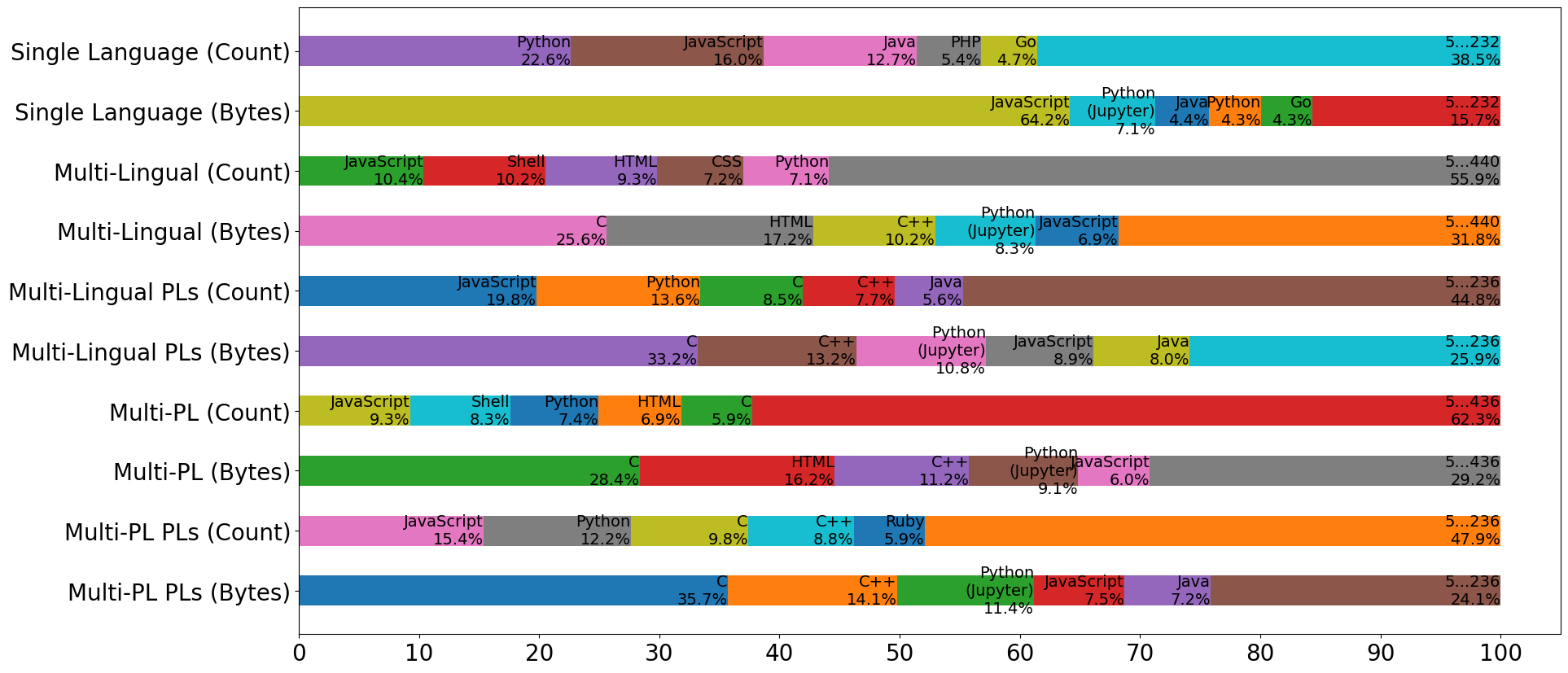}
    \caption{GitHub languages in multi-PL projects by size}
    \label{fig:github-prog-multi-lingual-language-bytes-size}
  \end{subfigure}
  \caption{GitHub Languages by size and count}
\end{figure}

\subsection{Stack Overflow}
\begin{figure}[H]
    \centering
    \captionsetup[subfigure]{width=2\linewidth}
    \begin{subfigure}[h]{0.45\textwidth}
        \includegraphics[scale=0.8,width=\textwidth]{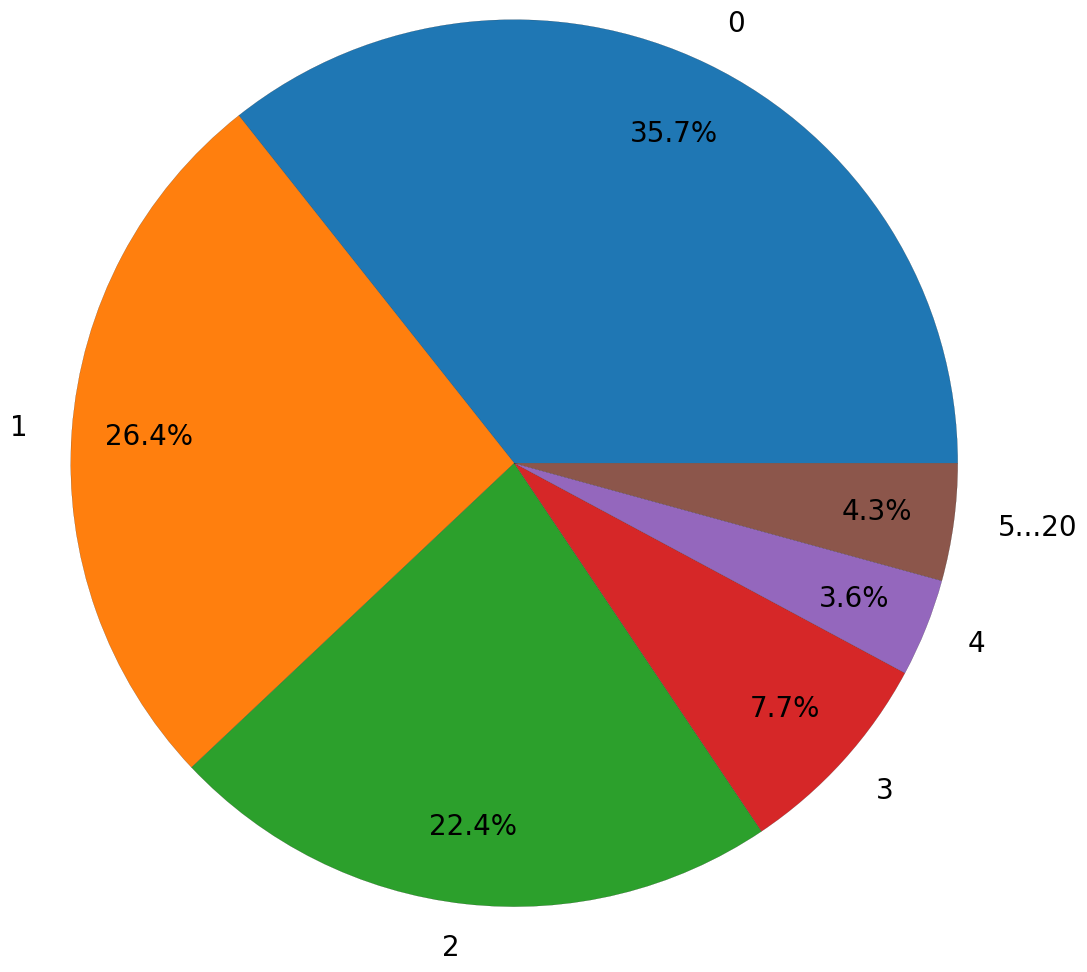}
        \caption{Questions count}
        \label{fig:stackoverflow-number-of-relevant}
    \end{subfigure}
    \begin{subfigure}[h]{0.45\textwidth}
        \includegraphics[scale=0.8,width=\textwidth]{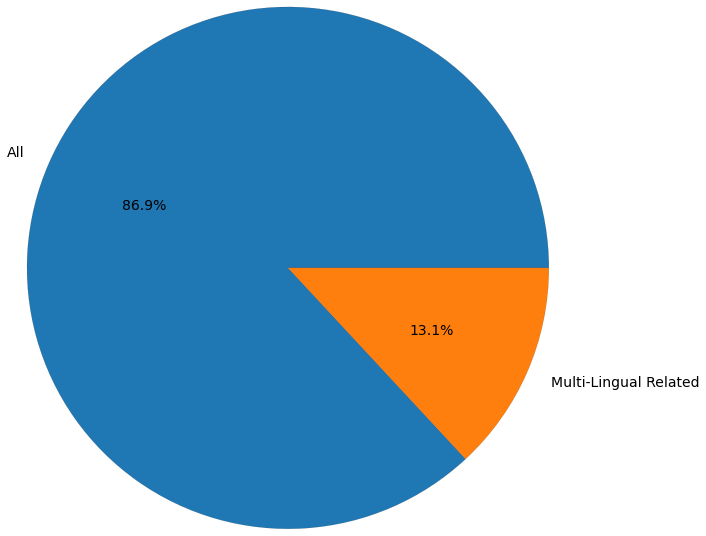}
        \caption{Questions views}
        \label{fig:stackoverflow-views-of-relevant}
    \end{subfigure}
    \caption{Stack Overflow Language Counts}
\end{figure}

\begin{figure}[h]
  \centering
  \begin{subfigure}[h]{0.35\textwidth}
    \includegraphics[scale=0.35,width=\textwidth]{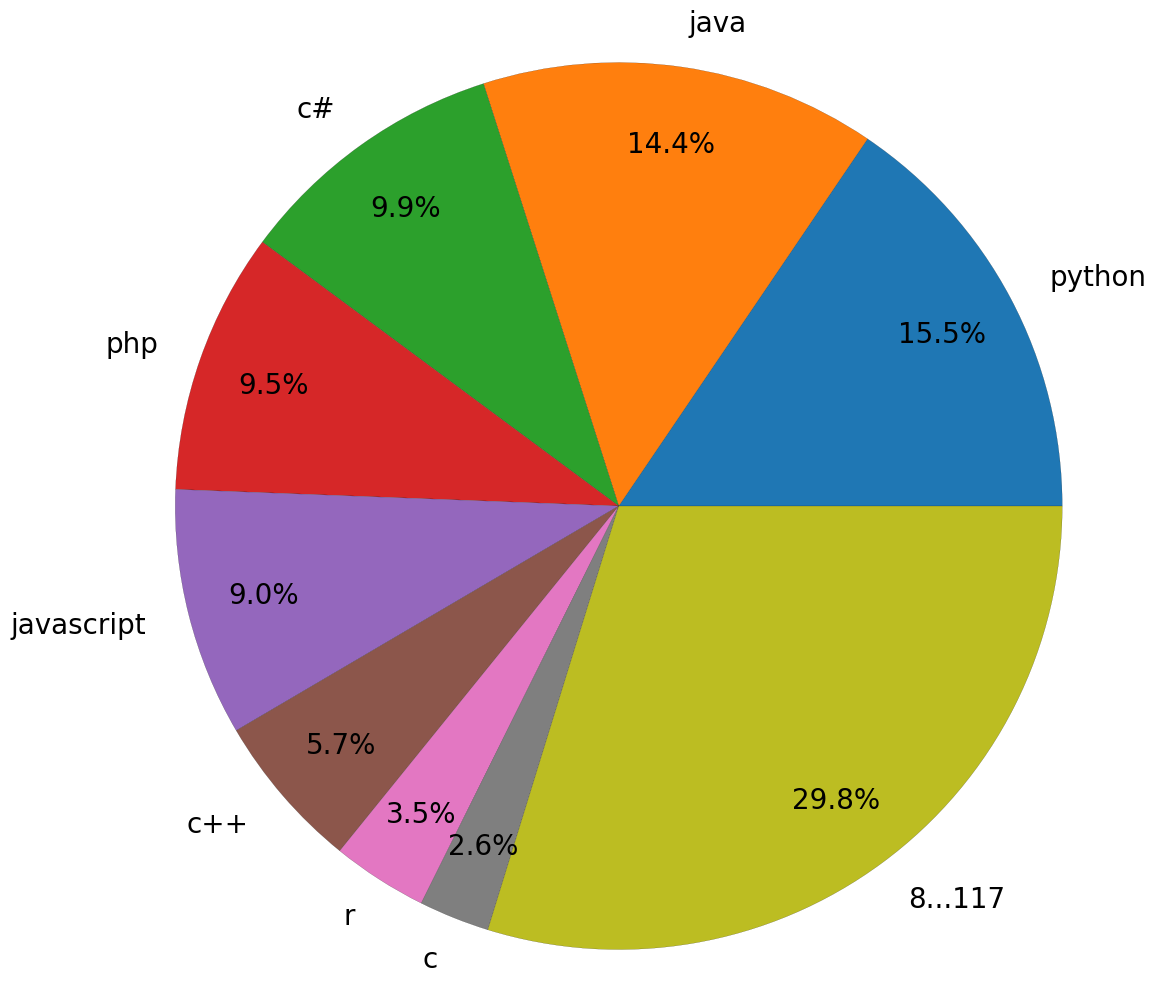}
    \caption{Stack Overflow languages in single-language questions by count}
    \label{fig:stackoverflow-single-language-count-percentage}
  \end{subfigure}
  \begin{subfigure}[h]{0.3\textwidth}
    \includegraphics[scale=0.3,width=\textwidth]{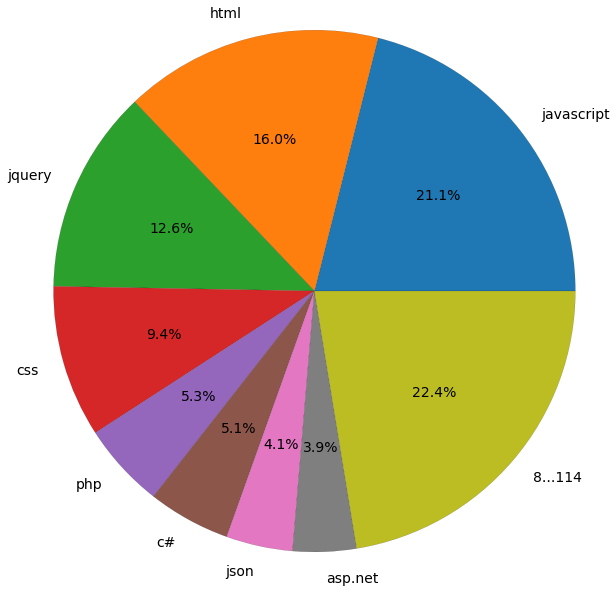}
    \caption{Stack Overflow languages in multi-lingual questions by count}
    \label{fig:stackoverflow-multi-lingual-count-percentage}
  \end{subfigure}
  \begin{subfigure}[h]{0.3\textwidth}
    \includegraphics[scale=0.3,width=\textwidth]{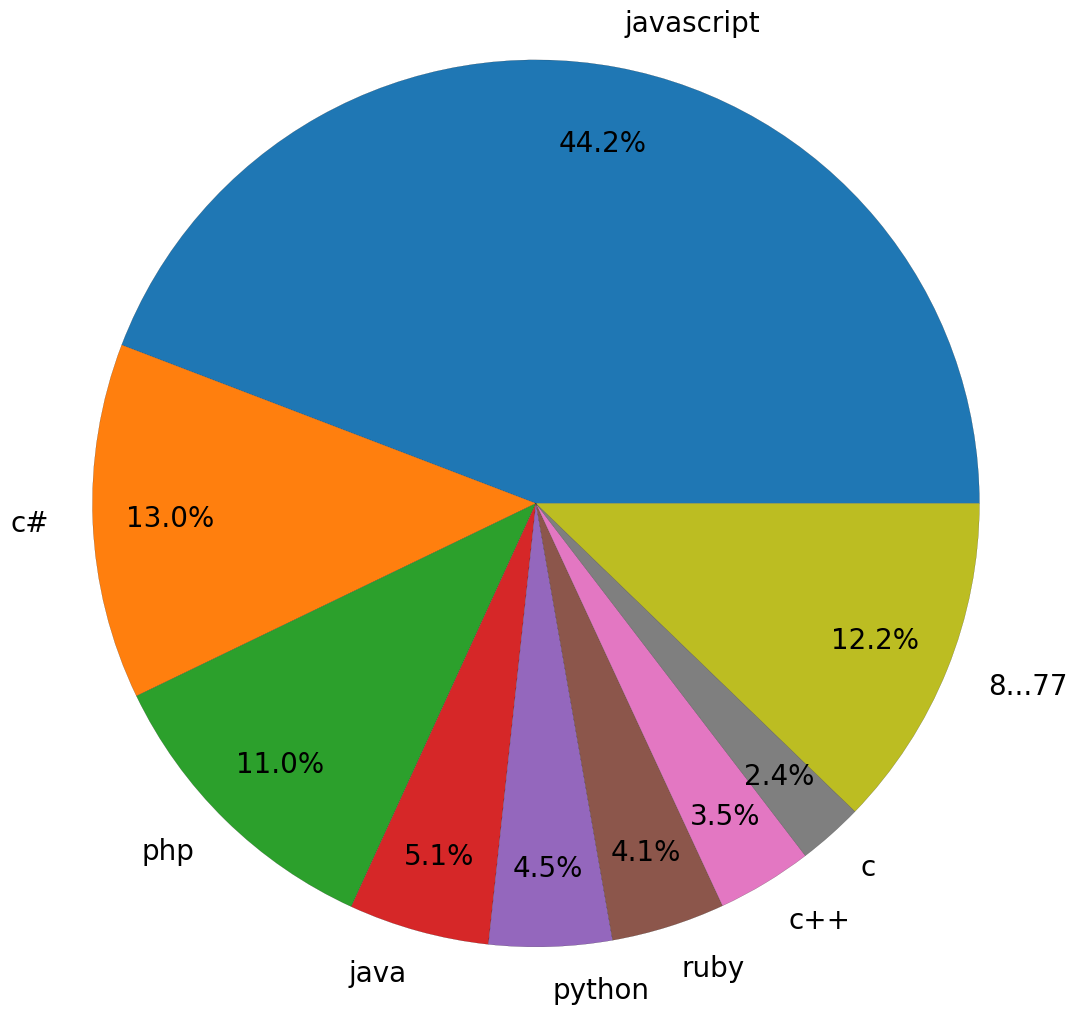}
    \caption{Stack Overflow languages in multi-PL questions by count}
    \label{fig:stackoverflow-prog-multi-lingual-count-percentage}
  \end{subfigure}
  \caption{Languages by size and count}
\end{figure}

\bibliography{supplement}